\documentclass[structabstract]{aa} 
\usepackage{comment}
\usepackage{graphicx}
\usepackage{txfonts}
\usepackage{natbib}
\usepackage{longtable}
\usepackage{lscape}
\usepackage{multirow}
\usepackage{pdfpages}
\usepackage[colorlinks=true, allcolors=blue]{hyperref}

\bibpunct{(}{)}{;}{a}{}{,}

\newcommand{\kms}{km\,s$^{-1}$}

\newcommand{\degree}{$^{\circ}$}

\begin{document}
\title{A global view on star formation: The GLOSTAR Galactic plane survey}
\subtitle{VIII. Formaldehyde absorption in Cygnus~X}
\author{Y.~Gong\inst{1}, G. N.~Ortiz-Le{\'o}n \inst{2,1}, M.~R.~Rugel\inst{1,3,4}\fnmsep\thanks{Jansky Fellow of the National Radio Astronomy Observatory.}, K.~M.~Menten\inst{1}, A.~Brunthaler\inst{1},  F.~Wyrowski\inst{1}, C.~Henkel\inst{1,5,6}, H.~Beuther\inst{7}, S.~A.~Dzib\inst{8,1}, J.~S.~Urquhart\inst{9}, A.~Y.~Yang\inst{10,11}, J.~D.~Pandian\inst{12}, R.~Dokara\inst{1}, V.~S.~Veena\inst{1}, H.~Nguyen\inst{1}, S.-N.~X.~Medina\inst{13,1}, W.~D.~Cotton\inst{14}, W.~Reich\inst{1}, B.~Winkel\inst{1}, P.~M{\"u}ller\inst{1}, I.~Skretas\inst{1}, T.~Csengeri\inst{15},  S.~Khan\inst{1}, A.~Cheema\inst{1}}

\institute{
Max-Planck-Institut f{\"u}r Radioastronomie, Auf dem H{\"u}gel 69, D-53121 Bonn, Germany
\and
Instituto Nacional de Astrof\'isica, \'Optica y Electr\'onica, Apartado Postal 51 y 216, 72000, Puebla, Mexico
\and
Center for Astrophysics $\mid$ Harvard \& Smithsonian, 60  Garden St., Cambridge, MA 02138, USA
\and
National Radio Astronomy Observatory, 1003 Lopezville RD, Socorro, NM 87801, USA
\and
Department of Astronomy, Faculty of Science, King Abdulaziz University, P.~O.~Box 80203, Jeddah 21589, Saudi Arabia
\and 
Xinjiang Astronomical Observatory, Chinese Academy of Sciences, 830011 Urumqi, PR China
\and
Max Planck Institute for Astronomy, K{\"o}nigstuhl 17, 69117 Heidelberg, Germany
\and 
IRAM, 300 rue de la piscine, 38406 Saint Martin d’H{\'e}res, France
\and 
Centre for Astrophysics and Planetary Science, University of Kent, Canterbury, CT2 7NH, UK
\and 
National Astronomical Observatories, Chinese Academy of Sciences, A20 Datun Road, Chaoyang District, Beijing 100101, P. R. China
\and
Key Laboratory of Radio Astronomy and Technology, Chinese Academy of Sciences, A20 Datun Road, Chaoyang District, Beijing, 100101, P. R. China
\and
Department of Earth and Space Science, Indian Institute for Space Science and Technology, Trivandrum 695547, India
\and
German Aerospace Center, Scientific Information, 51147 Cologne, Germany
\and
National Radio Astronomy Observatory, 520 Edgemont Road, Charlottesville, VA 22903, USA
\and 
Laboratoire d'astrophysique de Bordeaux, Univ. Bordeaux, CNRS, B18N, all{\'e}e Geoffroy Saint-Hilaire, 33615 Pessac, France
}

\date{Received date ; accepted date}

\abstract
{Cygnus~X is one of the closest and most active high-mass star-forming regions in our Galaxy, making it one of the best laboratories for studying massive star formation.}
{We aim to investigate the properties of molecular gas structures on different linear scales with 4.8~GHz formaldehyde (H$_{2}$CO) absorption line in Cygnus~X.}
{As part of the GLOSTAR Galactic plane survey, we performed large scale (7\degree$\times$3\degree) simultaneous H$_{2}$CO (1$_{1,0}$--1$_{1,1}$) spectral line and radio continuum imaging observations toward Cygnus~X at $\lambda\sim$6~cm with the Karl G. Jansky Very Large Array and the Effelsberg-100 m radio telescope. We used auxiliary HI, $^{13}$CO (1--0), dust continuum, and dust polarization data for our analysis.}
{Our Effelsberg observations reveal widespread H$_{2}$CO (1$_{1,0}$--1$_{1,1}$) absorption with a spatial extent of $\gtrsim$50~pc in Cygnus~X for the first time. On large scales of 4.4~pc, the relative orientation between local velocity gradient and magnetic field tends to be more parallel at H$_{2}$ column densities of $\gtrsim$1.8$\times 10^{22}$~cm$^{-2}$. On the smaller scale of 0.17~pc, our VLA+Effelsberg combined data reveal H$_{2}$CO (1$_{1,0}$--1$_{1,1}$) absorption only toward three bright H{\scriptsize II} regions. Our observations demonstrate that H$_{2}$CO (1$_{1,0}$--1$_{1,1}$) is commonly optically thin. Kinematic analysis supports the assertion that molecular clouds generally exhibit supersonic motions on scales of 0.17--4.4~pc. We show a non-negligible contribution of the cosmic microwave background radiation in producing extended absorption features in Cygnus~X. Our observations suggest that H$_{2}$CO ($1_{1,0}-1_{1,1}$) can trace molecular gas with H$_{2}$ column densities of $\gtrsim 5 \times 10^{21}$~cm$^{-2}$ (i.e., $A_{\rm V} \gtrsim 5$). The ortho-H$_{2}$CO fractional abundance with respect to H$_{2}$ has a mean value of 7.0$\times 10^{-10}$. 
A comparison of velocity dispersions on different linear scales suggests that the dominant $-3$~\kms\, velocity component in the prominent DR21 region has nearly identical velocity dispersions on scales of 0.17--4.4~pc, which deviates from the expected behavior of classic turbulence.}
{}


\titlerunning{Cygnus~X}

\authorrunning{Gong et al.}

\maketitle

\section{Introduction}
Stars are basic units of the Universe, but their formation is still one of the unsettled questions in modern astronomy. 
The Global View on Star Formation in the Milky Way (GLOSTAR\footnote{\url{https://glostar.mpifr-bonn.mpg.de/glostar/}}) survey is an unbiased survey of the interstellar medium (ISM) and  star formation regions in the Milky Way using the wide-band (4--8 GHz) C-band receivers of the Karl G. Jansky Very Large Array (VLA) and the Effelsberg-100~m telescope to simultaneously observe the radio continuum emission and selected spectral lines \citep{2021A&A...651A..85B}. So far, the survey data have been used to characterize radio continuum sources \citep{2019A&A...627A.175M,2021A&A...651A..88N,2022arXiv221000560D}, identify supernova remnants \citep{2021A&A...651A..86D,2022arXiv221113811D}, and search for methanol masers emitting in the 6.7 GHz transition, the strongest class II CH$_3$OH maser line \citep{2021A&A...651A..87O,2022arXiv220710548N}. This work is the first GLOSTAR study to investigate the interstellar medium (ISM) and star formation regions in the 4.8~GHz formaldehyde transition. 

\subsection{To what extent does formaldehyde trace molecular gas?}
Formaldehyde (H$_2$CO) was the first polyatomic organic molecule to be discovered in the interstellar medium \citep{1969PhRvL..22..679S}. Being a slightly asymmetric top molecule, its rotational energy levels are split into K$-$doublets. This molecule has ortho and para symmetry species, depending on whether the spins of the hydrogen nuclei are parallel (ortho) or antiparallel (para). In molecular clouds, formaldehyde can be formed in the gas-phase but more efficiently on the surface of dust grains by successive hydrogenation of CO \citep{2002ApJ...571L.173W}, and is then released to the gas phase by thermal and non-thermal desorption.

The discovery of formaldehyde was made in the $J_{\rm Ka,Kc}=1_{1,0} - 1_{1,1}$ doublet line of its ortho species near 4.8 GHz (6 cm) \citep{1969PhRvL..22..679S}.
This line is generally observed in absorption against strong continuum background sources such as the Galactic center \citep[Sgr A;][]{1969PhRvL..22..679S}, and even its hyperfine structure (HFS) components have been detected in dark clouds \citep[e.g.,][]{1973ApJ...183..441H}. The ubiquity of absorption in this line is explained by the ease with which the lowest energy level of ortho-H$_{2}$CO ($J_{\rm Ka,Kc}=1_{1,1}$) gets overpopulated by collisional pumping, which results in a very low excitation temperature of $<$2.73~K for the lowest K-doublet transition of ortho-H$_{2}$CO (1$_{1,0}$--1$_{1,1}$) \citep[e.g.,][]{1975ApJ...196..433E}. This ``overcooling'' results in absorption even against the Cosmic Microwave Background (CMB).  

The substantial electric dipole moment of H$_{2}$CO of 2.33 D \citep{Fabricant1977} makes its (sub)millimeter wavelength rotational transitions good probes of dense gas in star forming regions. These properties have inspired several studies that use the rotational transitions of H$_{2}$CO to probe the density and temperature of dense gas in the Milky Way and in external galaxies \citep[e.g.,][]{1993ApJS...89..123M, 2013A&A...550A.135A, 2015A&A...573A.106G, 2018A&A...609A..16T, 2018A&A...611A...6T}. On the other hand, H$_{2}$CO has been observed in absorption against extragalactic continuum sources indicating that H$_{2}$CO can survive in diffuse and translucent molecular clouds \citep{1990ApJS...72..303N, 1995A&A...299..847L, MentenReid1996, 2006ARA&A..44..367S, 2006A&A...448..253L}. This ease with which the ortho-H$_{2}$CO (1$_{1,0}$--1$_{1,1}$) transition is excited in diffuse and translucent clouds suggests that it traces the largest extent of molecular gas when compared with other H$_{2}$CO transitions. These facts trigger a question: to what extent can this H$_{2}$CO transition trace the general distribution of molecular gas? In order to address this question, one requires large-scale mapping observations of H$_{2}$CO, but such observations are still scarce. Large-scale mapping studies of H$_{2}$CO (1$_{1,0}$--1$_{1,1}$) absorption have been performed toward the central molecular zone \citep{1992A&AS...96..525Z}, W51 \citep{2015A&A...573A.106G}, and the Aquila molecular cloud \citep{2019ApJ...874..172K}, revealing a widespread distribution of H$_{2}$CO in these regions. However, these observations mainly focus on high H$_{2}$ column density (i.e., high extinction) molecular gas, and the maps cover less than 5 square degrees in total. 

Because the (1$_{1,0}$--1$_{1,1}$) and (2$_{1,1}$--2$_{1,2}$) pair of H$_{2}$CO lines has been proven to be a good densitometer \citep[e.g.,][]{1980A&A....82...41H, 1993ApJS...89..123M, 2008ApJ...673..832M}, large-scale H$_{2}$CO (1$_{1,0}$--1$_{1,1}$) mapping observations are able to pinpoint regions with appreciable absorption, which can facilitate follow-up H$_{2}$CO (2$_{1,1}$--2$_{1,2}$) imaging for density determinations of molecular clouds on large scales \citep[e.g.,][]{2015A&A...573A.106G}.

Thanks to the large coverage of the GLOSTAR observations, our survey can potentially reveal the distribution of H$_2$CO (1$_{1,0}$--1$_{1,1}$) absorption on a Galactic scale for the first time \citep{2021A&A...651A..85B}. In this work, we will present the first results based on GLOSTAR measurements of H$_{2}$CO in Cygnus~X.


\subsection{Cygnus~X as an excellent astrophysical laboratory}
Cygnus~X, named by \citet{1952AuSRA...5...17P}, is one of the closest and most active high-mass star-forming regions in the Milky Way \citep[e.g.,][]{2008hsf1.book...36R, 2014AJ....148...11K}. This region exhibits very extended bright Galactic radio continuum emission that arises from discrete H{\scriptsize II} regions, supernova remnants (SNRs), and diffuse thermal emission \citep{1984A&AS...58..291W, 1991A&A...241..551W, 2013A&A...559A..81X, 2022A&A...664A..88E}. Cygnus~X contains several OB associations that harbour a large number of massive stars \citep{2000A&A...360..539K, 2015MNRAS.449..741W, 2018A&A...612A..50B}.  Observations have also shown that a large X-ray bubble and very high energy $\gamma$-rays were found to surround one of the most prominent of these, Cyg OB2  \citep[e.g.,][]{1980ApJ...238L..71C, 2021NatAs.tmp...50A, 2021Natur.594...33C}.
Due to the irradiation from the Cyg OB2 association, pillars and globules are formed in ambient molecular clouds with an orientation toward the center of the Cyg OB2 association \citep{2016A&A...591A..40S, 2021A&A...653A.108S}, demonstrating the role of the feedback of OB stars on shaping molecular clouds. 

The distance of Cygnus-X has been a subject of considerable debate. Since molecular cloud complexes in Cygnus~X seem to be associated with each other due to the coherence in line-of-sight velocities \citep{2006A&A...458..855S}, a fixed distance of about 1.4--1.7 kpc is commonly assumed for all molecular clouds in this region by many previous studies. While this is consistent with the results of accurate trigonometric parallaxes of maser sources located in various parts of Cygnus~X and massive stars in the Cygnus~OB2 association \citep{2012A&A...539A..79R, 2013ApJ...769...15X, 2013ApJ...763..139D}, later studies also find sources at farther distances of 3.3--3.6 kpc \citep{2012A&A...539A..79R,2013ApJ...769...15X} and even $\gtrsim$9 kpc \citep[indicated by the radial velocity of $<-$60~\kms;][]{1989ApJS...71..469L, 2021A&A...651A..87O, 2021ApJ...918L...2L}. However, studies using the \textit{Gaia} parallax measurements and their line-of-sight extinctions suggest the bulk of molecular gas should be located at 1.3--1.5~kpc in Cygnus~X \citep{2020A&A...633A..51Z, 2020MNRAS.493..351C, 2022A&A...658A.166D}. For simplicity, we adopt a distance of 1.4~kpc in this work and the values of physical parameters obtained from previous studies scaled to this distance for comparison.



Cygnus~X harbors one of the most massive molecular cloud complexes ($\sim 2$--$3 \times 10^{6}$~M$_{\odot}$) identified in the Milky Way \citep{2006A&A...458..855S}. 
Previous CO and dust continuum observations show a highly structured distribution of molecular clouds \citep[e.g.,][]{2006A&A...458..855S,2012A&A...543L...3H,2019ApJS..241....1C}, and the presence of filamentary structures in Cygnus~X. The filaments, especially toward the prominent compact H{\scriptsize II} region DR21 and its environment, are found to show velocity gradients \citep{2010A&A...520A..49S, 2021ApJ...908...70H, 2022ApJ...927..106C,2023ApJ...948L..17L,2023arXiv230507785B}, indicating ongoing accretion flows on sub-parsec scales. 
A large number of dense cores and massive protostars are found in these filaments (including DR21) \citep{2007A&A...476.1243M,2011A&A...529A...1S, 2010A&A...524A..18B, 2011ApJ...727..114R, 2012A&A...543L...3H, 2019ApJS..241....1C, 2022MNRAS.512..960C,2022ApJ...941..122C}. Among these studies, \citet{2019ApJS..241....1C} built a large sample of 151 massive dense cores that have masses of $>$35~$M_{\odot}$ with a typical size of $\sim$0.1 pc. Their further efforts led to a sample of 8,431 dust cores being identified \citep{2021ApJ...918L...4C}. Based on these results, the large-scale density structure is studied with a triangulation-based method \citep{2021ApJ...916...13L}, implying that dense cores form through fragmentation controlled by scale-dependent turbulent pressure support. 

As part of the GLOSTAR survey, \citet{2021A&A...651A..87O} detected thirteen 6.7 GHz methanol masers, which are exclusively associated with high mass young stellar objects, in Cygnus~X. Further evidence of widespread star formation activity in such sources is provided by the discovery of about 60 molecular outflows in Cygnus~X in large-scale CO surveys \citep{2012A&A...541A..79G, 2013A&A...558A.125D, 2021MNRAS.503.1264D, 2022A&A...660A..39S}. These properties make Cygnus~X one of the best laboratories for studying massive star formation. The Cygnus~X region is therefore targeted by many ongoing large-scale  projects, including GLOSTAR \citep{2021A&A...651A..85B}, K-band focal plane array Examinations of Young STellar Object Natal Environments \citep[KEYSTONE;][]{2019ApJ...884....4K}, Surveys of Clumps, CorEs, and CoNdenSations in CygnUS-X \citep[CENSEUS;][]{2019ApJS..241....1C}, and the Cygnus Allscale Survey of Chemistry and Dynamical Environments \citep[CASCADE;][]{2022arXiv220710964B}.
Therefore, the study of H$_{2}$CO absorption in Cygnus X will pave the way toward understanding gas distribution, kinematics, chemistry, and evolutionary processes associated with high-mass star formation.  



\section{Observations and data reduction}\label{Sec:obs}
\subsection{Effelsberg-100 m observations}
As part of the GLOSTAR survey \citep{2021A&A...651A..85B}, we performed C band observations with the dual-polarization S45mm receiver of the 100-m telescope near Effelsberg/Germany\footnote{The 100-m telescope at Effelsberg is operated by the Max-Planck Institut f{\"u}r Radioastronomie (MPIfR) on behalf of the Max-Planck Gesellschaft (MPG).} between 2019 January 11 and 2020 December 23 (project codes: 22-15 and 102-20). The observations and data reduction have been described in \citet{2021A&A...651A..85B}, and the calibration quality of the Effelsberg spectral line observations will be discussed in Rugel et al. (in prep). In the following, we summarize the observations and data products relevant to this publication. Two different kinds of backends, the SPEctro-POLarimeter (SPECPOL) and fast Fourier transform spectrometers \citep[FFTSs;][]{2012A&A...542L...3K}, were used to record full Stokes continuum emission and spectral line signals, respectively. The on-the-fly (OTF) mode was used to map Cygnus~X with a scanning speed of 90\arcsec\,per second. Like all of the area covered by GLOSTAR, the region has been mapped both along Galactic longitude and latitude, 
in order to reduce striping artifacts in the image restoration. Our observations cover an area of 7\degree$\times$3\degree\,in size (i.e., 76\degree$\leq l \leq$83\degree, $-$1\degree$\leq b\leq$2\degree). The flux calibrators 3C 286 and NGC 7027 were used to establish the flux density scale of both our radio continuum and spectral line data. The system temperatures typically range from 28--42 K. Nearby pointing observations were carried out every 2 to 3 hours. The rms pointing uncertainty was found to be within 10\arcsec\,which is less than 1/10 of the half power beam width (HPBW) of 145\arcsec\,at 4.83~GHz.

In this paper, we have used the Effelsberg data to study the large-scale ISM structure of Cygnus~X in the 4.8~GHz formaldehyde transition, while the associated 4.89~GHz continuum data are used to derive the optical depth of the H$_{2}$CO line emission. In our setup, we simultaneously cover H$_{2}$CO (1$_{1,0}$--1$_{1,1}$) at 4.8296600 GHz and its isotopologue H$_{2}^{13}$CO (1$_{1,0}$--1$_{1,1}$) at 4.5930885 GHz \citep{2005JMoSt.742..215M, ENDRES201695}. The channel spacings for H$_{2}$CO (1$_{1,0}$--1$_{1,1}$) and H$_{2}^{13}$CO (1$_{1,0}$--1$_{1,1}$) are 0.19~\kms\,and 2.49~\kms \citep[see Table 2 in][]{2021A&A...651A..85B}, respectively. All the velocities are given with respect to the local standard of rest (LSR). Spectral data were pre-processed and calibrated with the standard Effelsberg pipeline, which includes bandpass and absolute intensity calibration \citep{2012A&A...540A.140W}, as well as correction of atmospheric attenuation based on a water vapor radiometer operating between 18 and 26 GHz. Since the spectra were regridded using Gaussian convolution in the pipeline, the actual spectral resolution corresponds to two channel widths (e.g., 0.38~\kms\,for the 4.8 GHz H$_{2}$CO line).  Further data reduction and mapping of the data was performed with the GILDAS\footnote{\url{https://www.iram.fr/IRAMFR/GILDAS/}} software \citep{2005sf2a.conf..721P}. Six out of 1946700 H$_{2}$CO spectra were affected by radio frequency interference (RFI), and were thus discarded in the data reduction. The spectral baseline subtraction has been carried out using a first order polynomial. 

Our spectral map was first convolved to an effective HPBW of 3\arcmin\,with a single pixel size of 30\arcsec $\times$ 30\arcsec. However, at this spatial resolution, the signal-to-noise ratio of the H$_{2}$CO image was not sufficient to detect the extended absorption. To improve the fidelity of the extended absorption, the data were further convolved to an effective angular resolution of 10$\rlap{.}$\arcmin8. 

\begin{figure*}[!htbp]
\centering
\includegraphics[width = 0.95 \textwidth]{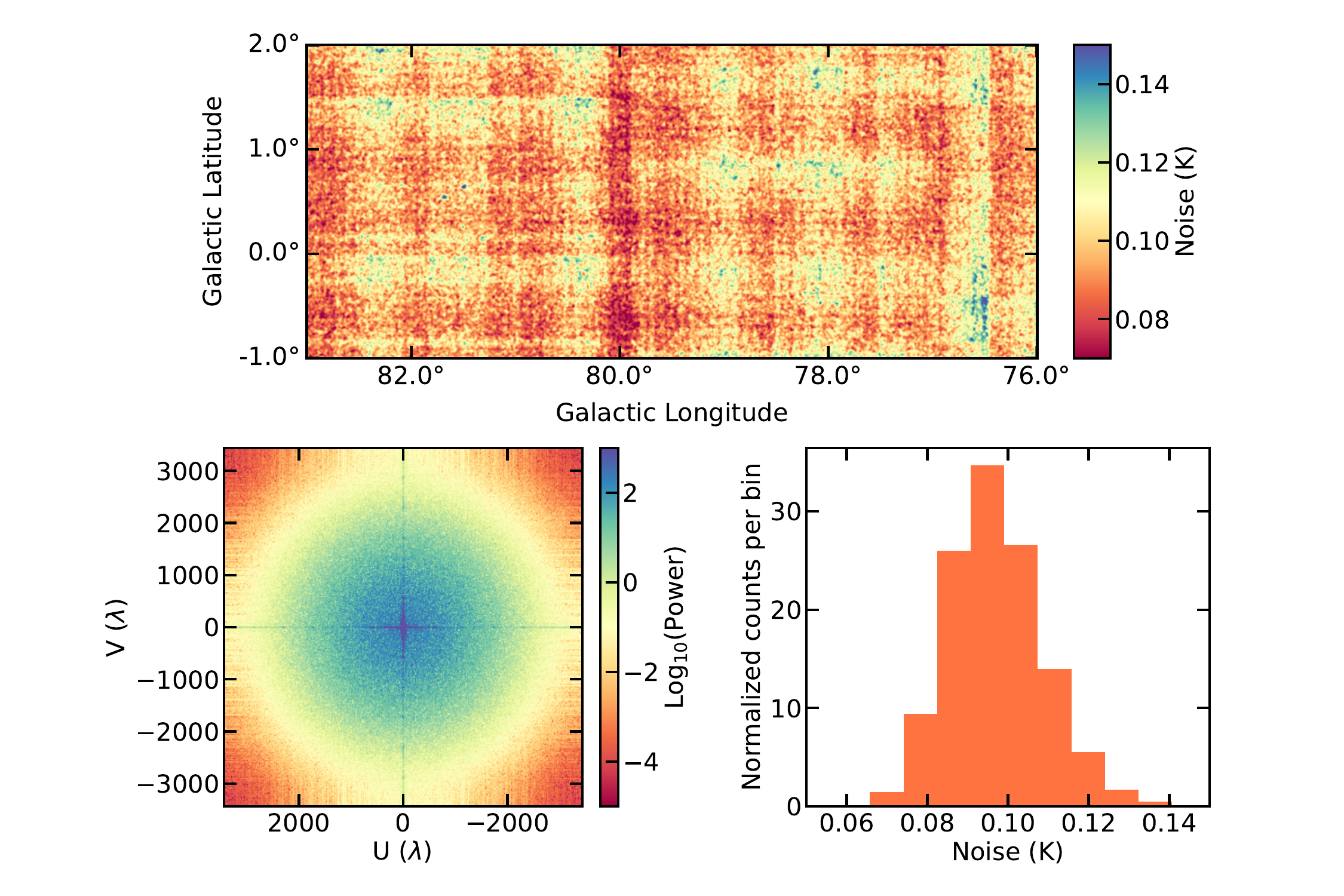}
\caption{{\textit{Top:} Spatial distribution of the rms noise of the Effelsberg H$_{2}$CO (1$_{1,0}$--1$_{1,1}$) observations at a channel width of 0.5~\kms\,and an HPBW of 3\arcmin. \textit{Lower left:} Power spectrum of the noise image. The very high power pixels in the cross are artifacts  that are caused by the Fourier transform of the sharp image (also known as the Gibbs phenomenon). \textit{Lower right:} Histogram of the rms noise values. The mean and standard deviation values are 0.10~K and 0.01~K, respectively.}\label{Fig:noise}}
\end{figure*}

The spatial distribution of rms noise values is shown in the top panel of Fig.~\ref{Fig:noise}. The rms noise can vary by a factor of 2 because of different effective integration times. We also illustrate the 2D power spectra in the lower left panel of Fig.~\ref{Fig:noise}. The power spectrum shows that there is no clear correlation at any specific spatial scale. The lower right panel of Fig.~\ref{Fig:noise} presents the histogram of the rms noise values which range from 0.07 to 0.24~K with a median value of 0.10~K at a channel width of 0.5~\kms. 

The simultaneously observed 4.89~GHz radio continuum data were reduced with the NOD3 software package \citep{2017A&A...606A..41M}. The typical rms noise levels are 5~mK at a narrow bandwidth of 120~MHz. As described in \citet{2021A&A...651A..85B}, the zero-level intensities of our Effelsberg data needed to be restored using the Urumqi 4.8 GHz continuum data \citep{2007A&A...463..993S, 2011A&A...527A..74S}. However, the zero-level intensities of the Urumqi 4.8 GHz continuum data of Cygnus~X need to be restored as well, because the radio continuum emission of Cygnus~X is very extended, reaching $|b|>$5\degree.
For this reason, we made use of model c from the WMAP foreground maps\footnote{\url{https://lambda.gsfc.nasa.gov/product/wmap/dr5/m_products.html}} \citep{2013ApJS..208...20B}. We derived the WMAP-based 4.8~GHz continuum emission by interpolating the free-free, synchrotron, and dust emission. Smoothing the WMAP-based and Urumqi 4.8 GHz continuum images to a common angular resolution of 1$\rlap{.}$\degree5, we derived the zero-level shift of the Urumqi 4.8 GHz data from the difference between the derived WMAP-based and Urumqi 4.8 GHz continuum images. The difference ranges from $-$0.065~K to 0.269~K, which is a small correction. These differences are then added back to the original Urumqi 4.8 GHz continuum image. Our Effelsberg 4.89 GHz data cover only $-$1\degree $<b<$ 2\degree, and thus have a zero-level shift due to the continuum baseline correction only covering the limited range of Galactic latitude. Following the same method introduced in \citet{2021A&A...651A..85B}, we used the restored Urumqi 4.8 GHz data to recover the zero-level shift of our Effelsberg 4.89 GHz data. Consequently, the restored Effelsberg 4.89 GHz continuum map is used in this study.

\subsection{VLA observations}
As part of the GLOSTAR survey \citep{2019A&A...627A.175M,2021A&A...651A..85B}, the Cygnus-X region was observed using the D configuration of the Karl G. Jansky Very Large Array (VLA) of the National Radio Astronomy Observatory\footnote{The National Radio Astronomy Observatory is a facility of the National Science Foundation operated under cooperative agreement by Associated Universities, Inc.} with the correlator configuration including the H$_2$CO line. Details about the observations have been presented in \citet{2021A&A...651A..87O}; here we give a brief summary. We observed 14 strips of 1\degree$\times$1.5\degree in size that cover the same area as the Effelsberg observations. The observations registered sixteen 128 MHz wide spectral windows for the continuum. The H$_{2}$CO (1$_{1,0}$--1$_{1,1}$) line was observed simultaneously with 4~MHz of bandwidth and 1024 channels, resulting in a channel spacing of 0.25~\kms\,and a total velocity coverage of 260~\kms.
The calibration of the spectral line data was performed using the Common Astronomy Software Applications (CASA) package \citep{2007ASPC..376..127M} using a customized version of the VLA pipeline\footnote{\url{https://science.nrao.edu/facilities/vla/data-processing/pipeline}}. The  line imaging was performed for each strip using the mosaic mode in CASA and a pixel size of 2$\rlap{.}\arcsec$5$\times$2$\rlap{.}\arcsec$5.
The synthesized beam is about 19\arcsec$\times$15\arcsec\, with a position angle of $-$34\degree\,for H$_{2}$CO (1$_{1,0}$--1$_{1,1}$). The largest angular scale structure that our VLA D array observations are sensitive to is about 4\arcmin.

The calibration and imaging of the radio continuum emission were done using the Obit package \citep{Cotton2008}; see \cite{2021A&A...651A..85B} for more details. For the analysis presented here, we only used a $\sim$200~MHz wide frequency sub-band centered at 4.9~GHz out of the 16 spectral windows. The continuum image has a circular beam of 19\arcsec and a pixel size of 2$\rlap{.}\arcsec$5$\times$2$\rlap{.}\arcsec$5. In order to match the angular resolution of the continuum and line emission, we smoothed both VLA data sets to a circular beam of 25\arcsec\, for our analysis.

\subsection{Combination of VLA and Effelsberg data}
Since the VLA D configuration data lack the short-spacing information, we combine the VLA and Effelsberg data in order to recover the extended emission and absorption. As illustrated in \citet{2021A&A...651A..85B}, there is no clear systematic offset between the VLA and Effelsberg flux calibration. Hence, no flux scaling was needed. Before the combination, we regridded both VLA and Effelsberg H$_{2}$CO data sets to the same channel width of 0.5~\kms. The combination was performed with the ``feather" task in CASA. The combined data have a circular beam of 25\arcsec\, for both the continuum and the H$_{2}$CO (1$_{1,0}$--1$_{1,1}$) spectral line images, which corresponds to a linear scale of $\sim$0.17~pc in Cygnus~X. The typical 1$\sigma$ noise level is about 20 mJy~beam$^{-1}$ (or 1.7~K in units of brightness temperatures) at a channel width of 0.5~\kms.

\begin{figure*}[!htbp]
\centering
\includegraphics[width = 0.9 \textwidth]{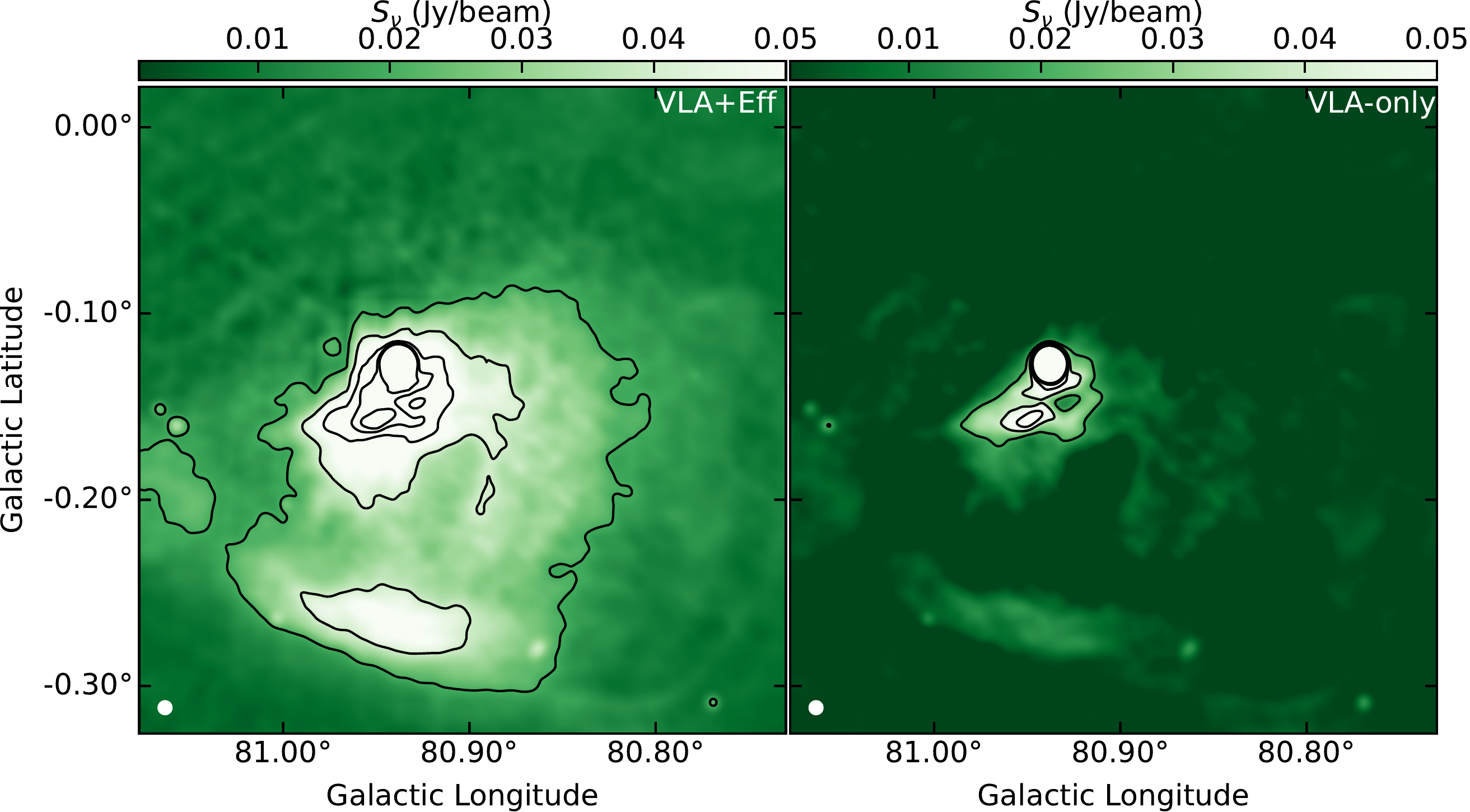}
\caption{{Comparison of the 4.9 GHz radio continuum  maps of DR22 from the VLA+Effelsberg combined and VLA-only data sets. In both panels, the contours start from 0.02~Jy~beam$^{-1}$ and increase by 0.02~Jy~beam$^{-1}$. The beam size is shown in the lower left corner of each panel. }\label{Fig:vla-comp}}
\end{figure*}

We compare the VLA-only and VLA+Effelsberg combined data toward DR22 in Fig~\ref{Fig:vla-comp}. While the distributions are similar in both data sets, it is evident that the VLA+Effelsberg combined data show more extended emission and higher flux densities, which confirms that the combined data recover the missing flux in the VLA D-configuration data. Therefore, we adopt the VLA+Effelsberg combined data for the following analysis on small scales.

\subsection{Archival data}\label{Sec:arc}
In order to determine the H$_{2}$ column density in Cygnus~X, we use the \textit{Planck} 353~GHz map of the dust optical depth that was derived by fitting the spectral energy distribution (SED) of dust emission inferred from continuum maps ranging from 353 to 3000~GHz \citep{2014A&A...571A..11P}. The conversion factor from the 353~GHz dust optical depth to the H$_{2}$ column density will be discussed in Sect.~\ref{sec.col}. The HPBW is 4$\rlap{.}$\arcmin9. The \textit{Planck} thermal dust polarization data at 353~GHz are used to study the polarization properties of molecular clouds in Cygnus~X \citep{2015A&A...576A.105P, 2015A&A...576A.104P}. The data are smoothed to 10\arcmin\,to achieve a signal-to-noise ratio greater than 3 in the amplitude of linear polarization. These maps are obtained from the public \textit{Planck} Legacy Archive\footnote{\url{http://pla.esac.esa.int/}}.

We also used $^{13}$CO (1--0) data obtained from the Five College Radio Astronomical Observatory (FCRAO), the details being described in \citet{2011A&A...529A...1S}. The HPBW is 46\arcsec, and the channel spacing is 0.066~\kms. The typical 1$\sigma$ rms noise level is 0.2~K per channel on an antenna temperature scale. A main beam efficiency of 0.48 is adopted in this study. 

A HI column density map was obtained from the Effelsberg-Bonn HI Survey \citep[EBHIS;][]{2011AN....332..637K, 2016A&A...585A..41W}. The HPBW is 10$\rlap{.}$\arcmin8 at 1.420 GHz. The rms noise level is 90 mK at a channel spacing of 1.29~\kms. 

\section{Results}\label{Sec:res}
\subsection{Widespread formaldehyde absorption}
\subsubsection{Overall distribution}\label{sec.mor}
\begin{figure}[!htbp]
\centering
\includegraphics[width = 0.45 \textwidth]{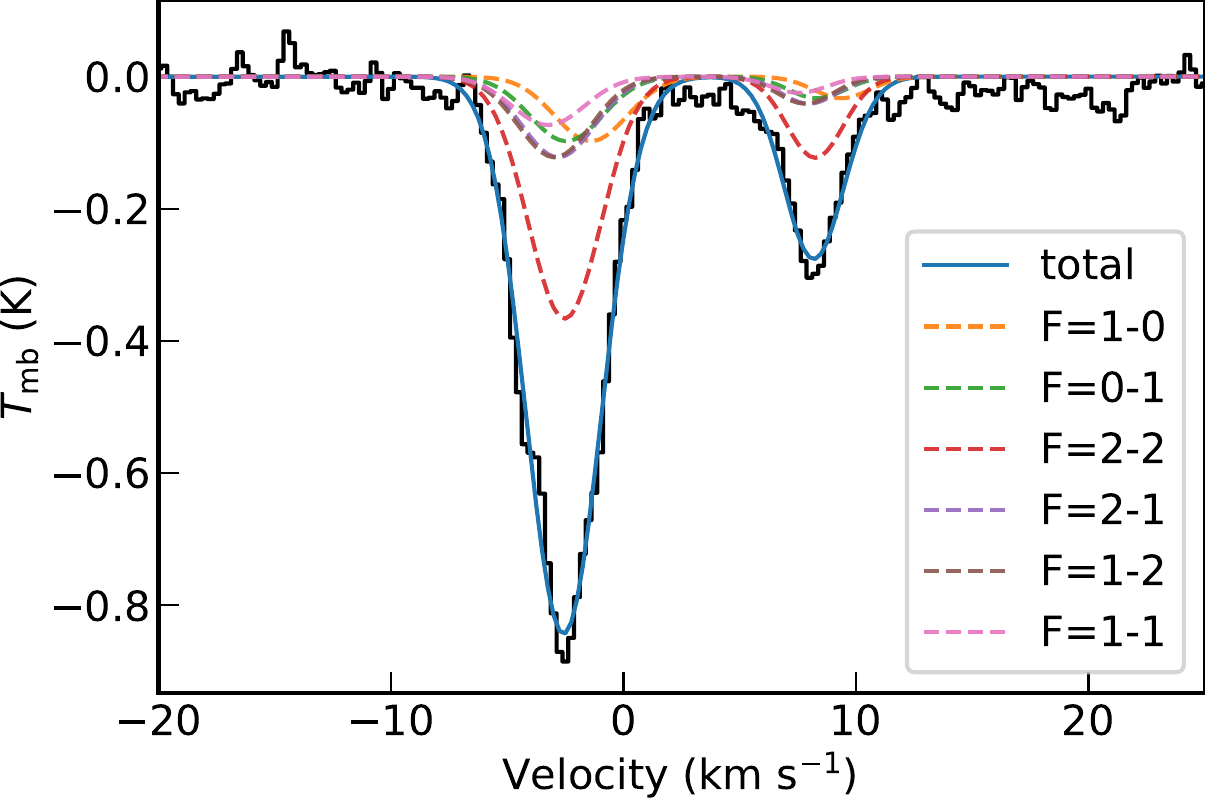}
\caption{{Observed H$_{2}$CO (1$_{1,0}$--1$_{1,1}$) spectrum at an HPBW of 10$\rlap{.}$\arcmin8 toward DR21 (the black solid line) overlaid with the fitted model (the blue solid line). The two velocity components at $-$3~\kms\, and 8~\kms\,correspond to two physically distinct velocity components which arise from the DR21 cloud and its foreground cloud associated with W75N. The fitted HFS components are indicated by the colored dashed lines in the legend.}\label{Fig:specfit}}
\end{figure}

\begin{figure*}[!htbp]
\centering
\includegraphics[width = 0.95 \textwidth]{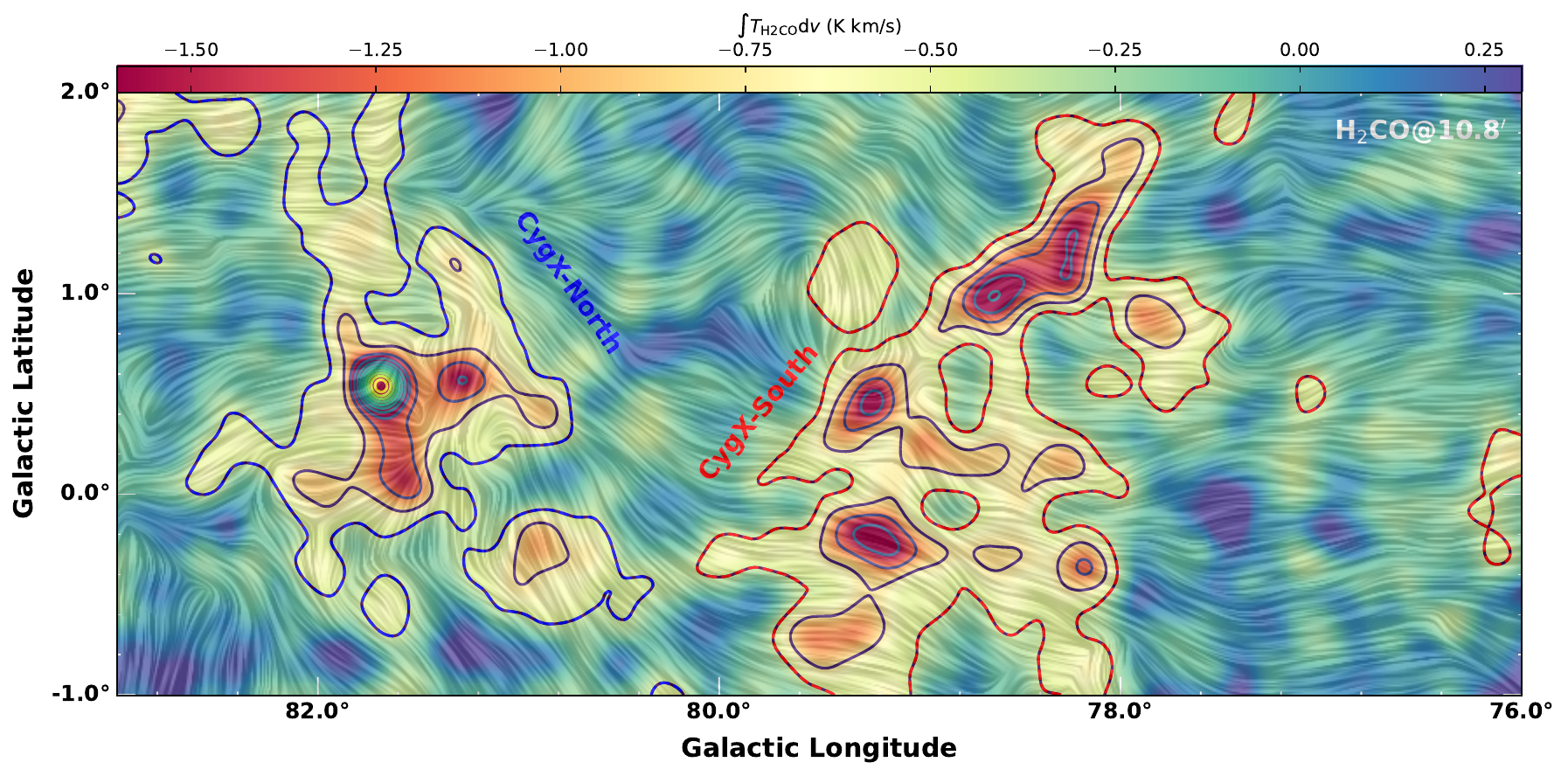}
\includegraphics[width = 0.95 \textwidth]{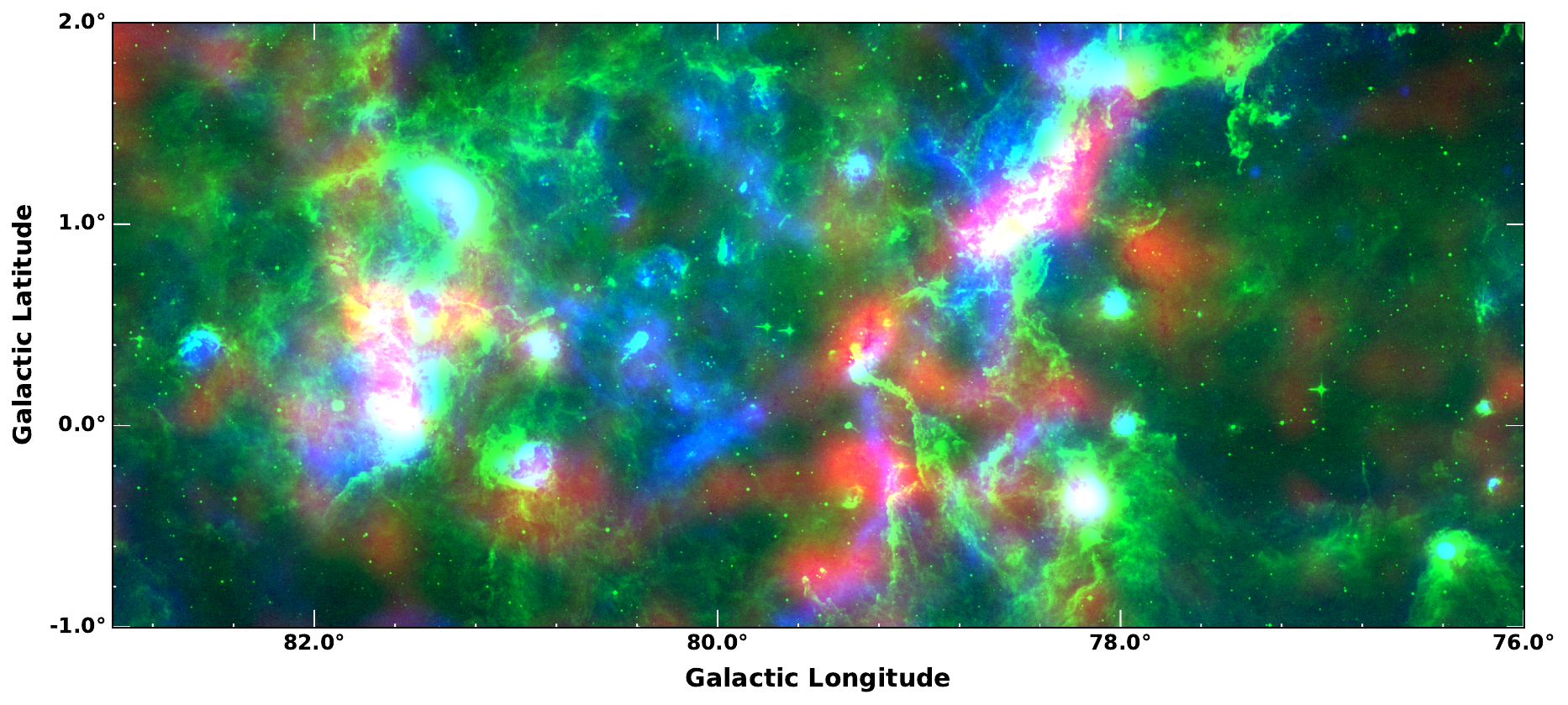}
\caption{{\textit{Top}: Integrated-intensity map of the Effelsberg H$_{2}$CO (1$_{1,0}$--1$_{1,1}$) absorption at an HPBW of 10$\rlap{.}$\arcmin8. The integrated velocity range is from $-$10 to 20~\kms. The contours start at $-$0.4~K~\kms\,(5$\sigma$), with each subsequent contour being twice the previous one. The overlaid pattern indicates the magnetic field direction based on the polarization measurements by the \textit{Planck} satellite \citep{2014A&A...571A..11P} created by the line integral convolution (LIC) method \citep{Cabral93}. \textit{Bottom:} Overview of the Cygnus~X region in a three-color composite image with the Effelsberg H$_{2}$CO (1$_{1,0}$--1$_{1,1}$) absorption at an HPBW of 10$\rlap{.}$\arcmin8 in red, \textit{MSX} 8~$\mu$m image in green, and the Effelsberg 4.89~GHz continuum emission in blue. }\label{Fig:overview}}
\end{figure*}

The 4.8~GHz formaldehyde transition is typically observed in absorption with a sample spectrum being shown in Fig.~\ref{Fig:specfit}, where the two velocity components at $-$3~\kms\, and 8~\kms\,arise from the DR21 cloud and its foreground cloud associated with W75N \citep[e.g.,][]{2003ApJ...596..344C, 2010A&A...520A..49S, 2019PASJ...71S..12D}, respectively. Figure~\ref{Fig:overview} shows the distribution of the H$_{2}$CO intensity integrated over the velocity range between $-$10~\kms\,and 20~\kms, at an angular resolution of 10$\rlap{.}$\arcmin8. Although H$_{2}$CO (1$_{1,0}$--1$_{1,1}$) has been investigated toward several positions in Cygnus~X \citep[e.g.,][]{1982A&AS...49..607B, 1983A&A...127..388H, 1988A&A...191..313P, 2019ApJ...877..154Y}, our Effelsberg-100 m observations reveal the widespread nature of H$_{2}$CO absorption in Cygnus~X for the first time. With an area coverage of 21 square degrees, this is the largest H$_{2}$CO (1$_{1,0}$--1$_{1,1}$) map of the region to date. 

\begin{figure*}[!htbp]
\centering
\includegraphics[width = 0.95 \textwidth]{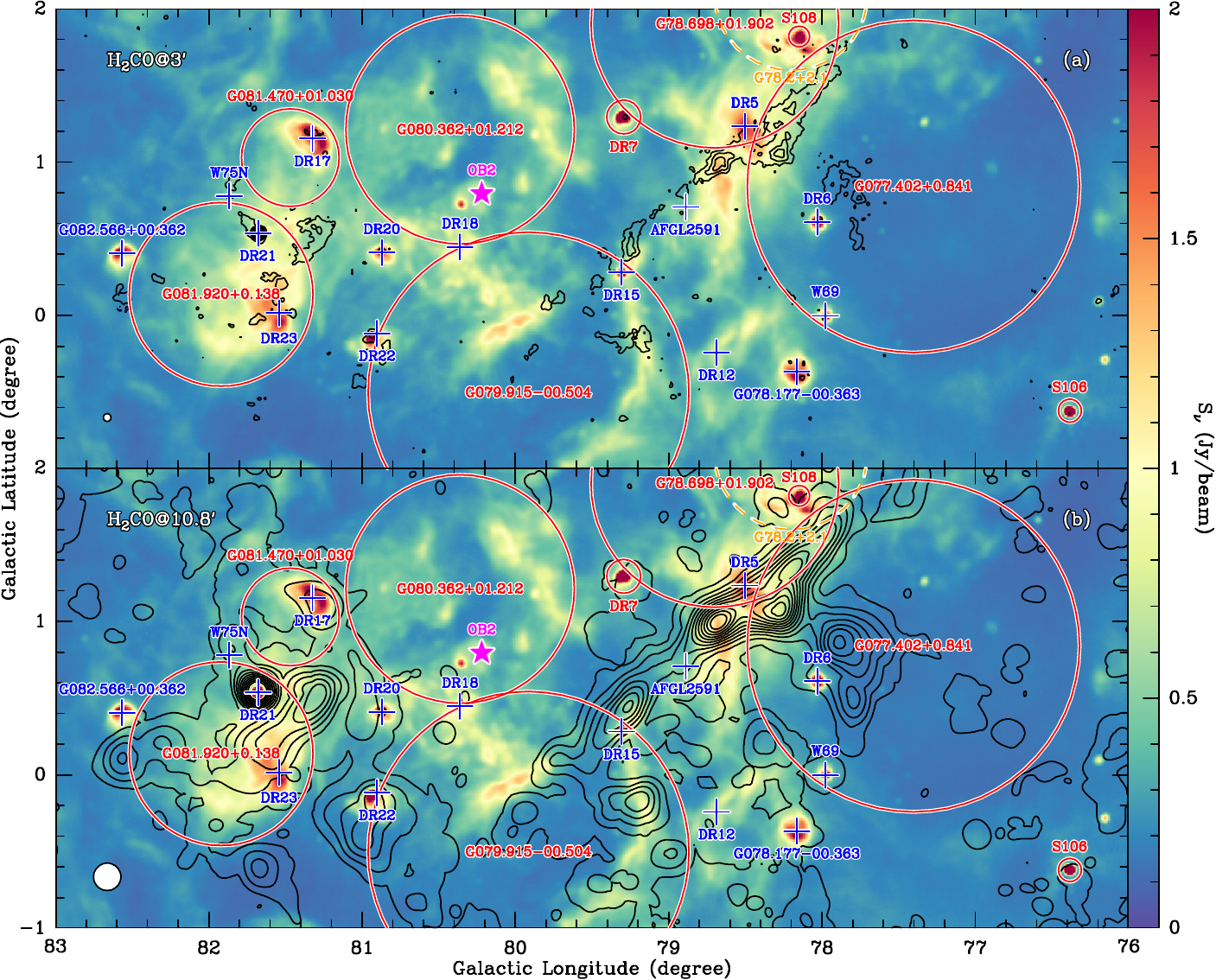}
\caption{{(a) Effelsberg 4.89~GHz radio continuum emission overlaid with the peak absorption contours of H$_{2}$CO (1$_{1,0}$--1$_{1,1}$). The corresponding HPBW of H$_{2}$CO (1$_{1,0}$--1$_{1,1}$) is 3\arcmin. The color bar represents the flux densities of the radio continuum emission. The H$_{2}$CO absorption contours start from $-$0.5~K (5$\sigma$), and decrease by 0.5~K. The developed H{\scriptsize II} regions from \citet{2014ApJS..212....1A} are marked with red solid circles, while SNR G78.2+2.1 is indicated by the orange dashed circle. Blue crosses represent the radio continuum sources and active star-forming objects, and the purple star represents the massive star cluster, Cygnus OB2. (b) Similar to Fig.~\ref{Fig:peak-abs}a but the corresponding HPBW of H$_{2}$CO (1$_{1,0}$--1$_{1,1}$) is 10$\rlap{.}$\arcmin8. The H$_{2}$CO absorption contours start from $-$0.08~K (4$\sigma$), and decrease by 0.06~K. In both panels, the beam size is shown in the lower left corner.}\label{Fig:peak-abs}}
\end{figure*}

Figure~\ref{Fig:peak-abs} shows the Effelsberg 4.89~GHz radio continuum image overlaid with the peak H$_{2}$CO absorption contours at two different angular resolutions (i.e., 3\arcmin\,and 10$\rlap{.}$\arcmin8). The widespread absorption is mainly attributed to two main structures known as CygX-North and CygX-South (labeled in the top panel of Fig.~\ref{Fig:overview}) following the nomenclature by \citet{2006A&A...458..855S, 2011A&A...529A...1S}.


Figure~\ref{Fig:peak-abs} shows that the absorption toward CygX-North covers an area $\sim 2\degr \times 2\degr$, in which several discrete absorption peaks are superimposed on more diffuse absorption. In CygX-South, the absorption covers a larger area with a similar diffuse morphology on which more compact peaks are superimposed. The extended absorption is more evident in Fig.~\ref{Fig:peak-abs}b compared to Fig.~\ref{Fig:peak-abs}a, since the former's larger beam size makes it much more sensitive to extended absorption. The lowest contour in Fig.~\ref{Fig:peak-abs}b suggests a total area of approximately 4700~pc$^{2}$ for the detectable H$_{2}$CO ($1_{1,0}-1_{1,1}$) absorption.

In Fig.~\ref{Fig:peak-abs}, strong H$_{2}$CO absorption features coincide with the bright radio continuum emission from (most prominently) DR21, DR17, DR22, DR6, W69, and G078.177-00.363. This is due to enhancements of the amplitude of absorption (in units of main beam temperature) against the bright continuum emission. The strongest H$_{2}$CO absorption arises from the line of sight toward DR21, which is largely due to the fact that DR21 is the brightest radio continuum source in this field (see Fig.~\ref{Fig:specfit}). This is consistent with previous pointed observations \citep{1988A&A...191..313P}. However, a large amount of extended H$_{2}$CO absorption features are not associated with bright compact radio continuum sources in Fig.~\ref{Fig:peak-abs}b.  

\subsubsection{Optical depth}\label{sec.exc}
Based on the radiative transfer equation in the Rayleigh-Jeans regime, the main beam temperature, $T_{\rm mb}$, can be expressed as: 
\begin{equation}\label{f.radiative}
 T_{\rm mb} = f_{\rm b}(T_{\rm ex}-T_{\rm bg}- T_{\rm c}) (1-{\rm exp}(-\tau))\;,
\end{equation}
where $T_{\rm ex}$ is the excitation temperature, $T_{\rm bg}$ is the temperature of the cosmic microwave background radiation that is taken to be 2.73~K \citep[e.g.,][]{2009ApJ...707..916F}, and $T_{\rm c}$ is the brightness temperature of the continuum emission behind the H$_{2}$CO gas, $f_{\rm b}$ is the beam dilution factor, and $\tau$ is the optical depth. 

Measurements of the HFS components of H$_{2}$CO (1$_{1,0}$--1$_{1,1}$) give an excitation temperature of $\sim$1.6 K in dark clouds \citep[e.g.,][]{1973ApJ...183..441H}. However, this method can only be applied to cases where the line widths are narrow enough. Our spectra are too broad to resolve the HFS components, so we have to assume excitation temperatures for our study. 
In addition to the HFS measurements \citep[e.g.,][]{1973ApJ...183..441H}, statistical equilibrium calculations suggest that the excitation temperatures are typically 1.2--1.8~K in massive star-forming regions \citep{1980A&A....82...41H,2019ApJ...877..154Y}. This suggests that the excitation temperature does not change significantly in different regions, and consequently, we assume a constant $T_{\rm ex}$ of 1.6 K for our analysis (see also Sect.~\ref{sec.col}). If the excitation temperature varies in the range of 1.2--2.0~K, this assumption will result in at most 25\%\,uncertainties in the derived optical depth. 

\begin{figure*}[!htbp]
\centering
\includegraphics[width = 0.95 \textwidth]{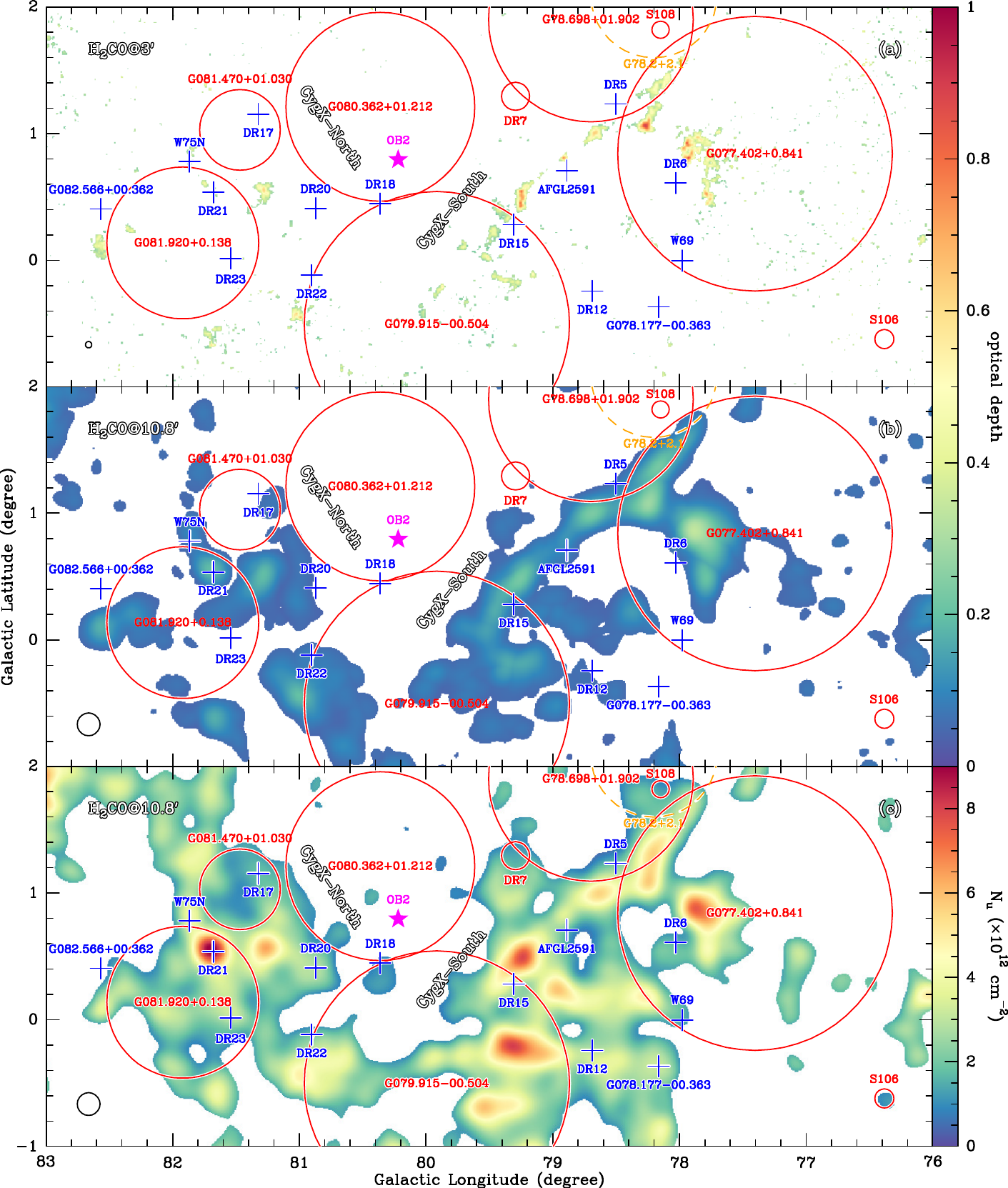}
\caption{{(a) Distribution of the peak optical depth of the H$_{2}$CO (1$_{1,0}$--1$_{1,1}$) line. The HPBW of the H$_{2}$CO image is 3\arcmin. The color bar represents the peak optical depth. The H{\scriptsize II} regions from \citet{2014ApJS..212....1A} are marked with red solid circles, while SNR G78.2+2.1 is indicated by an orange dashed circle. The blue crosses represent the radio continuum sources and active star-forming objects, and the purple star represents the massive star cluster, Cygnus OB2. (b) Similar to Fig.~\ref{Fig:peaktau}a but the  HPBW of the H$_{2}$CO (1$_{1,0}$--1$_{1,1}$) image is 10$\rlap{.}$\arcmin8. (c) Similar to Fig.~\ref{Fig:peaktau}b but for the H$_{2}$CO column density in the 1$_{1,0}$ level. 
In all panels, the beam size is shown in the lower left corner.}\label{Fig:peaktau}}
\end{figure*}

In order to estimate the H$_{2}$CO optical depth, we assume that all the observed radio continuum emission lies behind the molecular gas. Hence, the observed radio continuum emission in Fig.~\ref{Fig:peak-abs} contributes to $T_{\rm c}$ in Eq.~\ref{f.radiative}. Because of the widespread H$_2$CO distribution (see Sect.~\ref{sec.mor}), $f_{\rm b}$ is simply assumed to be unity to estimate $\tau$. Figure~\ref{Fig:peaktau} shows the derived peak optical depth distribution at angular resolutions of 3\arcmin\,and 10$\rlap{.}$\arcmin8. The peak optical depth values are found to be lower than unity at all locations. The optical depths at an angular resolution of 3\arcmin\,are generally higher than those at 10$\rlap{.}$\arcmin8 resolution. For the angular resolution of 10$\rlap{.}$\arcmin8, all the peak optical depth values are lower than 0.4. The maximum $\tau$ of $\sim$0.34 is located at the region centered at $l$=77.877\degree, $b$=0.865\degree. This suggests that H$_{2}$CO (1$_{1,0}$--1$_{1,1}$) is optically thin at almost all locations. We also note that the optical depth would be underestimated if only a small fraction of the radio continuum emission contributes to the background emission in Eq.~\ref{f.radiative}. However, optical observations have shown that Cygnus~X is seen as a dark patch in the sky \citep[see Fig.~1 in][for instance]{2006A&A...458..855S}, which suggests that most radio continuum emission from the H{\scriptsize II} regions should lie behind the molecular clouds. Toward the same line of sight, the H$_{2}$CO gas behind the radio continuum emission is likely weaker than the H$_{2}$CO gas in the front, because the absorption can be enhanced against radio continuum emission. Therefore, the derived optical depths are likely reliable.


The column density of H$_{2}$CO in the upper energy level, $N_{1_{1,0}}$,can be estimated from the optical depth using the following formula \citep[Eq. 30 in][]{2015PASP..127..266M}:
\begin{equation}\label{f.Nu}
    N_{1_{1,0}} = \frac{3h}{8 \pi^{3}\mu_{\rm lu}^{2} }\left[{\rm exp}\left(\frac{h\nu}{kT_{\rm ex}}\right)-1\right]^{-1}\int \tau~{\rm d}\varv\;,
\end{equation}
where $h$ is the Planck constant, $\mu_{\rm lu}$ is the dipole moment of 2.33 D \citep{Fabricant1977}, and $k$ is the Boltzmann constant. Assuming a constant $T_{\rm ex}$ of 1.6~K, Eq.~(\ref{f.Nu}) becomes
\begin{equation}\label{f.Nu_num}
    N_{1_{1,0}} = 9.45\times 10^{12}\int \tau~{\rm d}\varv\;{\rm cm}^{-2}\;.
\end{equation}
Integrating the optical depth over the velocity range from $-$10~\kms\,to 20~\kms, we derive the H$_{2}$CO column density in the $1_{1,0}$ level with Eq.~(\ref{f.Nu_num}), and the results are presented in Fig.~\ref{Fig:peaktau}c. The H$_{2}$CO column densities range from $3\times 10^{11}$ to $9.0\times 10^{12}$~cm$^{-2}$ with a median value of $2.9\times 10^{12}$~cm$^{-2}$ in the $1_{1,0}$ level. We note that variations in the  excitation temperatures can affect the accuracy of the column density determination. If the expected excitation temperatures vary from 1.2--2~K, the assumption of constant excitation temperature will lead to an uncertainty of a factor of $\sim$2 in the derived $N_{1_{1,0}}$. Based on the method introduced in Appendix~\ref{app.radex}, we derive the total ortho-H$_{2}$CO column density to range from $8\times 10^{11}$ to $2.3\times 10^{13}$~cm$^{-2}$ with a median value of $7.4\times 10^{12}$~cm$^{-2}$.

\subsubsection{Decomposition}\label{sec.decomp}
\begin{figure*}[!htbp]
\centering
\includegraphics[width = 0.95 \textwidth]{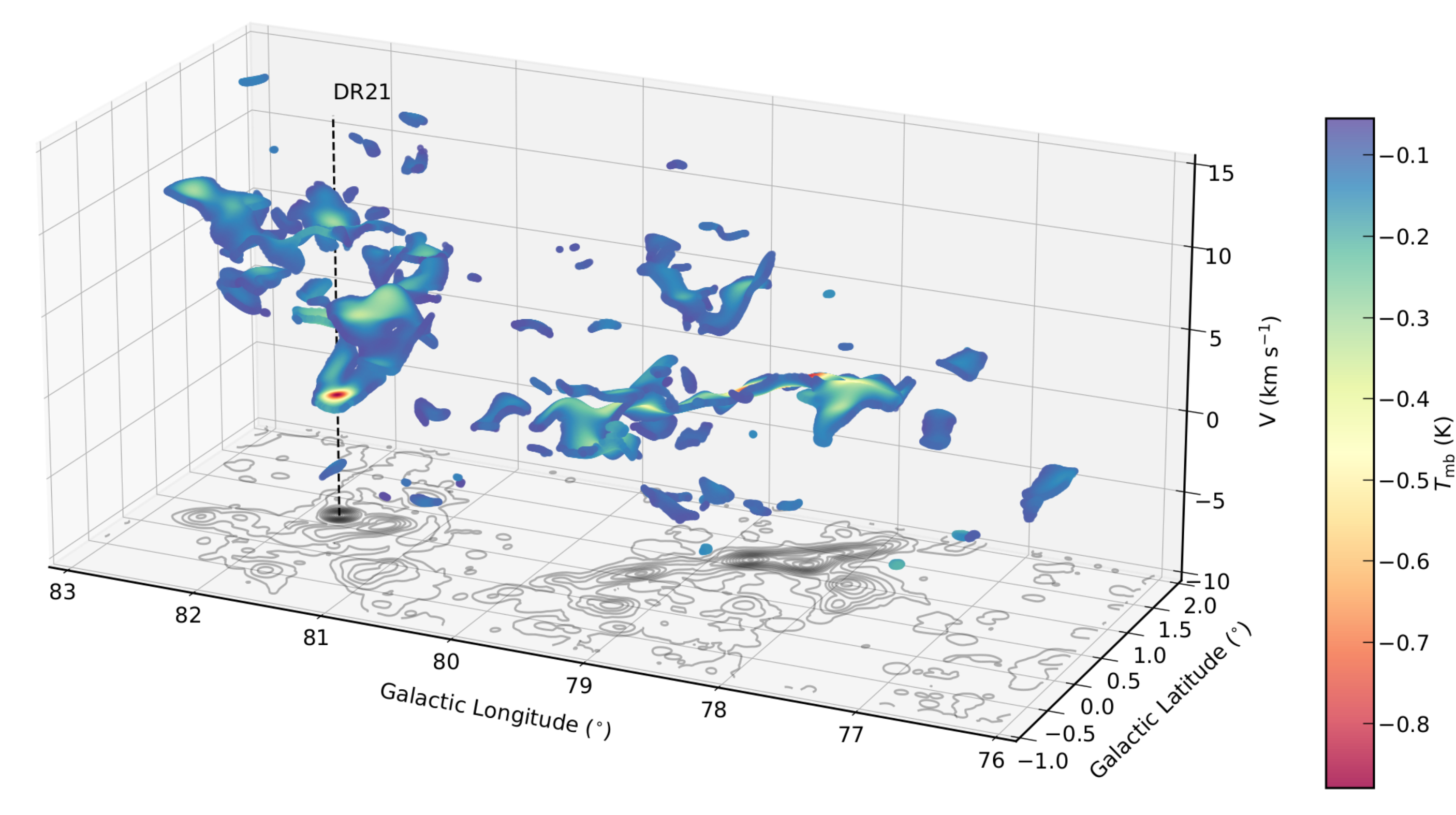}
\includegraphics[width = 0.95 \textwidth]{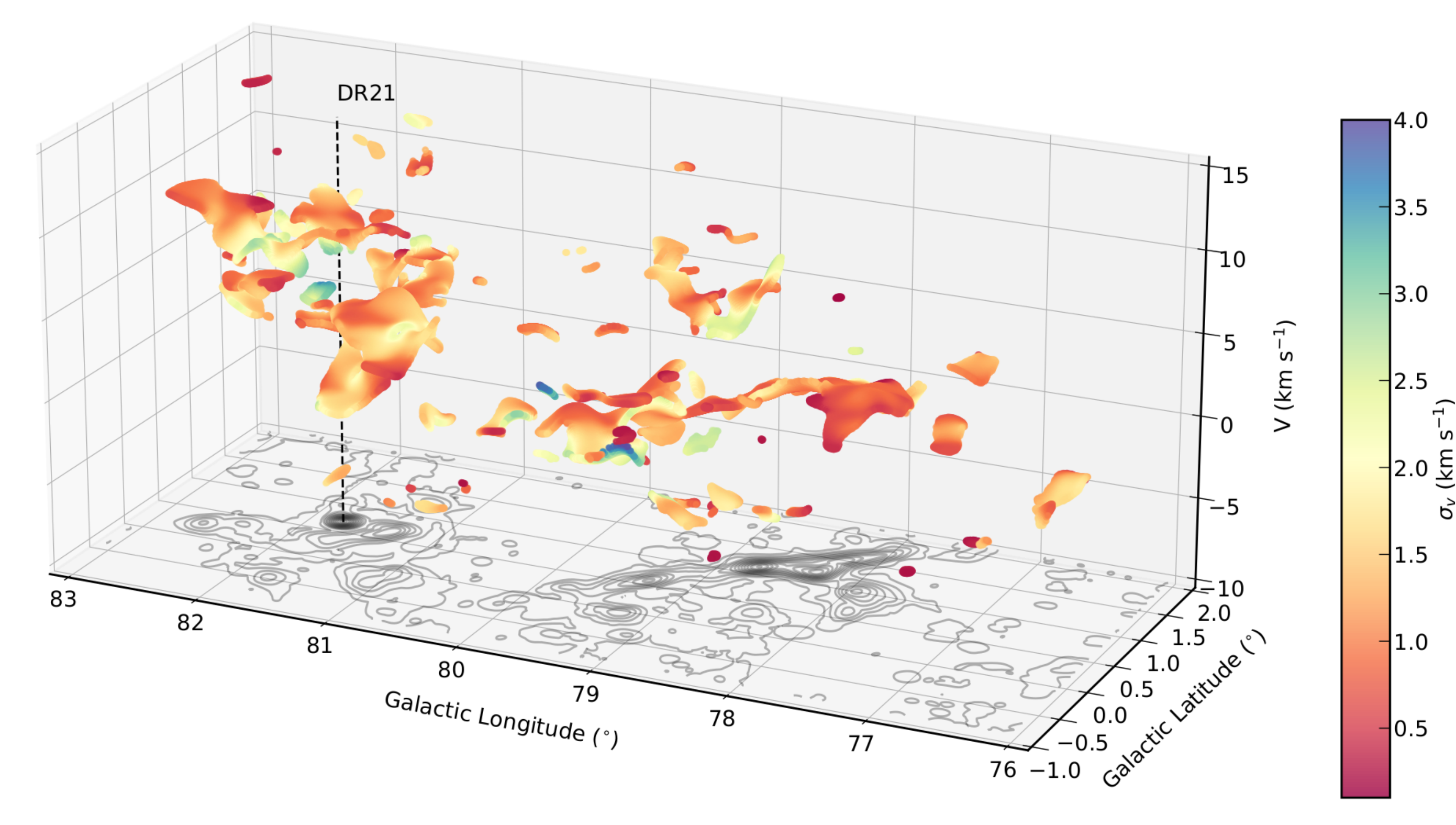}
\caption{{3D view of the decomposition of the observed H$_{2}$CO (1$_{1,0}$--1$_{1,1}$) spectra in Cygnus~X. The line of sight toward DR21 is indicated by the black dashed line. The upper panel shows the distribution of the fitted intensity. The H$_{2}$CO (1$_{1,0}$--1$_{1,1}$) peak absorption contours are shown at the base of the plot, and the contours are the same as shown in Fig.~\ref{Fig:peak-abs}. The lower panel shows the velocity dispersion distribution. The interactive version of this 3D view can be accessed via the links (Top: \href{https://gongyan2444.github.io/3D/cyg-h2co-amp.html}{high sampling}, \href{https://gongyan2444.github.io/3D/cyg-h2co-amp-low.html}{low sampling}; Bottom: \href{https://gongyan2444.github.io/3D/cyg-h2co-dis.html}{high sampling}, \href{https://gongyan2444.github.io/3D/cyg-h2co-dis-low.html}{low sampling}). }\label{Fig:decomp}}
\end{figure*}

The H$_{2}$CO (1$_{1,0}$--1$_{1,1}$) transition is comprised of six HFS lines \citep[e.g.,][]{1971ApJ...169..429T}. The overlapping HFS lines might introduce uncertainties in the fitted velocities and line widths. We therefore performed a simulation to study the impact of the HFS lines to test the effects. The results, which are presented in Appendix~\ref{app.hfs}, demonstrate that the velocity centroid derived by Gaussian fitting can have an intrinsic velocity shift of $-$0.12~\kms\,to 0.03~\kms\,and the line widths can be overestimated by approximately a factor of 1.5--2.5. In order to properly decompose the H$_{2}$CO spectra, we simultaneously fit six Gaussian components to the observed spectra on the basis of the rest frequencies and relative line strengths of the six HFS lines \citep[][]{2005JMoSt.742..215M} using the `LMFIT'\footnote{\url{https://lmfit.github.io/lmfit-py/}} python package \citep{newville_matthew_2014_11813}. Since most of the H$_{2}$CO (1$_{1,0}$--1$_{1,1}$) absorption is expected to be optically thin (see discussion in Sect.~\ref{sec.exc}), the method should be valid across the whole region. In regions with multiple velocity components, first, the brightest H$_{2}$CO absorption component along the line of sight is fitted, which is followed by fitting an additional component to the residual if significant. We repeat this process until the peak residual absorption is no brighter than 5$\sigma$. The chosen threshold allows us to avoid fit results with low confidence levels. As an example, Figure~\ref{Fig:specfit} presents the two-component fitting to the spectrum along the line of sight toward DR21. Figure~\ref{Fig:decomp} shows the fitted results for the complete H$_2$CO distribution in Cygnus X covered by us. From its upper panel, it is evident that the fitted velocities are between $-$7~\kms\, and 15~\kms. The lower panel of Fig.~\ref{Fig:decomp} suggests that the velocity dispersions are within the range of 0.16--4.04~\kms\, with a median value of 1.07~\kms. These are consistent with early statistical results of 34 positions in Cygnus~X \citep{1988A&A...191..313P}.

\begin{figure*}[!htbp]
\centering
\includegraphics[width = 0.95 \textwidth]{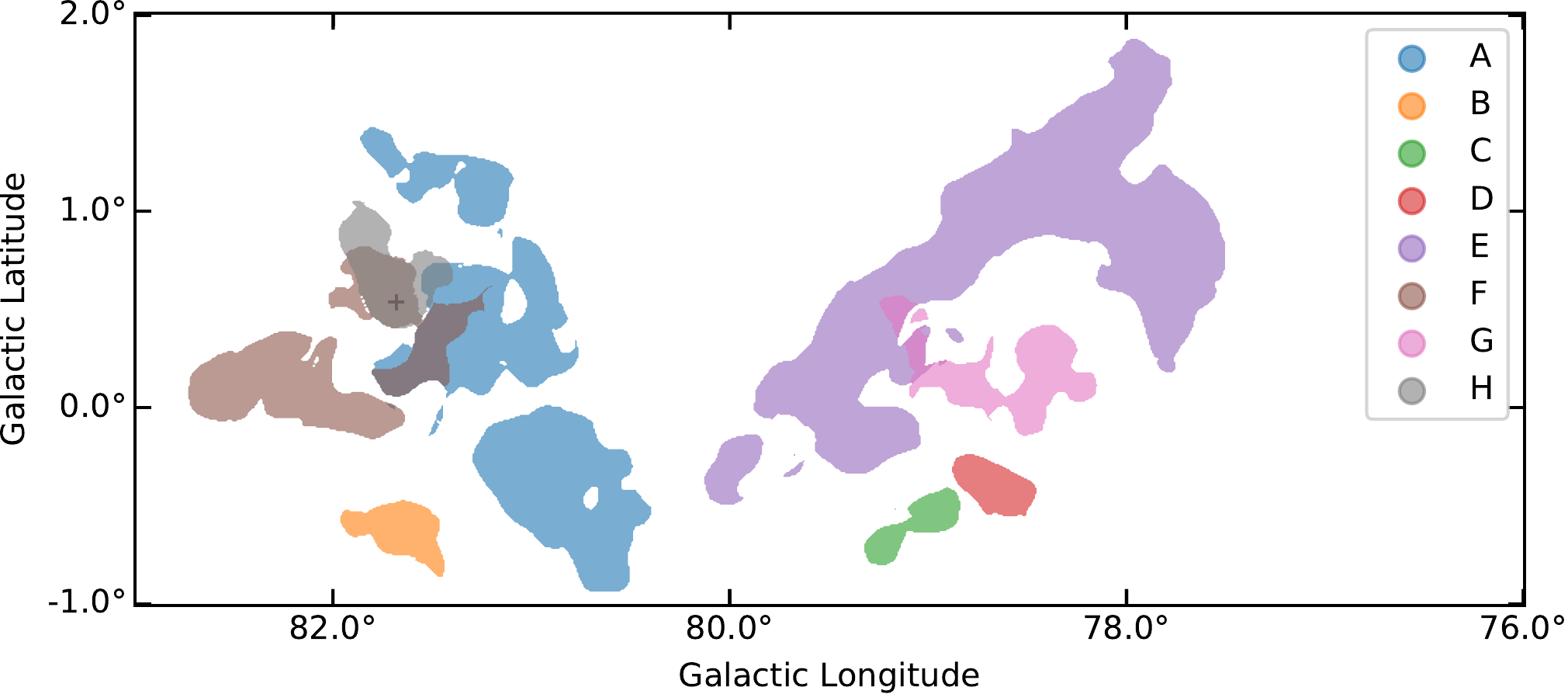}
\caption{{Eight coherent cloud structures derived from the DBSCAN algorithm. The different structures are labeled with different colors. The cross marks the position of DR21.}\label{Fig:cluster}}
\end{figure*}

Before investigating the velocity information, we first apply the clustering algorithm DBSCAN\footnote{\url{https://scikit-learn.org/stable/modules/generated/sklearn.cluster.DBSCAN.html}}  \citep[Density-based spatial clustering of applications with noise, e.g.,][]{Ester96, Schubert17, 2020ApJ...898...80Y}, to the fitted results (coordinates, LSR velocities, and velocity dispersions) in order to assign the observed absorption to different coherent cloud structures. The algorithm requires two parameters, $\epsilon$ and $p_{\rm min}$. $\epsilon$ corresponds to the maximum distance between two samples for one to be considered as being in the neighborhood of the other, while $p_{\rm min}$ represents the minimum number of points required to form a coherent region. We use twice the number of dimensions (i.e., 4 considering that the four dimensions correspond to Galactic longitude, Galactic latitude, LSR velocity, and velocity dispersion) of our data as $p_{\rm min}$. The results of the clustering algorithm depend sensitively on $\epsilon$, with higher values of $\epsilon$ leading to more extended cloud structures (see discussions in Appendix~\ref{app.dbscan}). 
Since we only intend to study extended cloud structures that are well resolved at an angular resolution of 10$\rlap{.}$\arcmin8, we manually increase $\epsilon$ to 0.25 (see discussions in Appendix~\ref{app.dbscan}). Consequently, we detect eight coherent and extended cloud structures that are well resolved at the angular resolution of 10$\rlap{.}$\arcmin8, the results being shown in Fig.~\ref{Fig:cluster}. The cloud structures, labelled A to H cover areas of 51--1055~pc$^{2}$ with cloud E in CygX-South being the most extended and coherent cloud structure.

In Fig.~\ref{Fig:cluster}, three cloud structures (i.e., A, F, H) overlap each other along the line of sight in CygX-North, while two cloud structures (i.e., E, G) overlap along the line of sight toward CygX-South. In CygX-North, the three cloud structures are characterized by LSR velocities of $-$3~\kms, 5~\kms, and 8~\kms. Based on previous studies \citep[e.g.,][]{2003ApJ...596..344C, 2010A&A...520A..49S, 2019PASJ...71S..12D}, the $-$3~\kms\,component (i.e., H) mainly stems from the molecular gas associated with DR21, while the 8~\kms\,component (i.e., F) arises from the W75N component in front of molecular clouds associated with DR21. Previous studies suggest that the interaction between the two components may trigger massive star formation in this region \citep[e.g.,][]{1978ApJ...223..840D, 2019PASJ...71S..12D}. The 5~\kms\,cloud (i.e., A) is connected to clouds F and H in both spatial and velocity spaces, which implies that cloud F is also interacting with the other two clouds. Toward CygX-South, clouds E and G also overlap along the line of sight although we do not see signatures of massive star formation at the intersection. 

\begin{figure*}[!htbp]
\centering
\includegraphics[width = 0.95 \textwidth]{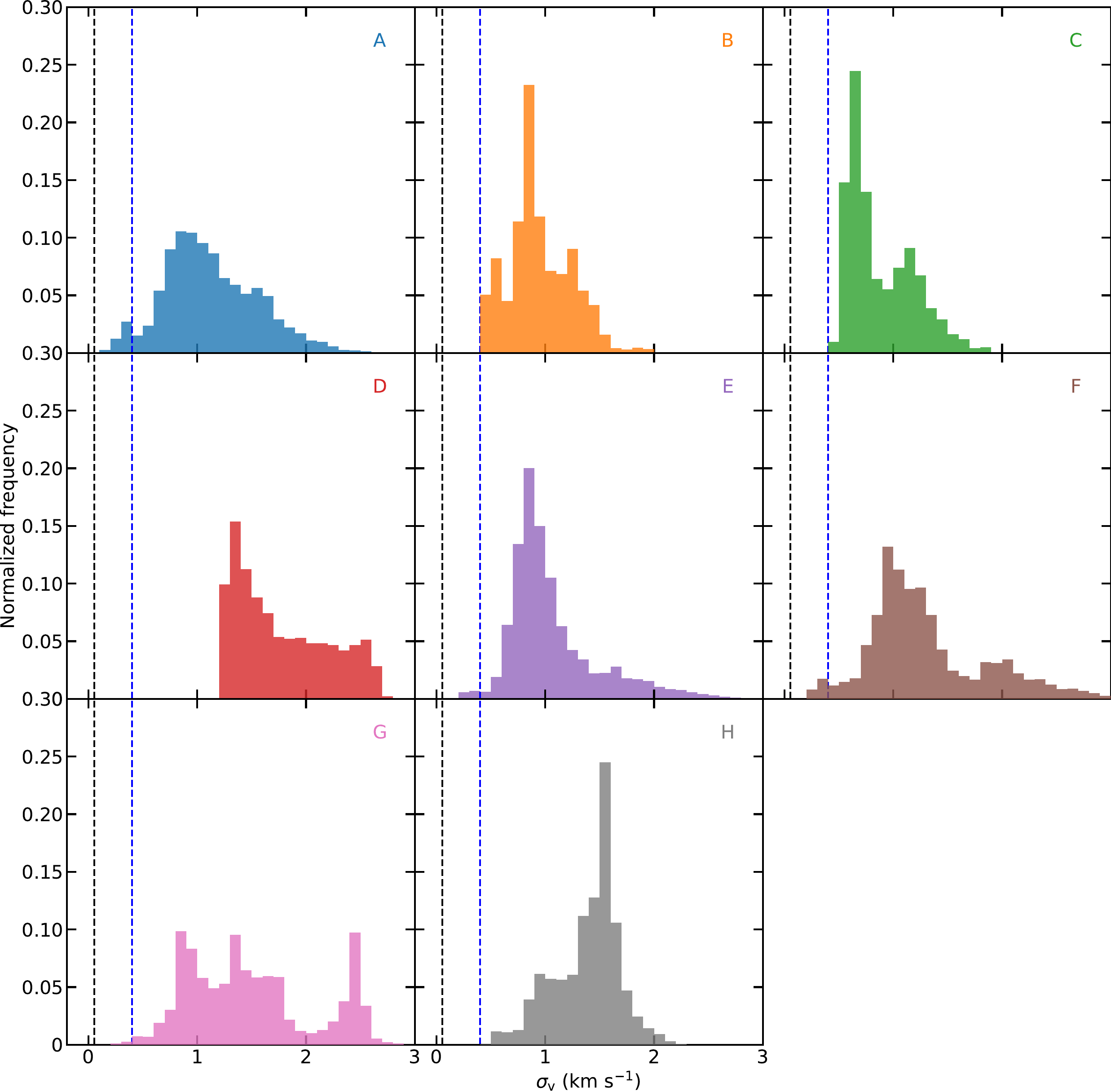}
\caption{{Histogram of observed velocity dispersions of the eight cloud structures derived from the Effelsberg H$_{2}$CO data. The black and blue vertical dashed lines represent the thermal velocity dispersion of H$_{2}$CO and twice the sonic speed of 0.19~\kms\,at a kinetic temperature of 10~K, respectively.}\label{Fig:eff-vdis}}
\end{figure*}

The fitted LSR velocity centroids appear to show ordered velocity gradients, and these gradients will be further discussed in Sect.~\ref{dis:vg}. Figure~\ref{Fig:eff-vdis} presents the statistics of the velocity dispersions for the eight cloud structures that have a median value of 1.04~\kms. The observed velocity dispersions, $\sigma_{\rm v}$, consist of contributions from thermal and non-thermal motions, $\sigma_{\rm t}$ and $\sigma_{\rm nt}$:
\begin{equation}
    \sigma_{\rm v} = \sqrt{\sigma_{\rm t}^{2} +\sigma_{\rm nt}^{2}}\;.
\end{equation} 
The thermal velocity dispersion can be estimated from the following relation:
\begin{equation}
    \sigma_{\rm t} = \sqrt{\frac{kT_{\rm k}}{m_{i}}}\;,
\end{equation}
where $k$ is the Boltzmann constant, $T_{\rm k}$ is the kinetic temperature, $m_{i}$ is the mass of the molecule (e.g., $m_{i}=$30 for H$_{2}$CO). Since molecular clouds have typical kinetic temperatures of 10~K, the characteristic thermal velocity dispersion is 0.05~\kms. As is evident from Fig.~\ref{Fig:eff-vdis}, the observed velocity dispersions are much higher than 0.05~\kms, suggesting that the molecular gas in Cygnus~X is dominated by non-thermal motions on a 4.4~pc (i.e., 10$\rlap{.}$\arcmin8) scale. The Mach number is defined as 
\begin{equation}
    \mathcal{M} = \sigma_{\rm nt}/c_{\rm s}\;,
\end{equation}
where $c_{\rm s}$ is the sound speed of molecular gas. $c_{\rm s}$ is 0.19~\kms\,at $T_{\rm k}$=10~K where the mean molecular weight is taken to be 2.37 \citep{2008A&A...487..993K}. Figure~\ref{Fig:eff-vdis} suggests that most of the molecular gas traced by H$_{2}$CO absorption has $\mathcal{M} >2$, which is indicative of nearly ubiquitous supersonic motions in Cygnus~X.

The minimum Mach number could be slightly overestimated because of the spectral dilution. Our minimum velocity dispersion is about 0.5~\kms\,which corresponds to a Gaussian line width of $\sim$1.2~\kms. Taking the spectral dilution caused by the 0.5~\kms\,channel into account, the broadening line width becomes 1.3~\kms, which indicates about 8\% broader than the intrinsic value. Therefore, the spectral dilution does not play a crucial role.  

\subsection{Formaldehyde absorption on small scales}\label{sec.vla}
Our GLOSTAR VLA D array observations provide the first unbiased H$_{2}$CO (1$_{1,0}$--1$_{1,1}$) absorption survey toward Cygnus~X on a scale of $\sim 0.17$~pc. This has led to the robust detection ($\geq 5\sigma$) of H$_{2}$CO ($1_{1,0} - 1_{1,1}$) absorption toward three compact radio continuum sources, DR21, DR22, and G76.1883+0.0973 (also known as IRAS~20220+3728), which are known to be H{\scriptsize II} regions \citep[e.g.,][]{1991ApJS...75.1011G, 1994ApJS...91..659K, 2007A&A...476.1243M}. The sparse number of absorption detections toward compact sources is mainly attributed to our sensitivity -- given our $1\sigma$ sensitivity of about 0.02~Jy~beam$^{-1}$ (or 1.7~K) at a channel width of 0.5~\kms, only strong absorption features can be detected by our observations. 

\begin{figure*}[!htbp]
\centering
\includegraphics[height = 0.50 \textwidth]{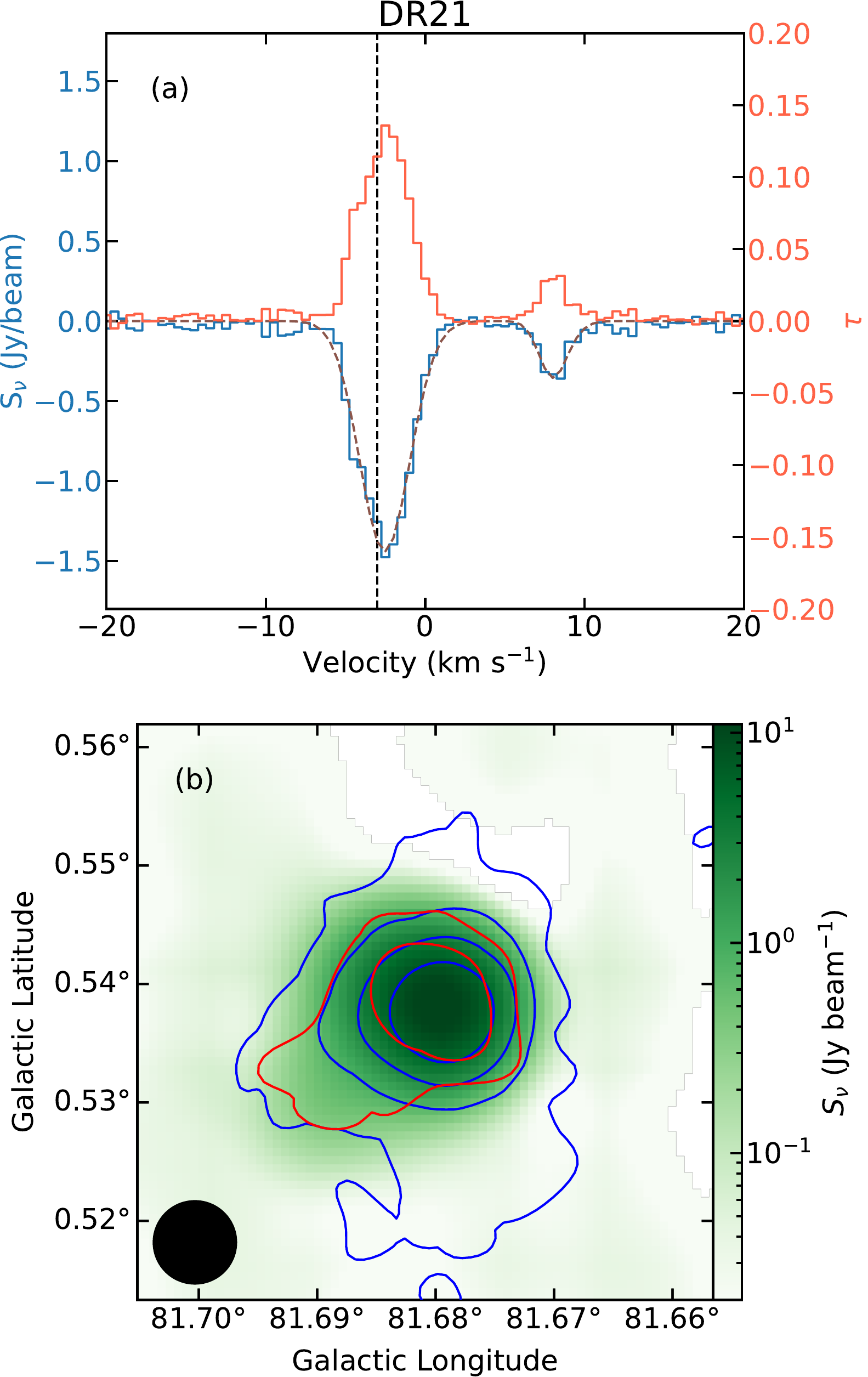}
\includegraphics[height = 0.50 \textwidth]{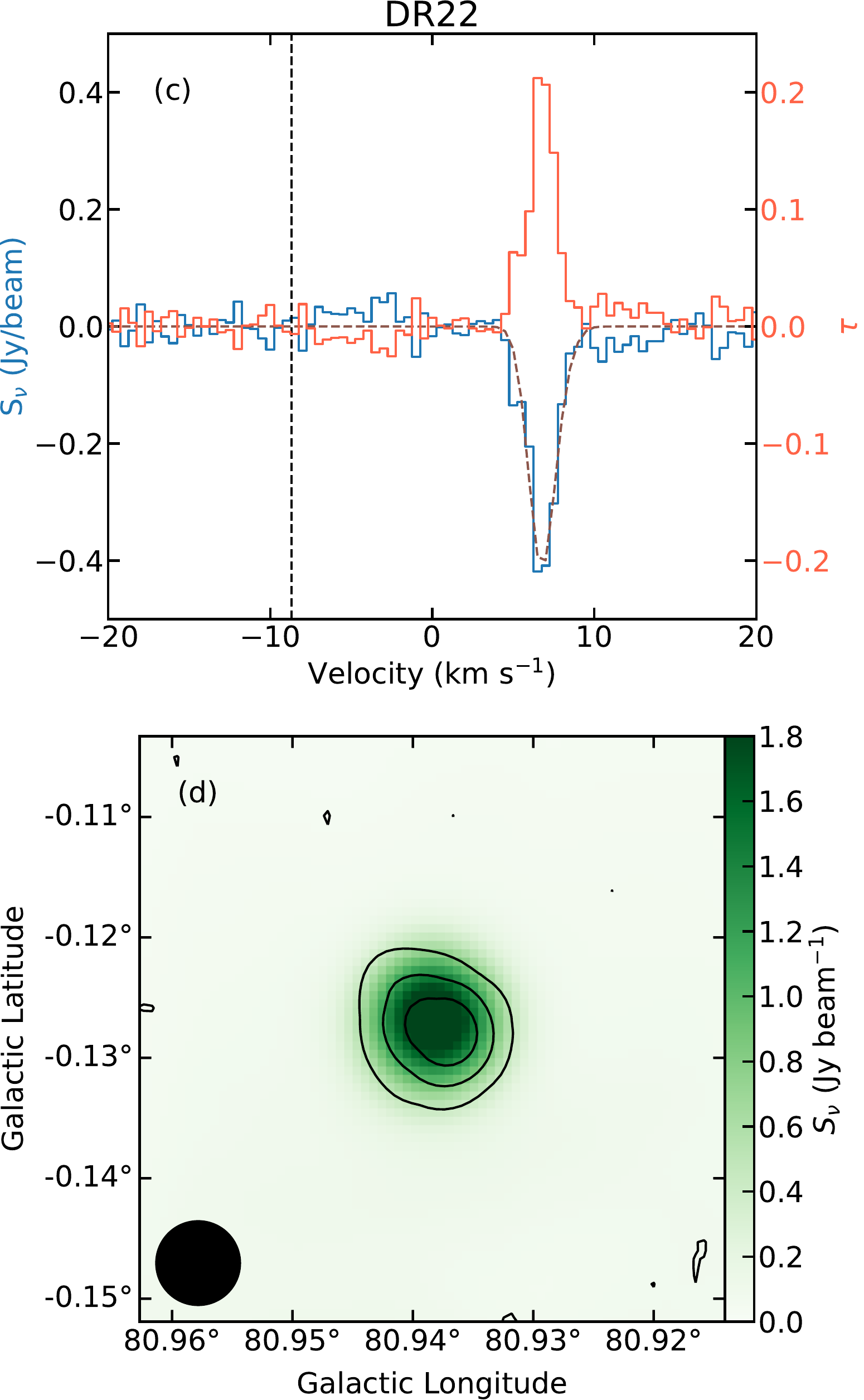}
\includegraphics[height = 0.50 \textwidth]{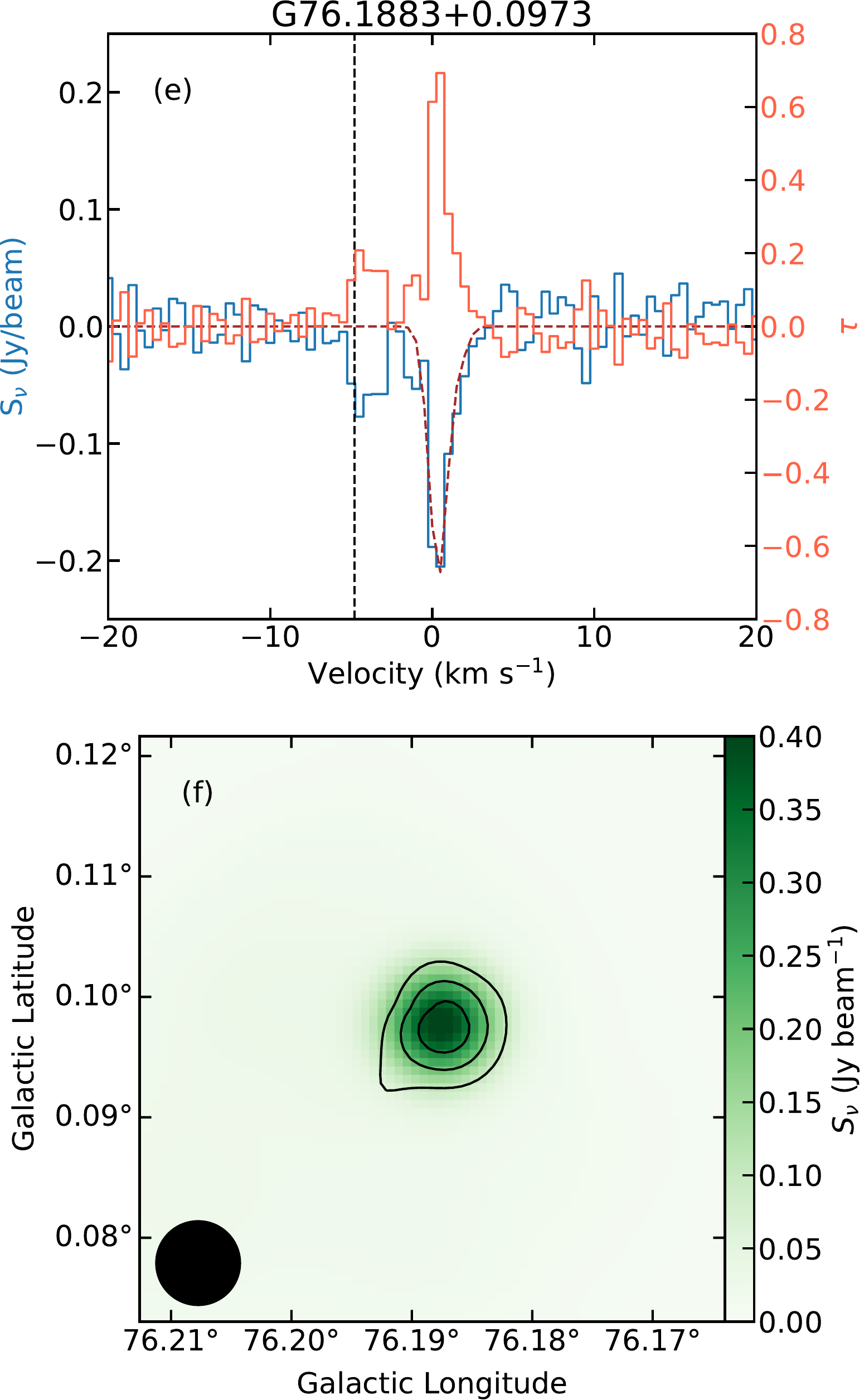}
\caption{{\textit{Top}: Observed H$_{2}$CO (1$_{1,0}$--1$_{1,1}$) spectra of DR21 (a), DR22 (c), and G76.1883+0.0973 (e) overlaid on the fit results indicated by the brown dashed lines. The derived optical depth spectra are shown by the red lines. In panels (a), (c), (e), the black dashed vertical lines represent the LSR velocities of the H{\scriptsize II} regions obtained from the radio recombination line measurements (Khan et al. in prep). \textit{Bottom}: VLA+Effelsberg 4.9~GHz radio continuum emission of DR21 (b), DR22 (d), and G76.1883+0.0973 (f) overlaid with the H$_{2}$CO (1$_{1,0}$--1$_{1,1}$) absorption contours. For DR21, the blue and red contours represent the H$_{2}$CO (1$_{1,0}$--1$_{1,1}$) absorption peak for the $-$3~\kms\,and 8~\kms\,components, respectively. The contours start at $-$0.1~Jy~beam$^{-1}$ (5$\sigma$), with each subsequent contour being twice the previous one. For DR22 and G76.1883+0.0973, the contours start at $-$0.1~Jy~beam$^{-1}$ (5$\sigma$) and decrease by 0.04~Jy~beam$^{-1}$. The synthesized beam is shown in the lower left corner of each panel. All the continuum and spectral line data are from the combination of the VLA D configuration and the Effelsberg single-dish observations.}\label{Fig:vla-image}}
\end{figure*}

As shown in Fig.~\ref{Fig:vla-image}, these bright H$_{2}$CO (1$_{1,0}$--1$_{1,1}$) absorption distributions match the distributions of the 4.9~GHz radio continuum emission, which strongly supports that these features are due to absorption of continuum emission (as opposed to the CMB). DR21, DR22, and G76.1883+0.0973 are well known massive star formation regions \citep[e.g.,][]{2007A&A...476.1243M, 2021A&A...651A..87O}. Therefore, all the absorption features detected by the high angular resolution data are in the direction of massive star forming regions.

As is evident in Fig.~\ref{Fig:vla-image}a, the spectrum toward DR21 exhibits two velocity components at $-$3~\kms\,and 8~\kms. Their distributions show that both components are against the radio continuum emission of DR21 (see Fig.~\ref{Fig:vla-image}b). The $-$3~\kms\,component has a higher optical depth than the 8~\kms\,component, which can explain the fact that the 8~\kms\,component was not detected by previous NH$_{3}$ (1,1) and 6.7~GHz methanol line observations \citep{2003ApJ...596..344C, 2021A&A...651A..87O}. The $-$3~\kms\,component was also detected in absorption in the 1667 MHz OH and 6.7~GHz CH$_{3}$OH lines \citep[see Fig.~8 in][]{2021A&A...651A..87O}, but the blueshifted wing-like features detected in the 1667 MHz OH and 6.7~GHz CH$_{3}$OH lines are absent in H$_{2}$CO (1$_{1,0}$--1$_{1,1}$). 

In Fig.~\ref{Fig:vla-image}, we also compare the H$_{2}$CO (1$_{1,0}$--1$_{1,1}$) LSR velocities with the LSR velocities of the three H{\scriptsize II} regions derived from radio recombination line (RRL) observations at a similar angular resolution (Khan et al. in prep). The velocity differences are $0.5 \pm 0.3$~\kms, $15.5 \pm 0.2$~\kms, and $5.2 \pm 0.8$~\kms\,for DR21, DR22, and G76.1883+0.0973, respectively. We find that the molecular gas is redshifted with respect to the RRL velocities at least toward DR22 and G76.1883+0.0973. The presence of H$_{2}$CO (1$_{1,0}$--1$_{1,1}$) absorption suggests that the molecular gas lies in front of the H{\scriptsize II} regions. The large velocity differences indicate that the molecular gas is likely not associated with the ionized gas for the two H{\scriptsize II} regions.


Following the same method used in Sect.~\ref{sec.exc}, we also derived the H$_{2}$CO optical depths in the VLA data (see Figure~\ref{Fig:vla-image}). All the derived optical depth values are lower than 0.8. Among the  detections, the line of sight toward G76.1883+0.0973 has the highest optical depth of $\sim 0.7$. It is worth noting that G76.1883+0.0973 does not reside in the bright cloud structures of CygX-North and CygX-South (see Fig.~\ref{Fig:peak-abs}). We also derive the H$_{2}$CO column densities in the $1_{1,0}$ level, which range from $4.2\times 10^{12}$ to $7.3\times 10^{13}$~cm$^{-2}$ with a median value of $6.8\times 10^{12}$~cm$^{-2}$.

\begin{figure}[!htbp]
\centering
\includegraphics[width = 0.45 \textwidth]{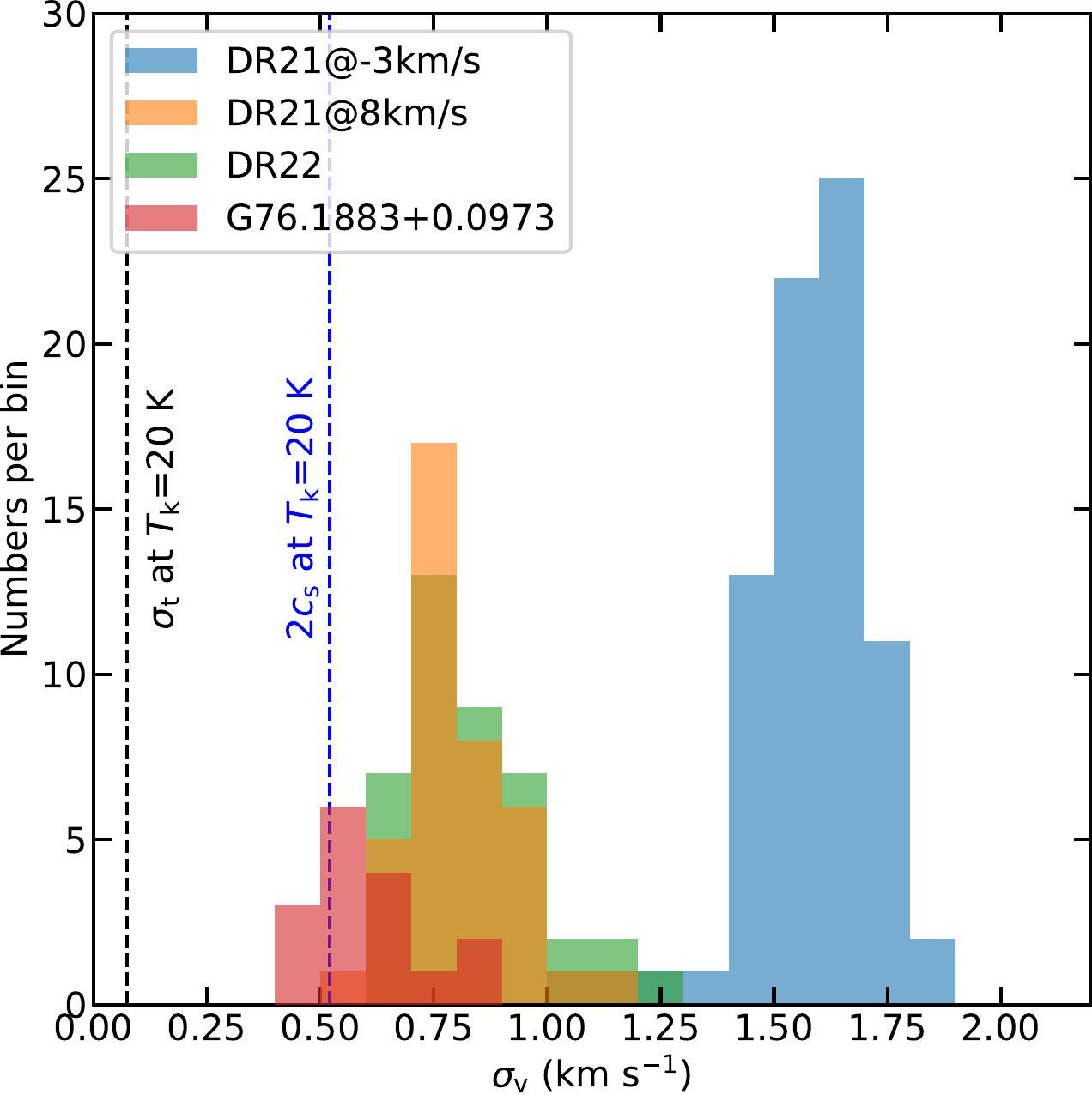}
\caption{{Histogram of the observed velocity dispersions derived from the VLA+Effelsberg combined H$_{2}$CO data. The black and blue vertical dashed lines represent the thermal velocity dispersion of H$_{2}$CO and twice the sonic speed at a kinetic temperature of 20 K, respectively.}\label{Fig:vla-vdis}}
\end{figure}

In order to study the kinematics, we also perform a decomposition of the VLA+Effelsberg data as in Sect.~\ref{sec.decomp}. Because two velocity components at $-$3~\kms\,and 8~\kms\,are evident toward DR21 in Fig.~\ref{Fig:vla-image}, we investigate them separately.
Figure~\ref{Fig:vla-vdis} shows a histogram of the velocity dispersions for the four components in the three regions. The distributions have mean values of 1.61~\kms, 0.91~\kms, 0.68~\kms, and 0.60~\kms\,for the $-$3~\kms\,component of DR21, the 8~\kms\,component of DR21, DR22, and G76.1883+0.0973, respectively. Based on previous ammonia observations, kinetic temperatures range from $17-28$~K around DR21 and DR22 \citep{2019ApJ...884....4K}, which corresponds to thermal H$_{2}$CO velocity dispersions of 0.07--0.09 \kms. It is evident that the observed velocity dispersions are much higher than what is expected from thermal motion. Hence, they are dominated by non-thermal motion (i.e., turbulence). Furthermore, the $-$3~\kms\,component toward DR21 appears to have higher velocity dispersions than the other regions by a factor of $\sim 2$, and is thus more turbulent. It is worth noting that the $-$3~\kms\,component toward DR21 appears to be the only component associated with an H{\scriptsize II} region. In order to estimate Mach numbers, we use a kinetic temperature of 20~K as a fiducial case (i.e., $c_{\rm s}=0.26$~\kms). Figure~\ref{Fig:vla-vdis} suggests that most of detected H$_{2}$CO absorption has Mach numbers of $> 2$, indicating that supersonic turbulence commonly exists in Cygnus~X on scales of 0.17~pc (statistical results on the 4.4~pc scale are presented in Sect.~\ref{sec.decomp}). 

\subsection{Nondetections}
Although our Effelsberg observations also cover the H$_{2}^{13}$CO (1$_{1,0}$--1$_{1,1}$) line, it is not detected in the Cygnus~X region. At the HPBW of 3\arcmin\,and the channel width of 2.5~\kms, the 1$\sigma$ noise levels range from 0.02~K to 0.14 K with a median value of 0.06 K. Based on Eq.~\ref{f.radiative}, we obtain a 3$\sigma$ upper limit of 0.18~K at the position of the peak continuum emission position ($T_{\rm c}$=29.7~K, i.e., DR21), which corresponds to an upper limit of 0.006 for the optical depth. Previous observations have detected H$_{2}^{13}$CO (1$_{1,0}$--1$_{1,1}$) absorption toward DR21 \citep{1976A&A....51..303W, 1980A&A....82...41H, 2019ApJ...877..154Y}, but with intensities that correspond to signals that are below our detection limit.

H$_{2}$CO (1$_{1,0}$--1$_{1,1}$) masers are known to be associated with massive star formation in our Galaxy \citep[e.g.,][]{2004ApJS..154..579A}, but these masers have only been detected in 11 massive star forming regions of the Milky Way to date \citep{1980A&A....84L...1F, 1983MNRAS.205P..27W, 1994ApJ...430L.129P, 2004ApJS..154..579A, 2007ApJS..170..152A, 2008ApJS..178..330A, 2015A&A...584L...7G, 2017ApJ...851L...3C, 2019ApJS..244...35L, 2022MNRAS.509.1681M}. With both our Effelsberg-100 m and VLA observations, we did not detect any H$_{2}$CO maser in Cygnus~X. The 3$\sigma$ upper limits for the masers are $\sim 0.09-0.3$~Jy at a channel width of 0.19~\kms\,for the Effelsberg observations and $\sim$0.07~Jy at a channel width of 0.25~\kms\,for the VLA D-configuration observations.

\section{Discussion}\label{Sec:dis}
\subsection{Absorbed photons on different scales}
As mentioned in Section 1.1, the H$_{2}$CO (1$_{1,0}$--1$_{1,1}$) line can be seen in absorption both against radio continuum sources and the CMB. Given the extent of bright radio continuum emission in Cygnus~X, it is not well known as to which source plays the dominant role as a background for absorption on different scales. In order to address this question, we investigate the relationship between the peak intensities of H$_{2}$CO absorption and the brightness temperatures of the 4.9~GHz radio continuum emission.

For our GLOSTAR VLA+Effelsberg results on a scale of $\sim$0.17~pc, all the detected absorption features are against bright continuum sources with brightness temperatures $> 20$~K (see Sect.~\ref{sec.vla}), suggesting that they are caused by absorption of photons mainly from the H{\scriptsize II} regions rather than the CMB. At the sensitivity limit of our observations, we cannot estimate the relative importance of radio continuum and the CMB for weak absorption features on this scale.  

\begin{figure}[!htbp]
\centering
\includegraphics[width = 0.45 \textwidth]{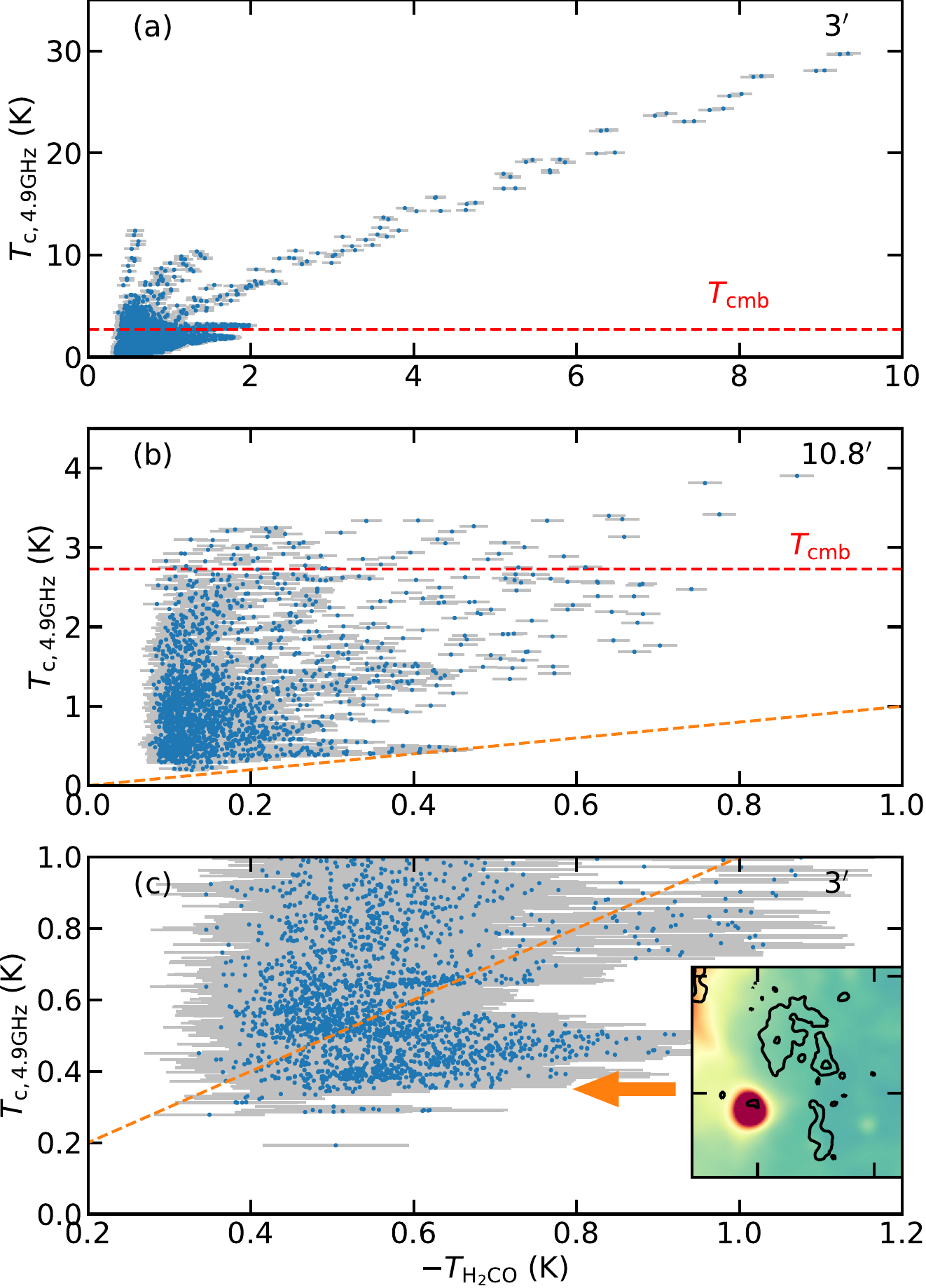}
\caption{{Comparison between the peak intensities of H$_{2}$CO absorption and the temperatures of the 4.89~GHz radio continuum emission at an angular resolution of 3\arcmin\,(a) and 10$\rlap{.}$\arcmin8 (b). All data points have signal-to-noise ratios of $>$5. In both panels, the red dashed line represents the CMB temperature of 2.73~K. (c) Same as panel (a) but zoomed into a narrower intensity range. The panel in the lower right corner is the same as Fig.~\ref{Fig:peak-abs}a but zoomed into the region where the absorption dip is greater than the radio continuum temperature. In panels (b) and (c), the orange dashed line marks the equality between H$_{2}$CO peak absorption and radio continuum temperature. In all panels, the error bars represent a 1$\sigma$ uncertainty.}\label{Fig:cmb-abs}}
\end{figure}

We further carry out a pixel-by-pixel comparison between the peak intensities of H$_{2}$CO absorption and the brightness temperatures of the 4.9~GHz radio continuum emission of the Effelsberg data, in which only the peak intensities of H$_{2}$CO absorption with at least 5$\sigma$ are taken into account. The results for two large scales (3\arcmin\,and 10$\rlap{.}$\arcmin8, i.e., 1.2~pc and 4.4~pc) are shown in Fig.~\ref{Fig:cmb-abs}. As mentioned above, Figure~\ref{Fig:cmb-abs}a is dominated more by compact sources at the high intensity end, and the points in the diagonal line arise from DR21. In contrast, Figure~\ref{Fig:cmb-abs}b shows the relationship for the extended H$_{2}$CO absorption features. In both panels, we find that a significant fraction of the Galactic 4.9~GHz continuum emission has brightness temperatures that are lower than the CMB \citep[2.73~K, e.g.,][]{2009ApJ...707..916F}. Especially for the extended H$_{2}$CO absorption (see Fig.~\ref{Fig:cmb-abs}b), about 97\% of points have Galactic 4.9~GHz radio continuum emission brightness temperatures $<$2.73~K. We also find that 9.1\% and 0.6\% of the H$_{2}$CO absorption dips can be greater than the Galactic radio continuum temperature in Fig.~\ref{Fig:peak-abs}c at the 1$\sigma$ and 3$\sigma$ significance levels, respectively. Such extreme cases are seen at around $l$=77.877\degree, $b$=0.865\degree. This unambiguously attributes the absorption primarily to the CMB. Overall, we expect that absorption of CMB photons contributes to extended H$_2$CO absorption features in addition to radio continuum emission.


\subsection{Formaldehyde abundance}\label{sec.col}
Based on the derived total column density of ortho-H$_{2}$CO (see Sect.~\ref{sec.exc} and Appendix~\ref{app.radex}), we can estimate its molecular fractional abundance with respect to H$_{2}$. Assuming, as usual, that the dust emission traces the H$_{2}$ column density \citep{2009ApJ...692...91G}, we use the \textit{Planck} 353~GHz map of the dust optical depth \citep{2014A&A...571A..11P} and the HI column density map from the Effelsberg-Bonn HI survey (EHBIS) \citep{2016A&A...585A..41W} to estimate the H$_{2}$ column densities in this study. The dust optical depth at 353~GHz, $\tau_{353}$ consists of contributions from both molecular (H$_{2}$) and atomic gas (HI). The column density of atomic gas, $N_{\rm HI}$, is related to $\tau_{353}$ by $N_{\rm HI} = 8.3\times 10^{25} \tau_{353}$~cm$^{-2}$ \citep{2014A&A...571A..11P}. Comparing the EHBIS HI column densities and the $\tau_{353}$-based HI column densities, we find that HI contributes to at least 26\% of the dust-based HI column densities for pixels with the detection of H$_{2}$CO absorption. Thus, the HI column density needs to be subtracted from the dust-based HI column density to calculate the H$_{2}$ column density which is determined as $N_{\rm H_{2}} = 0.5(8.3\times 10^{25} \tau_{353}-N_{\rm HI})$~cm$^{-2}$, where $N_{\rm HI}$ is based on the EBHIS HI column density map. All images are convolved to 10$\rlap{.}$\arcmin8 to calculate the H$_{2}$ column densities and the ortho-H$_{2}$CO fractional abundance. 

\begin{figure}[!htbp]
\centering
\includegraphics[width = 0.45 \textwidth]{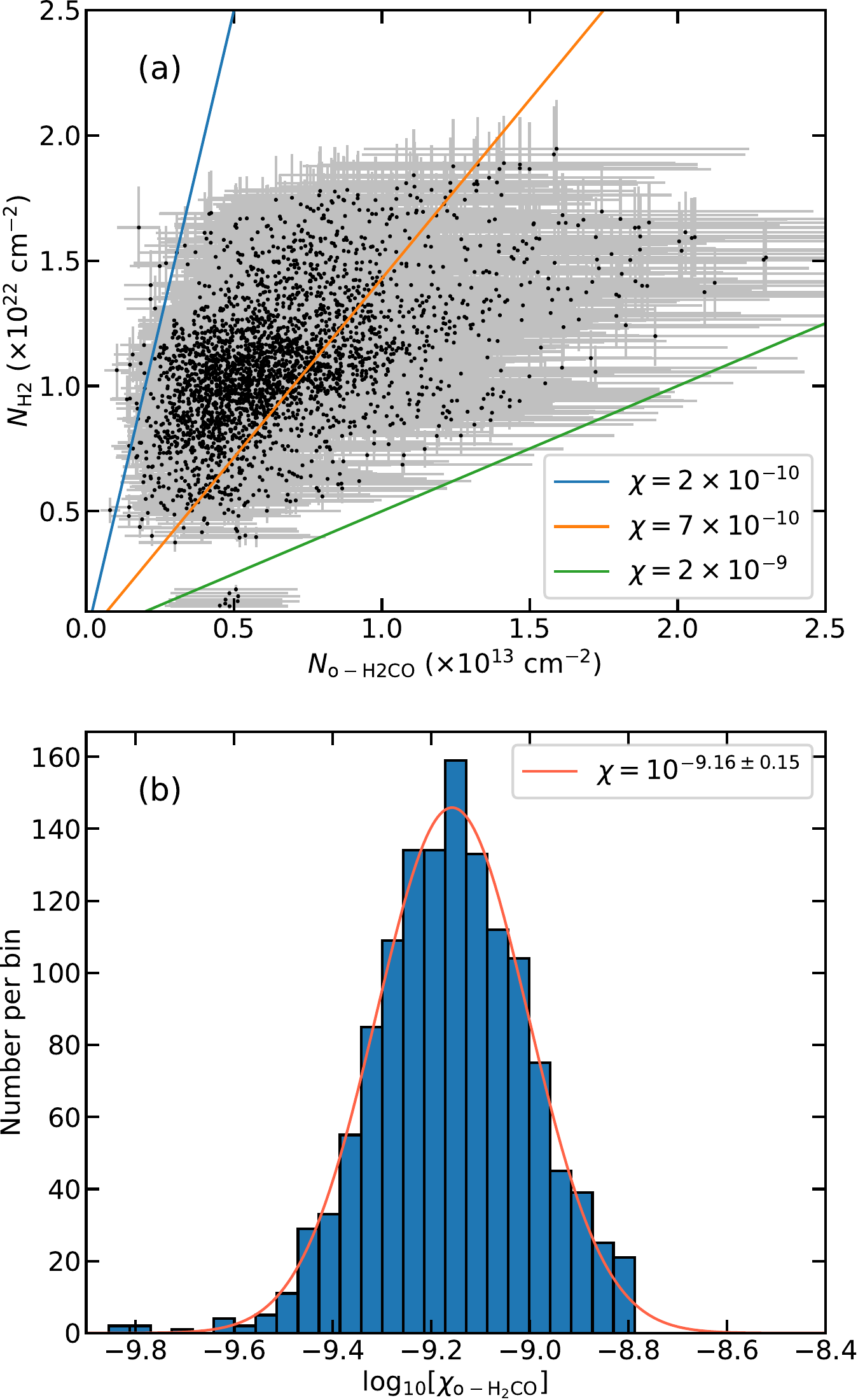}
\caption{{(a) H$_{2}$ column density as a function of the ortho-H$_{2}$CO column density. The three lines represent the fractional abundances of $2\times 10^{-10}$, $7\times 10^{-10}$, and $2\times 10^{-9}$, respectively. (b) Statistic histogram of the ortho-H$_{2}$CO abundance. The red line represents the Gaussian fit to the histogram.}\label{Fig:abun}}
\end{figure}

Figure~\ref{Fig:abun}a presents a comparison between the derived H$_{2}$ column densities and the ortho-H$_{2}$CO column densities. It is expected that the ortho-H$_{2}$CO column density increases with increasing H$_{2}$ column density, and the Pearson correlation coefficient is 0.49. The molecular fractional abundances are found to range from 1.4$\times 10^{-10}$ to 1.6$\times 10^{-9}$ with a median value of 6.9$\times 10^{-10}$ on a scale of 4.4~pc. On the other hand, the fractional abundances of ortho-H$_{2}$CO appear to be unaffected by the H$_{2}$ column densities on this scale, because the Pearson correlation coefficient between the ortho-H$_{2}$CO fractional abundances and the H$_{2}$ column densities is only $-$0.17.

As shown in Figure~\ref{Fig:abun}b, the histogram of the ortho-H$_{2}$CO abundances displays a Gaussian-like behavior in the logarithmic space. We perform a Gaussian fit to the histogram, which results in a mean abundance of 7.0$\times 10^{-10}$ with a dispersion of 0.15 dex (i.e., $10^{-9.16\pm 0.15}$). Our values are roughly consistent with the ortho-H$_{2}$CO abundances in the Galactic center and the W51 complex \citep{1983A&A...125..136G, 2015A&A...573A.106G}. This suggests that the abundances do not vary significantly for the different environments on the cloud scale. This also agrees with previous para-H$_{2}$CO studies that the fractional abundance of H$_{2}$CO is rather stable and the variation is usually within an order of magnitude in different environments \citep[e.g.,][]{2014A&A...563A..97G, 2020MNRAS.499.6018Z, 2021A&A...655A..12T}. 



\subsection{Comparison with other tracers}\label{sec.comp}
\begin{figure*}[!htbp]
\centering
\includegraphics[width = 0.95 \textwidth]{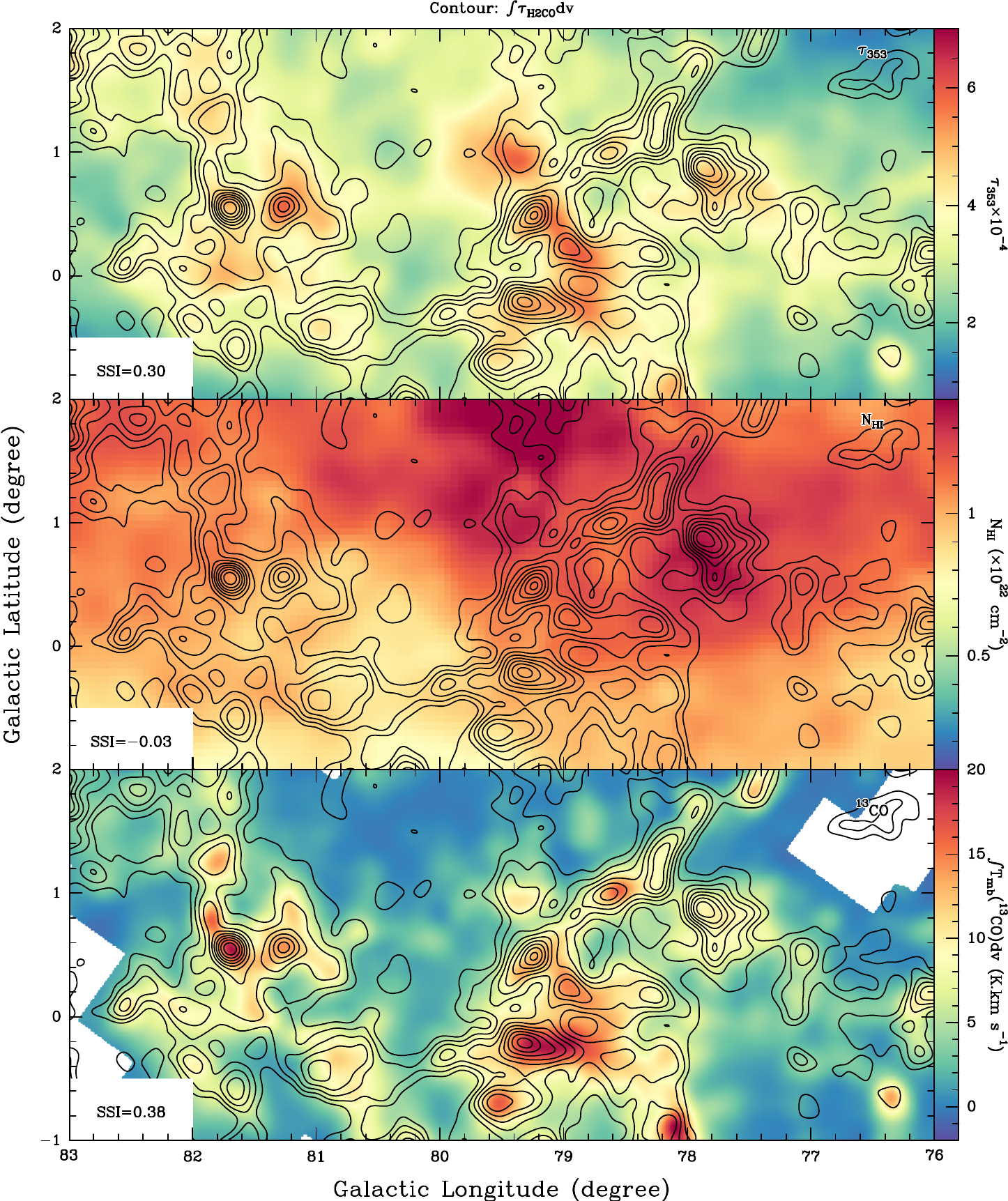}
\caption{{Comparison between the H$_{2}$CO distribution with that of $\tau_{353}$, $N_{\rm HI}$, and $^{13}$CO. Spatial distribution of $\tau_{353}$ derived from the \textit{Planck} measurements (top), HI column density from EBHIS \citep{2016A&A...585A..41W}, and $^{13}$CO (1--0) integrated intensity from \citet{2011A&A...529A...1S} at an angular resolution of 10$\rlap{.}$\arcmin8. In all panels, the contours correspond to the smoothed integrated optical depth map of H$_{2}$CO (1$_{1,0}$--1$_{1,1}$) integrated from $-$10~\kms\,to 20~\kms, and they start from 0.09~\kms, and increase by 0.09~\kms. The structure similarity index (SSI; see Sect.~\ref{sec.comp}) of the two corresponding tracers is indicated in the lower left corner of each panel.}\label{Fig:mor}}
\end{figure*}

In order to compare the distribution of H$_{2}$CO with that of other tracers including the 353 GHz dust optical depth, H{\scriptsize I} column density and the $^{13}$CO (1--0) line, we use the integrated optical depth maps rather than the integrated intensity map since the column density of H$_{2}$CO is directly related to the optical depth. The different data sets were convolved to a common angular resolution of 10$\rlap{.}$\arcmin8 and projected onto the same grid as the H$_{2}$CO absorption. Figure~\ref{Fig:mor} shows the comparison between the distribution of different tracers at the same angular resolution of 10$\rlap{.}$\arcmin8. 

We use the structural similarity index (SSI\footnote{\url{https://scikit-image.org/docs/dev/auto_examples/transform/plot_ssim.html}}) to quantify the similarity between the distributions of different tracers \citep{ssim2004}. The SSI has a value between $-$1 and 1, where an SSI of 1 implies perfect similarity, an SSI of 0 implies no similarity, and an SSI of $-$1 implies a perfect anti-correlation. Before making this comparison, we perform the quantile transformer of our data to make sure that the pixel values follow the Gaussian distribution. The SSI is then estimated for different pairs of tracers. We find that the comparison between H$_{2}$CO and $^{13}$CO (1--0) results in an SSI of 0.38, higher than found for the other two pairs (the SSI for each pair is shown in the lower left corner in Fig.~\ref{Fig:mor}). The results show that the best overall morphological agreement is between H$_{2}$CO (1$_{1,0}$--1$_{1,1}$) and $^{13}$CO (1--0). In contrast, the distribution of $\tau_{353}$ is more extended, while that of $N_{\rm HI}$ is even more extended and often uncorrelated with both H$_{2}$CO (1$_{1,0}$--1$_{1,1}$) and $^{13}$CO (1--0). This strongly suggests that our Effelsberg data of the H$_{2}$CO (1$_{1,0}$--1$_{1,1}$) absorption traces the bulk of molecular gas, also seen in $^{13}$CO (1--0) emission (their relationship is further investigated in Appendix~\ref{app.h2covsco}).

Previous observations have suggested that H$_{2}$CO can exist in diffuse and translucent molecular clouds \citep[e.g.,][]{1990ApJS...72..303N, 1995A&A...299..847L, MentenReid1996, 2006ARA&A..44..367S, 2006A&A...448..253L}. Hence, H$_{2}$CO (1$_{1,0}$--1$_{1,1}$) can potentially be used to investigate the so-called ``CO-dark'' molecular gas \citep[e.g.,][]{2005Sci...307.1292G, 2010ApJ...716.1191W}. However, our survey data appear to only trace molecular gas seen in $^{13}$CO (1--0). Previous observations suggest that the ``CO-dark'' molecular gas is prevalent over the visual extinction range $0.4 \lesssim A_{\rm V} \lesssim 2.5$ \citep{2011A&A...536A..19P}, while numerical simulations indicate that the ``CO-dark'' molecular gas can be present in gas with $A_{\rm V} \lesssim 5$ \citep{2020MNRAS.492.1465S}. Given our sensitivity, our H$_{2}$CO (1$_{1,0}$--1$_{1,1}$) observations can only probe molecular gas with H$_{2}$ column densities $\gtrsim 5 \times 10^{21}$~cm$^{-2}$ (i.e., $A_{\rm V} \gtrsim 5$) in Cygnus X (see Fig.~\ref{Fig:abun}). We also stacked the H$_{2}$CO spectra for regions where $^{13}$CO (1--0) integrated intensities are lower than 0.15~K~\kms\,(3$\sigma$), but H$_{2}$CO absorption is not detected in the stacked spectrum. Therefore, we conclude that our observations do not reach the regime of the ``CO-dark'' molecular gas, and more sensitive observations are needed to address whether H$_{2}$CO (1$_{1,0}$--1$_{1,1}$) can trace the ``CO-dark'' molecular gas or not. 

\subsection{Local velocity gradient}\label{dis:vg}
The distribution of velocity centroids seems to show ordered LSR velocity gradients rather than random motions. To study the LSR velocity gradients within the individual cloud structures, we follow the definition of the local velocity gradients, $\nabla \varv$, given by \citet{1993ApJ...406..528G}:
\begin{equation}\label{f.grad}
    \varv_{\rm lsr} = \varv_{0} + x \Delta l + y \Delta b \;,
\end{equation}
where $\varv_{\rm lsr}$ is the observed velocity centroid; $\varv_{0}$ is the systemic velocity centroid; $\Delta l$ and $\Delta$b are the offsets in the Galactic longitude and latitude; $x$ and $y$ are the components of $\nabla \varv$ along the directions of the Galactic longitude and latitude. The magnitude of the local velocity gradient is defined as $|\nabla \varv| = \sqrt{x^{2} +y ^{2}}$. The position angle is $\theta_{\rm vg} = {\rm arctan}(x/y)$, and $\theta$ increases counter-clockwise with respect to the Galactic northern direction. In order to derive the $\nabla \varv$ distribution, we fit Eq.~\ref{f.grad} using the Levenberg–Marquardt algorithm toward each block of $3\times 3$ pixels \citep[e.g.,][]{2021A&A...646A.170G}. 

\begin{figure*}[!htbp]
\centering
\includegraphics[width = 0.95 \textwidth]{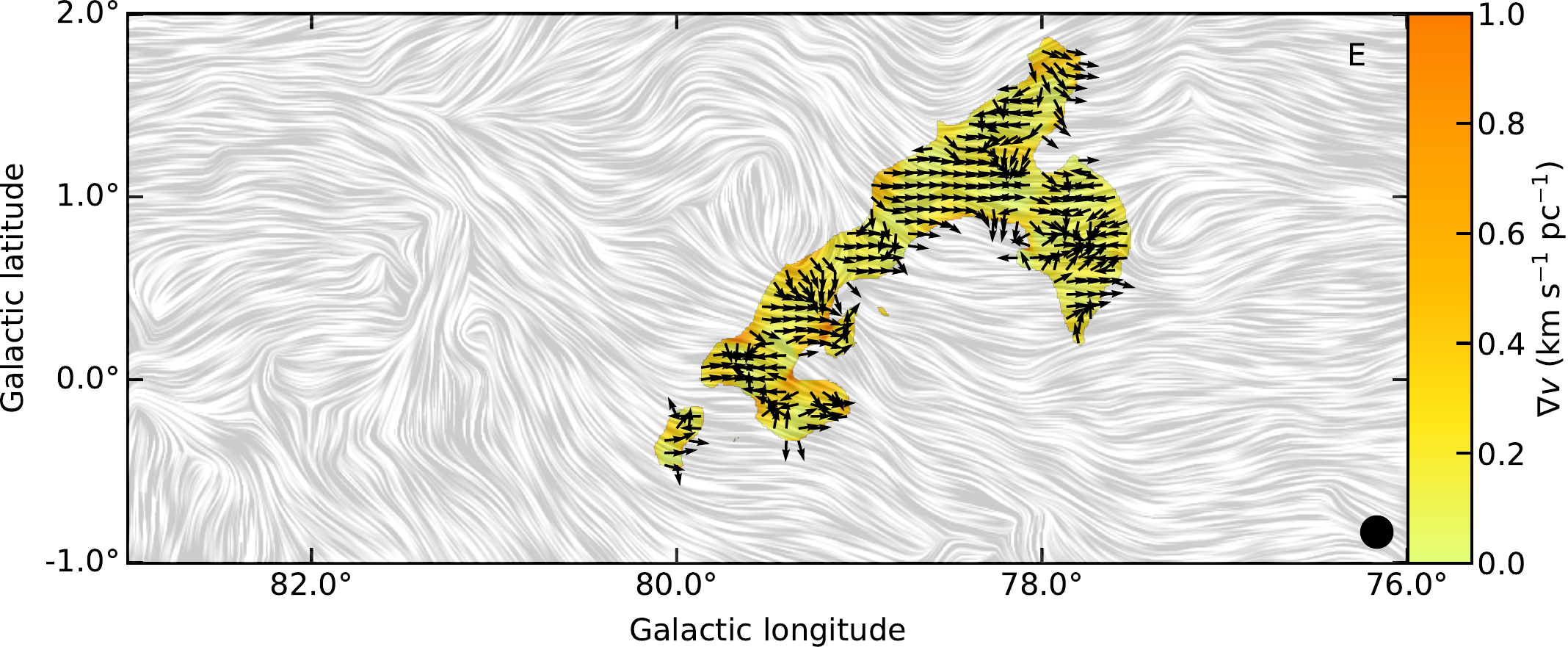}
\caption{{Local velocity gradient map overlaid with the magnetic field pattern derived from the \textit{Planck} 353~GHz dust polarization. The arrows represent the direction of normalized local velocity gradients, and the color bar represents the magnitude of the local velocity gradients in units of \kms~pc$^{-1}$. The beam size is shown in the lower right corner. The coherent structure is labeled in the top right corner. }\label{Fig:eff-vg}}
\end{figure*}


Figure~\ref{Fig:eff-vg} shows the derived $\nabla \varv$ distribution of cloud E, and the results for the other cloud structures are presented in Fig.~\ref{Fig:app-vg} of Appendix~\ref{app.vg}. The magnitude $|\nabla \varv|$ lies in the range from 0 to 2.38~\kms\,pc$^{-1}$ with a median value of 0.14~\kms\,pc$^{-1}$. More than 80\% of the $|\nabla \varv|$ values are lower than 0.3~\kms\,pc$^{-1}$. The high $|\nabla \varv|$ values mainly arise from two cloud structures (clouds A and H) close to  DR21 (see also Fig.~\ref{Fig:app-vg}). The velocity gradients toward DR21 are believed to be caused by cloud-cloud collisions \citep{1978ApJ...223..840D}. Furthermore, these plots confirm the presence of anisotropic velocity fields at least in parts of clouds on the large scale of 4.4~pc. 


Since gravity and turbulence can also affect the relative orientations between local velocity gradients and magnetic fields, we use the alignment measure (AM) to further investigate their relationship. Following previous studies \citep[e.g.,][]{2018ApJ...853...96L,2022arXiv221100152L}, AM is defined as:
\begin{equation}\label{f.am}
AM = \langle {\rm cos(2\phi)} \rangle \,,
\end{equation}
where $\phi$ is the relative orientation between local velocity gradients and magnetic fields in the range of 0\degr--90\degr. $AM$ has a value between $-$1 and 1, where $AM=-1$ implies perpendicular and $AM=1$ implies parallel alignment. 

\begin{figure*}[!htbp]
\centering
\includegraphics[height = 0.45 \textwidth]{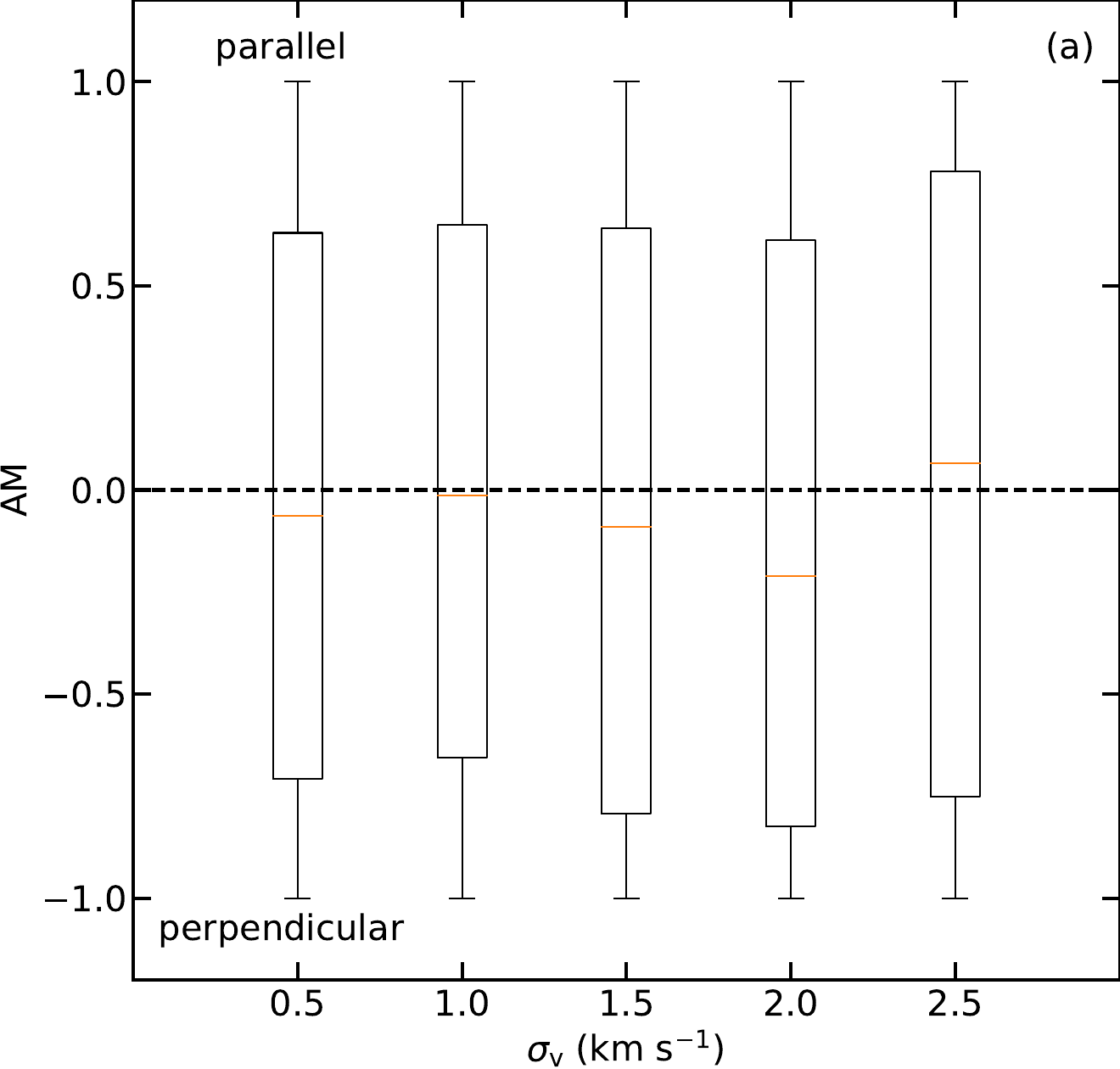}
\includegraphics[height = 0.45 \textwidth]{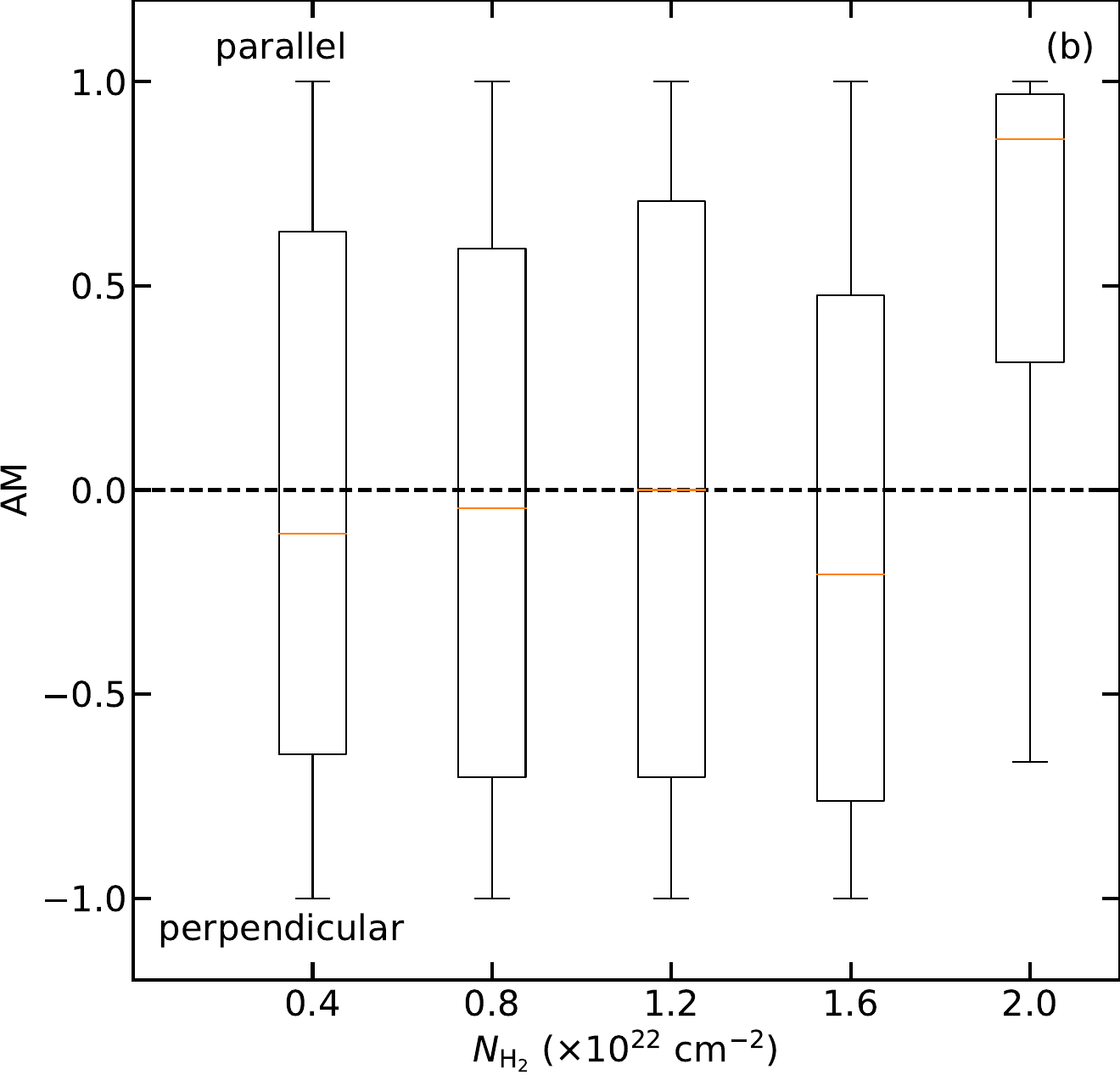}
\caption{{Alignment measure as a function of velocity dispersion (panel a) and H$_{2}$ column density (panel b). In each box plot, the median value is indicated by an orange line, and the box represents the data within the 25th and 75th percentiles.}\label{Fig:AM}}
\end{figure*}

Figure~\ref{Fig:AM} presents $AM$ as a function of velocity dispersion and H$_{2}$ column density on the 4.4~pc scale. In Fig.~\ref{Fig:AM}a, $AM$ appears to be uncorrelated with velocity dispersion. This is different from previous study on the Taurus cloud where strongly parallel or perpendicular alignments are restricted to regions with low levels of turbulence \citep{2020MNRAS.496.4546H}. This is because Cygnus X is more turbulent than the Taurus cloud and the correlation is weak in case of strong turbulence \citep{2017ApJ...835...41G,2018ApJ...853...96L}. 


Figure~\ref{Fig:AM}b shows that $AM$ tends to be more parallel at high H$_{2}$ column densities of $\gtrsim1.8\times 10^{22}$~cm$^{-2}$. In high-density regions, in which gravity becomes dominant, molecular gas tends to flow along magnetic field lines because of the resisting Lorentz force in the perpendicular direction \citep[e.g.,][]{2014prpl.conf..101L}. Our observations thus support that gas motions are channeled by gravity and magnetic fields in Cygnus X when the H$_{2}$ column densities are $\gtrsim1.8\times 10^{22}$~cm$^{-2}$ on the 4.4~pc scale. The critical H$_{2}$ column density for the transition from being perpendicular to being parallel appears to be higher than the values (10$^{21-21.5}$~cm$^{-2}$) predicted by previous simulations \citep{2020MNRAS.497.4196S}. However, the plane-of-sky magnetic field strengths in the ambient gas surrounding DR21 have been estimated to be $\sim$0.1~mG  \citep{2022ApJ...941..122C}, which is much higher than the values ($\lesssim 10$~$\mu$G) adopted in the simulations \citep{2020MNRAS.497.4196S}. The critical H$_{2}$ column density should depend on the magnetic field strength \citep[e.g.,][]{2014prpl.conf..101L,2020MNRAS.497.4196S}. Therefore, the higher critical H$_{2}$ column density in Cygnus X can be explained by its stronger magnetic fields.

\subsection{Comparison of multi-scale motions}
Molecular clouds are known to show hierarchical structures \citep[e.g.,][]{2008ApJ...679.1338R}. However, the connection between the large-scale and small scale structures is not well established. Our Effelsberg and VLA observations allow us to study the relationship between the large-scale and small scale properties. 

\begin{figure}[!htbp]
\centering
\includegraphics[width = 0.45 \textwidth]{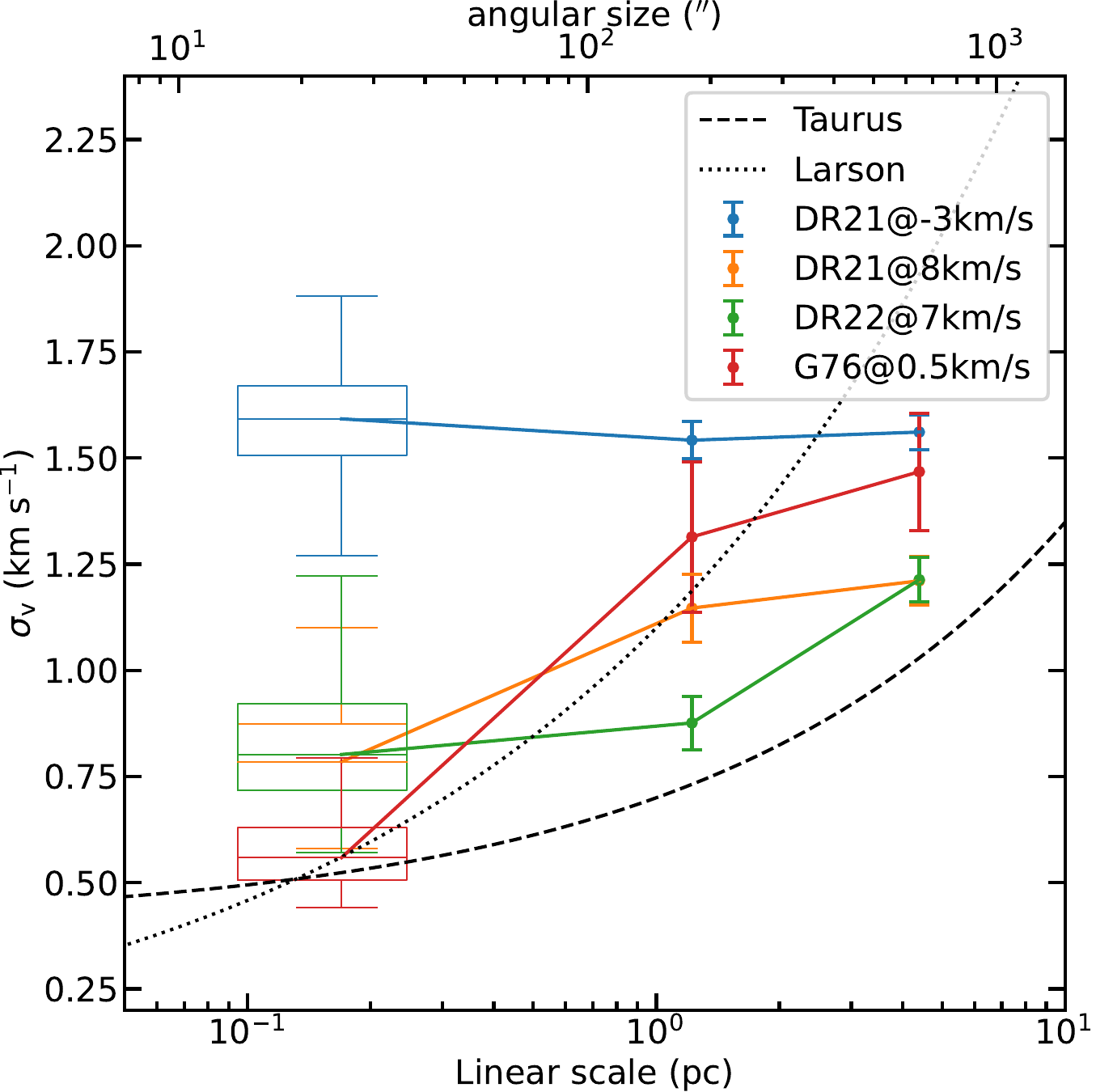}
\caption{{Velocity dispersions of H$_{2}$CO (1$_{1,0}$--1$_{1,1}$) as a function of different linear scales toward the targeted sources. The classic Larson relation is indicated by the black dotted line \citep{1981MNRAS.194..809L}, while the relation between the velocity dispersion and linear scales in the Taurus molecular cloud is shown by the black dashed line \citep{2012ApJ...760..147Q}.}\label{Fig:multiscale}}
\end{figure}

Since H$_{2}$CO absorption is only detected toward three bright H{\scriptsize II} regions (i.e., DR21, DR22, and G76.1883+0.0973) by our high angular resolution observations, we therefore can only meaningfully compare the gas properties toward these regions on different scales. Figure~\ref{Fig:multiscale} shows the comparison of the derived velocity dispersions on different scales. According to the classical turbulence cascade theory  \citep[e.g.,][]{2004ARA&A..42..211E, 2004ARA&A..42..275S} and previous observational studies \citep[e.g.,][]{1981MNRAS.194..809L, 2012ApJ...760..147Q, 2017A&A...601A.124S},  gas motions should decrease toward small scales. Our observations are in agreement with this scenario except for the $-$3~\kms\,component of DR21, which has nearly identical velocity dispersions on different scales (0.17--4.4~pc). This contradicts the expected behavior of this classic turbulence, which is thought to be externally driven on large scales of $\gtrsim$10 pc with turbulent energy cascading down to small scales \citep[e.g.,][]{2004ARA&A..42..211E,2004ARA&A..42..275S}. 

In order to explain this behavior, we speculate that the fact that we find nearly identical velocity dispersions on different scales indicates that the turbulence in the DR21 region is driven on 
scales $<$4.4~pc. Previous studies have proposed cloud--cloud collisions between the $-3$~\kms\,and 8~\kms\,components of DR21 \citep{1978ApJ...223..840D, 2019PASJ...71S..12D}, which can drive the additional turbulence. If the small-scale turbulence is due to a cloud--cloud collision, we expect to see an enhancement of turbulent motions in both velocity components. However, the 8~\kms\,component seems to follow the turbulence cascade picture (see Fig.~\ref{Fig:multiscale}). Instead, the additional turbulence in the $-$3~\kms\,component can be induced by locally convergent flows that result from self-gravity \citep{2010A&A...520A..49S}. Theoretical studies also suggest that the shallower relation between velocity dispersion and linear scales is indicative of gravitational collapse \citep[e.g.,][]{2015ApJ...804...44M, 2019MNRAS.490.3061V}. Alternatively, the additional turbulence can be driven by the powerful protostellar outflow in the region. DR21 is known to host one of the most powerful outflows in Cygnus X, whose lobes extend over $\sim$1.6~pc  \citep[][Skretas et al. in prep]{1992ApJ...392..602G}. Because the $-$3~\kms\,component of DR21 is physically associated with the molecular outflow, the outflow-driven turbulence can affect nearly all the physical scales probed in Fig.~\ref{Fig:multiscale}. While molecular outflows may also exist close to DR22 and G76.1883+0.0973 \citep[e.g.,][]{1996ApJ...457..267S, 2022A&A...660A..39S}, these may be too weak 
to drive additional turbulence comparable to that toward DR21. Another possible source of turbulence could come from the associated H{\scriptsize II} regions. As shown in Sect.~\ref{sec.vla}, the comparison between RRL and H$_{2}$CO velocity suggests that the $-$3~\kms\,component of DR21 is likely the only molecular gas that is associated with H{\scriptsize II} regions. Hence, the feedback of H{\scriptsize II} regions could also lead to the difference behavior of the $-$3~\kms\,component of DR21. Therefore, we suggest that the nearly identical velocity dispersions toward the $-3$~\kms\,component of DR21 on different scales can be caused by internally driven turbulence from convergent flows, YSO outflows, and H{\scriptsize II} regions. 

We also note that our H$_{2}$CO absorption measurements probe all of the gas along the line of sight. The probed lengths along the line of sight are completely unknown, which could bias the representative scales in Fig.~\ref{Fig:multiscale}. 

\section{Summary and conclusion}\label{Sec:sum}
As part of the GLOSTAR survey project, we carried out Effelsberg and VLA observations toward the Cygnus~X region to study multi-scale structure properties of its molecular gas. The main findings are summarized as follows:

\begin{itemize}
    \item[1.] Our Effelsberg observations reveal widespread H$_{2}$CO ($1_{1,0}-1_{1,1}$) absorption with a typical spatial extent of $\gtrsim$50~pc in Cygnus~X. Most of the observed H$_{2}$CO absorption is optically thin. Based on a decomposition of the spectra to a scale of 4.4~pc and the DBSCAN clustering method, we assign the observed H$_{2}$CO absorption into eight velocity-coherent cloud structures which are dominated by supersonic turbulent motions. \\
    
    \item[2.] The GLOSTAR VLA+Effelsberg combined data result in the robust detection of H$_{2}$CO ($1_{1,0}-1_{1,1}$) absorption toward three H{\scriptsize II} regions (i.e., DR21, DR22, and G76.1883+0.0973). The observed velocity dispersions suggest that supersonic turbulence commonly exists in the three H{\scriptsize II} regions on the 0.17~pc scale. \\
    
    \item[3.] While the compact absorption features are mainly due to absorption against the radio continuum in Cygnus~X,  extended absorption features are also seen where the radio continuum is weak. This suggests a non-negligible contribution of the cosmic microwave background in producing extended absorption features in Cygnus~X. \\
    

    \item[4.] On a large scale, our comparison of different tracers shows a high degree of similarity in the distributions of the H$_{2}$CO ($1_{1,0}-1_{1,1}$) absorption and $^{13}$CO (1--0)
    emission, indicating that H$_{2}$CO (1$_{1,0}$--1$_{1,1}$) can trace the bulk of the molecular gas seen in $^{13}$CO (1--0). Making use of the \textit{Planck} 353~GHz dust optical depth map, H{\scriptsize I} column density map, and our H$_{2}$CO observations, we find that H$_{2}$CO ($1_{1,0}-1_{1,1}$) can trace molecular gas with H$_{2}$ column densities of $\gtrsim 5 \times 10^{21}$~cm$^{-2}$ (i.e., $A_{\rm V} \gtrsim 5$) and the ortho-H$_{2}$CO fractional abundances with respect to H$_{2}$ has a mean abundance of 7.0$\times 10^{-10}$ with a dispersion of 0.15 dex (i.e., $10^{-9.16\pm 0.15}$).\\
    
    \item[5.] Local velocity gradients are investigated on scales of 4.4~pc and 0.17~pc. On a 4.4~pc scale, most of the magnitudes of the local velocity gradients are as low as $<$0.3~\kms~pc$^{-1}$. We find that the relative orientation between local velocity gradient and magnetic field tends to be more parallel at H$_{2}$ column densities of $\gtrsim$1.8$\times 10^{22}$~cm$^{-2}$, which could be caused by the scenario that gas motions are channeled by magnetic fields.\\ 
    
    \item[6.] Multi-scale comparisons of velocity dispersions show that the $-3$~\kms\,component of DR21 has nearly identical velocity dispersions on scales of 0.17--4.4~pc, which might deviate from the expected behavior of classic turbulence. This could be caused by internally driven turbulence from convergent flows, YSO outflows, and H{\scriptsize II} regions. 
    
\end{itemize}

Our GLOSTAR observations reveal widespread H$_{2}$CO ($1_{1,0}-1_{1,1}$) absorption and pinpoint the bright absorption regions in Cygnus~X, demonstrating that the GLOSTAR data can probe the H$_{2}$CO ($1_{1,0}-1_{1,1}$) absorption on cloud ($\gtrsim$4~pc) down to core ($\sim$0.17~pc) scales. Follow-up H$_{2}$CO ($2_{1,1}-2_{1,2}$) observations of regions showing appreciable H$_{2}$CO ($1_{1,0}-1_{1,1}$) absorption will allow for determinations of the density distributions of the covered molecular clouds.  

\section*{ACKNOWLEDGMENTS}\label{sec.ack}
We thank the Effelsberg-100 m telescope staff for their assistance with our observations. HB acknowledges support from the European Research Council under the Horizon 2020 Framework Programme via the ERC Consolidator Grant CSF-648505. HB also acknowledges support from the Deutsche Forschungsgemeinschaft in the Collaborative Research Center (SFB 881) ``The Milky Way System" (subproject B1). AYY acknowledges support from the NSFC grants No. 11988101 and No. NSFC 11973013. We thank Nicola Schneider for sharing her FCRAO data cubes. YG thanks Xuyang Gao for helpful discussions on the zero level restoration of the continuum data in Cygnus X. YG thanks Tao-Chung Ching for sharing the JCMT POL-2 data on DR21. This work is based on observations with the 100-m telescope of the MPIfR (Max-Planck-Institut f{\"u}r Radioastronomie) at Effelsberg. The National Radio Astronomy Observatory is a facility of the National Science Foundation operated under cooperative agreement by Associated Universities, Inc. This research has made use of NASA's Astrophysics Data System. This work also made use of Python libraries including Astropy\footnote{\url{https://www.astropy.org/}} \citep{2013A&A...558A..33A}, NumPy\footnote{\url{https://www.numpy.org/}} \citep{5725236}, SciPy\footnote{\url{https://www.scipy.org/}} \citep{jones2001scipy}, Matplotlib\footnote{\url{https://matplotlib.org/}} \citep{Hunter:2007}, LMFIT \citep{newville_matthew_2014_11813}, APLpy \citep{2012ascl.soft08017R}, plotly\footnote{\url{https://plotly.com/}}, and magnetar\footnote{\url{https://github.com/solerjuan/magnetar}} \citep{Soler2013}. We would like to thank the anonymous referee for the valuable comments which improve our draft.

\bibliographystyle{aa}
\bibliography{references}

\begin{appendix}
\section{Total formaldehyde column density}\label{app.radex}
Because of an unusual collisional pumping process, the level populations of ortho-H$_{2}$CO corresponding to the 1$_{1,0}$--1$_{1,1}$ transition often deviate from what is expected under conditions of local thermodynamic equilibrium (LTE). Hence, we do not derive the total ortho H$_{2}$CO column density from observations of this single line under the assumption of LTE. Instead, we use the following approach. First, we explore the level populations of ortho H$_{2}$CO for a range of physical conditions (that can be expected on scales $> 4.4$~pc) using a standard non-LTE radiative transfer model, and determine the fractional population in the $1_{1,0}$ level. We then determine the total ortho H$_{2}$CO column density by scaling the column density in the $1_{1,0}$ level by a factor at fiducial values of the physical conditions in the Cygnus~X region.

In order to investigate the level populations of ortho H$_2$CO, we make use of the non-LTE RADEX\footnote{\url{https://home.strw.leidenuniv.nl/~moldata/radex.html}} code \citep{2007A&A...468..627V}. The molecular data of ortho-H$_{2}$CO are obtained from the Leiden Atomic and Molecular Database \citep[LAMDA\footnote{\url{https://home.strw.leidenuniv.nl/~moldata/}};][]{2005A&A...432..369S}, where the energy levels, transition frequencies and Einstein A coefficients are taken from the CDMS catalog \citep{2005JMoSt.742..215M,ENDRES201695} and the collisional rates are taken from \citet{2013MNRAS.432.2573W}. A total of 40 energy levels are considered in our calculations. Based on the previous temperature measurements toward Cygnus~X \citep{2019ApJ...884....4K, 2019ApJS..241....1C}, most of the molecular gas lies in the kinetic temperature range of 10--30~K. Hence, we perform our calculations at kinetic temperatures ranging from 5 K to 30 K with a step size of 1 K. Because previous observations have shown H$_{2}$ number densities of 500--5000 cm$^{-3}$ on a scale of $\gtrsim$0.3~pc \citep[e.g.,][]{2015ApJS..216...18N}, we expect that the H$_{2}$ number density range of 10$^{2}$--10$^{4}$~cm$^{-3}$ on a scale of 4.4~pc should be suitable for our cases. Therefore,
the H$_{2}$ number density, in units of cm$^{-3}$, is varied logarithmically, with $\log[n(\rm{H}_{2})]$ from 2 to 4 with a step size of 0.1. The ortho-to-para ratio of H$_{2}$ is fixed to be 0.25 according to previous studies \citep[e.g.,][]{2006ApJ...649..816N}. For the specific column density (defined as the ratio between the column density and line width), we adopt a range from $1\times 10^{12}$ cm$^{-2}$~(\kms)$^{-1}$ to $1\times 10^{13}$ cm$^{-2}$~(\kms)$^{-1}$. 

\begin{figure*}[!htbp]
\centering
\includegraphics[width = 0.95 \textwidth]{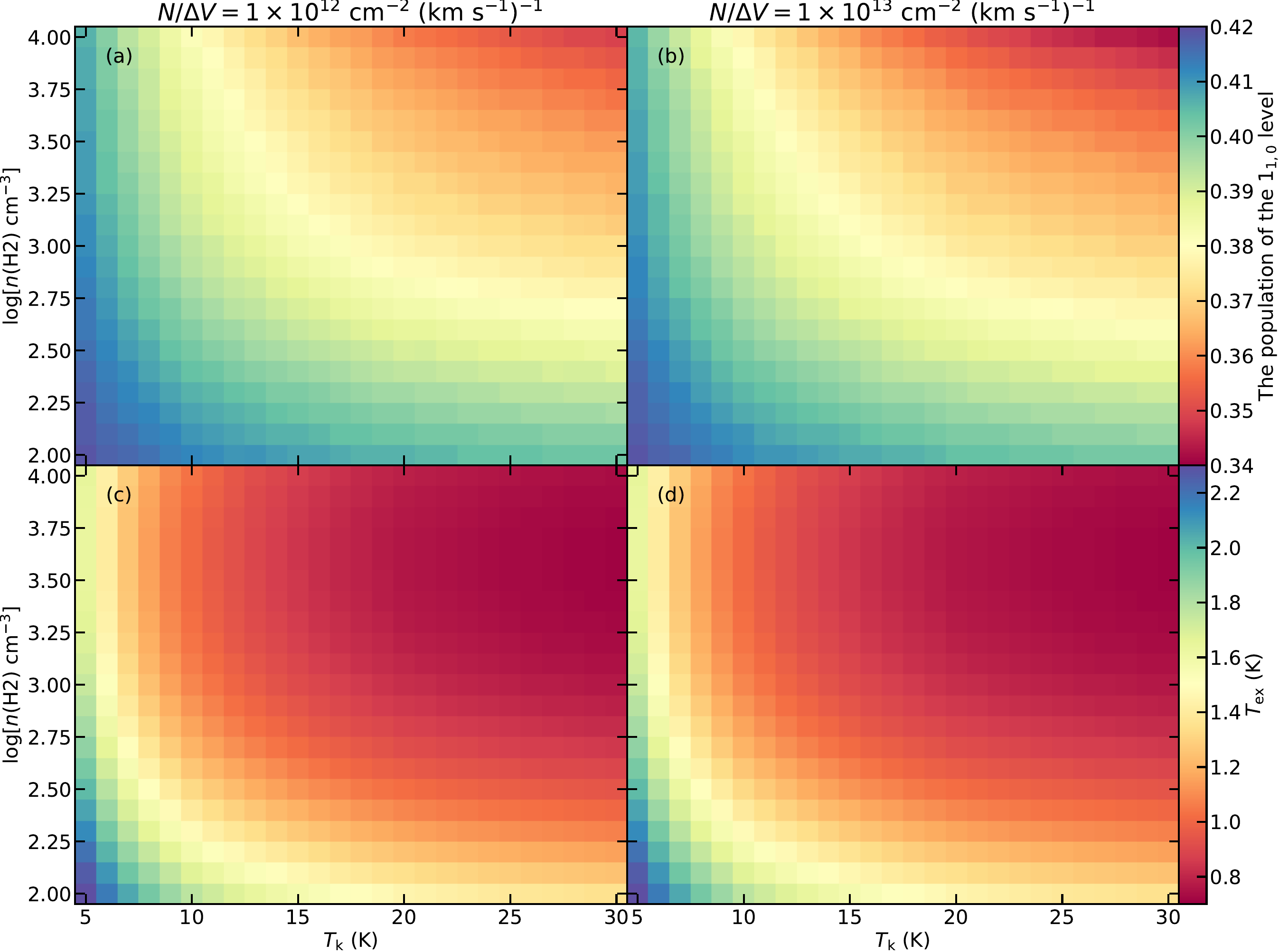}
\caption{{RADEX calculations of the fractional population of the $1_{1,0}$ level (top panels) and excitation temperatures (bottom panels). The left panels and right panels correspond to the two different specific column densities of $1\times 10^{12}$ cm$^{-2}$~(\kms)$^{-1}$ and $1\times 10^{13}$ cm$^{-2}$~(\kms)$^{-1}$, respectively.}\label{Fig:popu}}
\end{figure*}

Figures~\ref{Fig:popu}a--\ref{Fig:popu}b present the modeling results of the fractional population of the $1_{1,0}$ level. The modeling results suggest that the population in the $1_{1,0}$ level accounts for 34.2\%--41.9\% of the total population within the specific column density range of $1\times 10^{12}$ cm$^{-2}$--$1\times 10^{13}$ cm$^{-2}$~(\kms)$^{-1}$. This shows that the fractional population does not change significantly for the range of physical conditions expected in Cygnus~X. Furthermore, the modeling results show that the excitation temperatures range from 0.7 to 2.3 K (see Fig.~\ref{Fig:popu}c--\ref{Fig:popu}d), which suggests that adopting an excitation temperature of 1.6 K (see Section~\ref{sec.exc}) is a reasonable assumption.

To determine the total H$_2$CO column density, we adopt a kinetic temperature of 10~K, and H$_{2}$ density of $10^{3}$~cm$^{-3}$, and a specific H$_{2}$CO column density of 5$\times 10^{12}$~cm$^{-2}$~(\kms)$^{-1}$ as fiducial physical conditions in Cygnus~X on a scale of $\sim 4.4$~pc. For these parameters, the fractional  column density of H$_2$CO in the ${1_{1,0}}$ level is $\approx 0.39$. Adopting this ratio and using Eq.~\ref{f.Nu_num}, the total ortho H$_{2}$CO column density can be simply estimated by scaling the integrated optical depths of the H$_{2}$CO (1$_{1,0}$--1$_{1,1}$) line:
\begin{equation}\label{f.h2co_tot}
    N_{\rm ortho-H_{2}CO} = 2.41\times 10^{13}\int \tau {\rm d}\varv\;{\rm cm}^{-2}\;.
\end{equation}
We also note that adopting a constant ratio of 39\%\ implies at most an additional $\sim 13$\%\,uncertainty in the derived column densities. We conclude that Eq.~(\ref{f.h2co_tot}) can provide a reasonable estimate of the total ortho-H$_{2}$CO column density when only the single transition H$_{2}$CO (1$_{1,0}$--1$_{1,1}$) is observed. 

\section{Overlap effects of the hyperfine structure lines of H$_{2}$CO (1$_{1,0}$--1$_{1,1}$)}\label{app.hfs}
 The H$_{2}$CO (1$_{1,0}$--1$_{1,1}$) transition is known to have six HFS lines \citep[e.g.,][]{1971ApJ...169..429T}, which overlap on account of line broadening. This can bias the measurements of the line widths and velocity centroids. In order to study the overlap effects of the HFS lines of H$_{2}$CO (1$_{1,0}$--1$_{1,1}$), we follow the method introduced in Appendix D of \citet{2021A&A...646A.170G}. We specify the rest frequencies and relative line strengths of the six HFS lines based on the CDMS \citep{2005JMoSt.742..215M}. For the fiducial case, we assume an excitation temperature, $T_{\rm ex}$, of 1.6~K, the microwave background radiation temperature, $T_{\rm bg}$, to be 2.73~K, that no continuum emission arises from behind the H$_{2}$CO-bearing gas (i.e. $T_{\rm c}$ =0~K), and the systemic LSR velocity, $\varv_{0}$, to be 0~km~s$^{-1}$.  These parameters will not affect the velocity information of synthetic spectral line profiles, but only affect the amplitude of synthetic spectra. The peak optical depths can vary within the range of 0--0.5 (see Fig.~\ref{Fig:peaktau}). We use different peak optical depths, $\tau_{0}$, and velocity dispersion, $\sigma_{0}$, to create synthetic spectra to test the overlapping effects. 
 
\begin{figure*}[!htbp]
\centering
\includegraphics[width = 0.95 \textwidth]{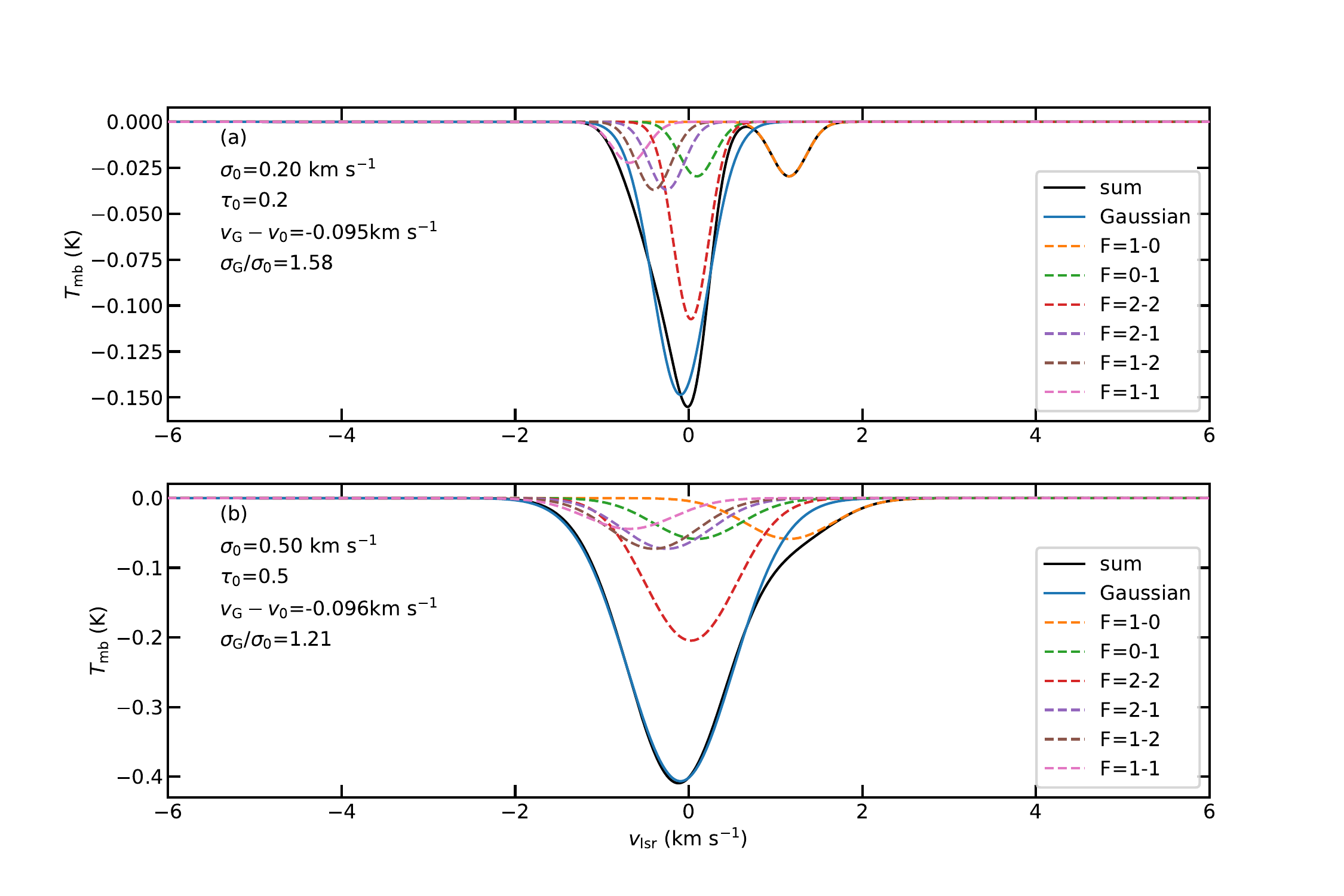}
\caption{{Synthetic H$_{2}$CO (1$_{1,0}$--1$_{1,1}$) spectra (black solid lines) with the modeled $\tau_{0}$ and $\sigma_{\varv}$ shown in the upper left corners. The Gaussian fitting results are indicated by the blue solid lines. The different HFS components are indicated by the colored dashed lines in the legend. }\label{Fig:hfs}}
\end{figure*}

Figure~\ref{Fig:hfs} presents two synthetic spectra for different values of velocity dispersion. It is evident that the absorption intensity still peaks at $\varv_{\rm lsr}\sim$0~\kms\,for a low velocity dispersion (see Fig.~\ref{Fig:hfs}a), but shifts to the blueshifted side at a higher velocity dispersion (see Fig.~\ref{Fig:hfs}b). For large velocity dispersions, the $F$=1--0 line can create a redshifted wing-like profile which should not be misinterpreted as an indication of molecular outflows (see Fig.~\ref{Fig:hfs}b).

\begin{figure*}[!htbp]
\centering
\includegraphics[width = 0.95 \textwidth]{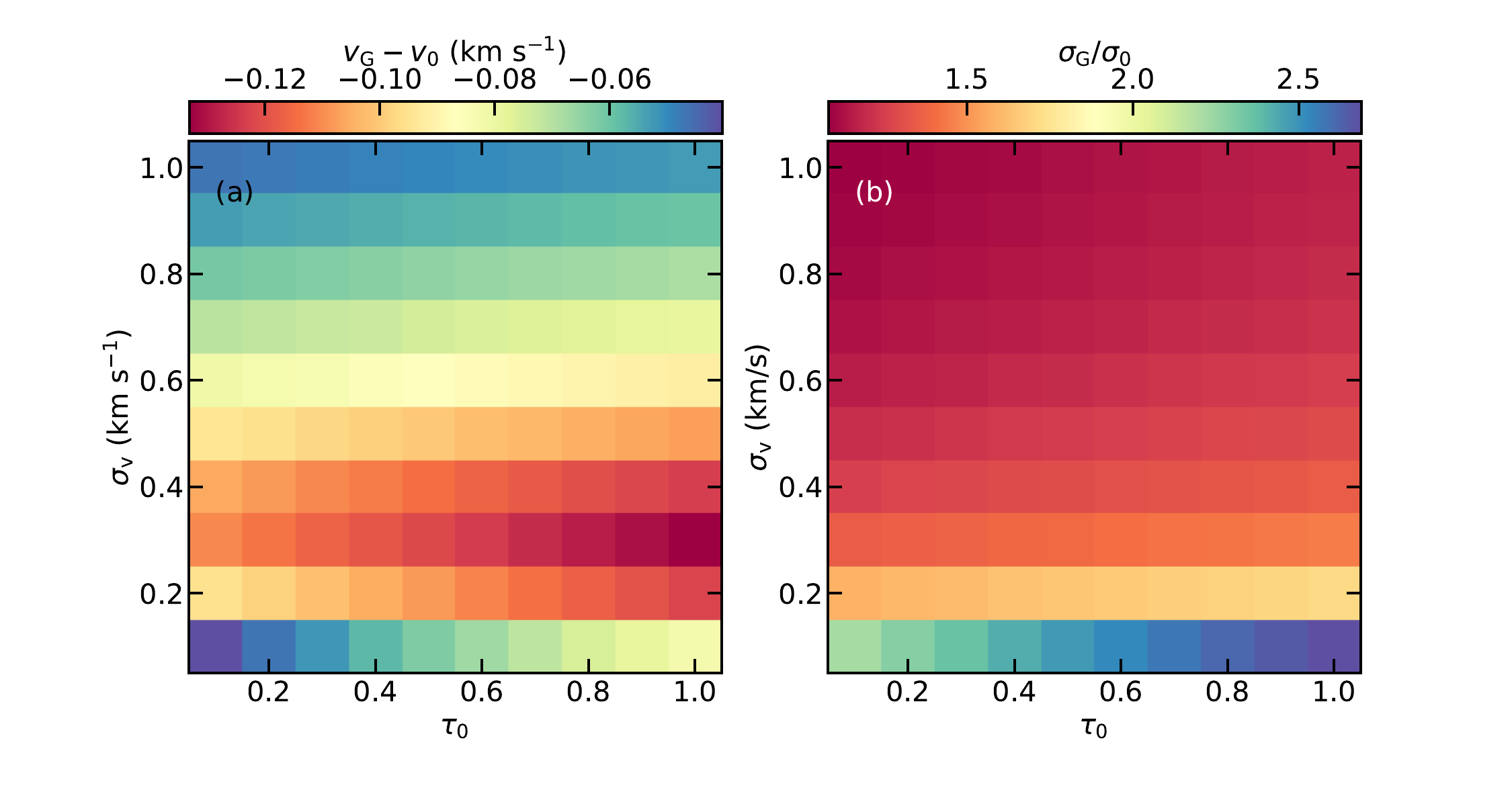}
\caption{{(a) $\varv_{\rm G}-\varv_{0}$ as a function of optical depth and velocity dispersion, where $\varv_{\rm G}$ and $\varv_{0}$ are the velocity centroid derived from the Gaussian fitting and the intrinsic velocity centroid. (b) $\sigma_{\rm G}/\sigma_{0}$ as a function of optical depth and velocity dispersion, where $\sigma_{\rm G}$ and $\sigma_{0}$ are the velocity dispersion derived from the Gaussian fitting and the intrinsic velocity dispersion.}\label{Fig:hfs-grid}}
\end{figure*}

We also note that these profiles deviate from the typical Gaussian profile. Especially when the synthetic spectral line profiles are fit with the Gaussian function to derive the observed parameters, the derived velocity centroids and line widths can deviate from their true values. In order to understand this effect, we adopt different velocity dispersions from 0.1~\kms\,to 1~\kms\,with a step of 0.1~\kms\,and different peak optical depths from 0.1 to 1 with a step of 0.1. The deviation is characterised by the velocity difference, $\varv_{\rm G}-\varv_{0}$, and the ratio of velocity dispersions, $\sigma_{\rm G}/\sigma_{0}$, where $\varv_{\rm G}$ and $\sigma_{\rm G}$ are the fit velocity centroid and dispersion obtained from Gaussian fitting. In order to reduce the broadening effects caused by the $F$=1--0 line (see Fig.~\ref{Fig:hfs}), we only perform the Gaussian fit to the spectra within the velocity range from $-$2~\kms\,to 1~\kms. The results are shown in Fig.~\ref{Fig:hfs-grid}. The results demonstrate that both $\varv_{\rm G}-\varv_{0}$ and $\sigma_{\rm G}/\sigma_{0}$ are largely regulated by velocity dispersions. When $\sigma_{0}$=0.3--0.4~\kms, $\varv_{\rm G}-\varv_{0}$ can be as large as $-$0.1~\kms. On the other hand,  $\sigma_{\rm G}/\sigma_{0}$ increases with decreasing $\sigma_{0}$ with the highest ratio of $>$2 at $\sigma_{0}$=0.1~\kms. Therefore, these overlapping effects should not be neglected in studying the kinematics with this transition. 

We also note that H$_{2}$CO (1$_{1,0}$--1$_{1,1}$) is typically not in LTE, which can lead to the line ratios of HFS lines being different from what is expected under LTE. However, the available collisional rate coefficients do not include the HFS lines \citep{2013MNRAS.432.2573W}, and non-LTE effects for the HFS lines are beyond the scope of this work. 

\section{Parameterizing DBSCAN}\label{app.dbscan}
\begin{figure*}[!htbp]
\centering
\includegraphics[width = 0.49 \textwidth]{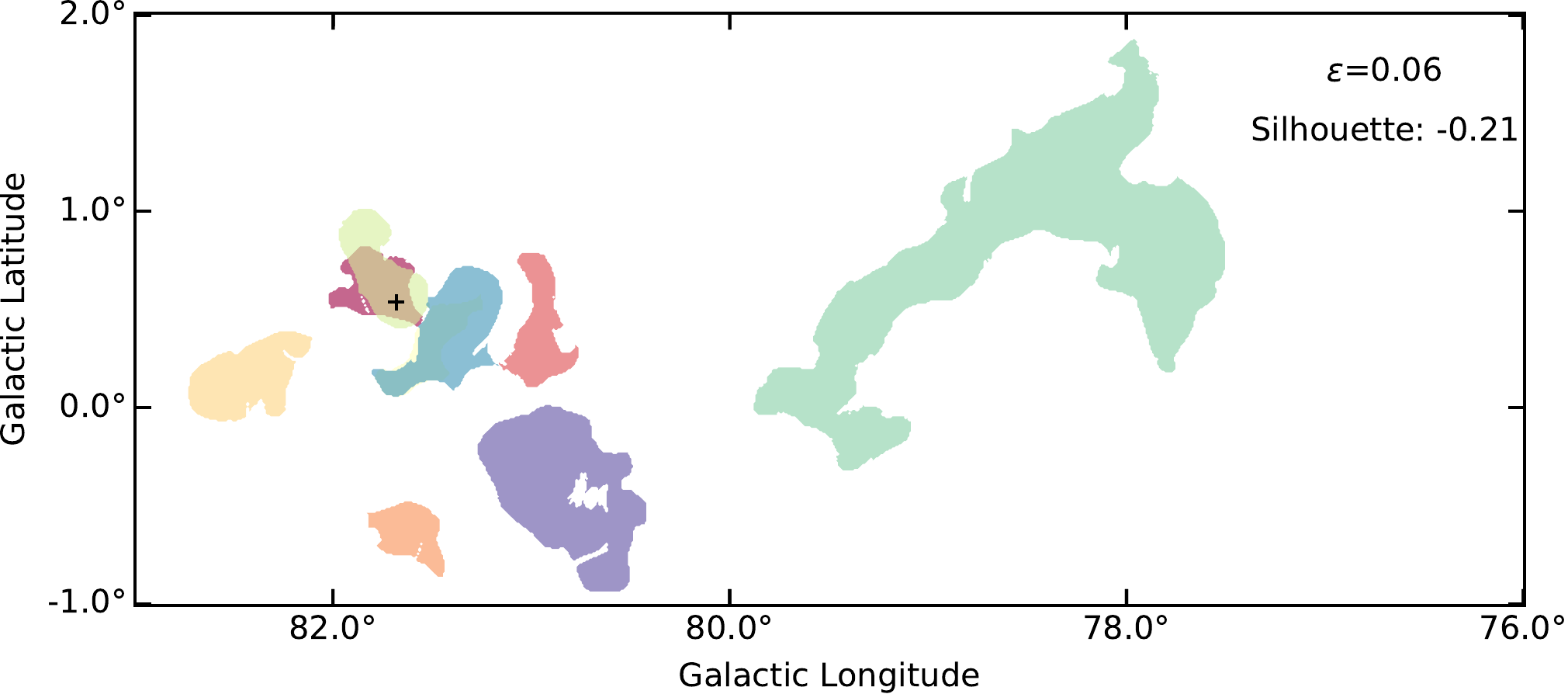}
\includegraphics[width = 0.49 \textwidth]{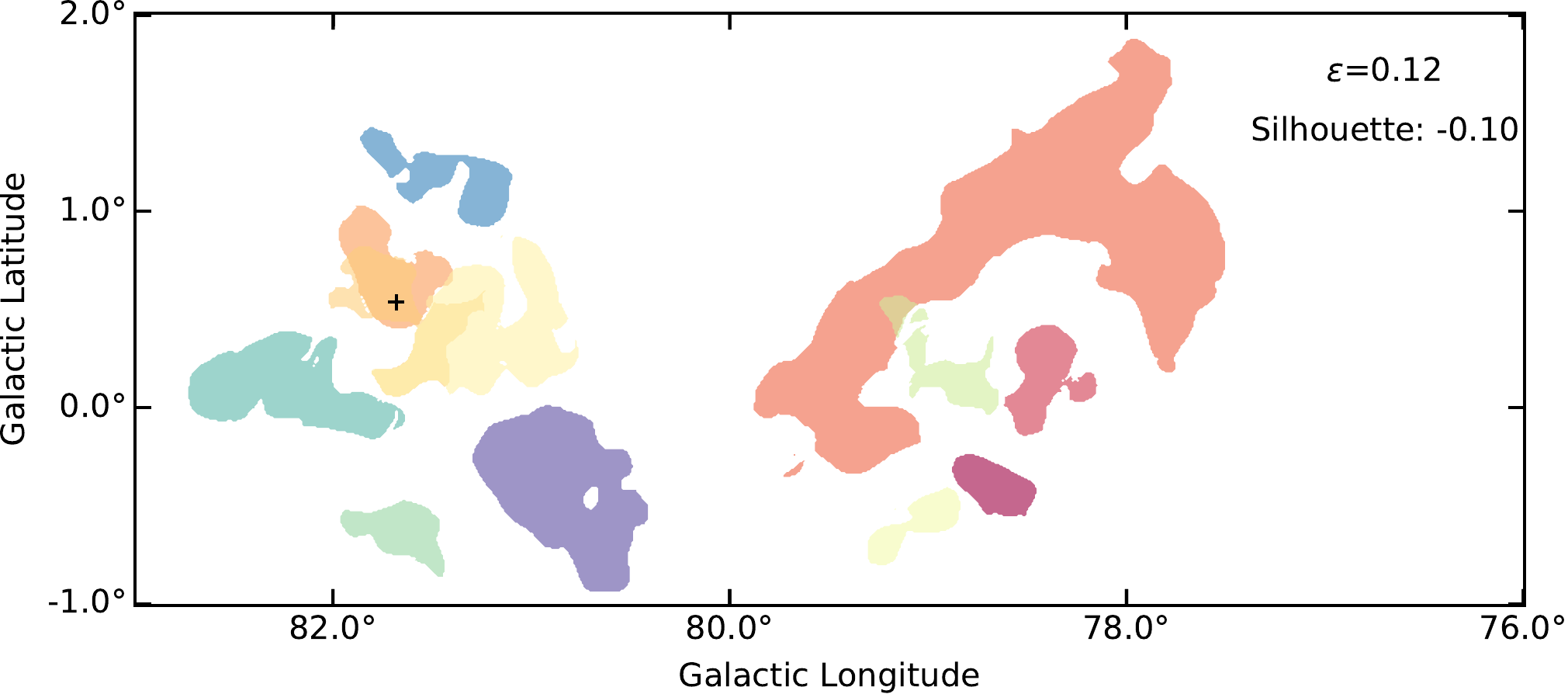}
\includegraphics[width = 0.49 \textwidth]{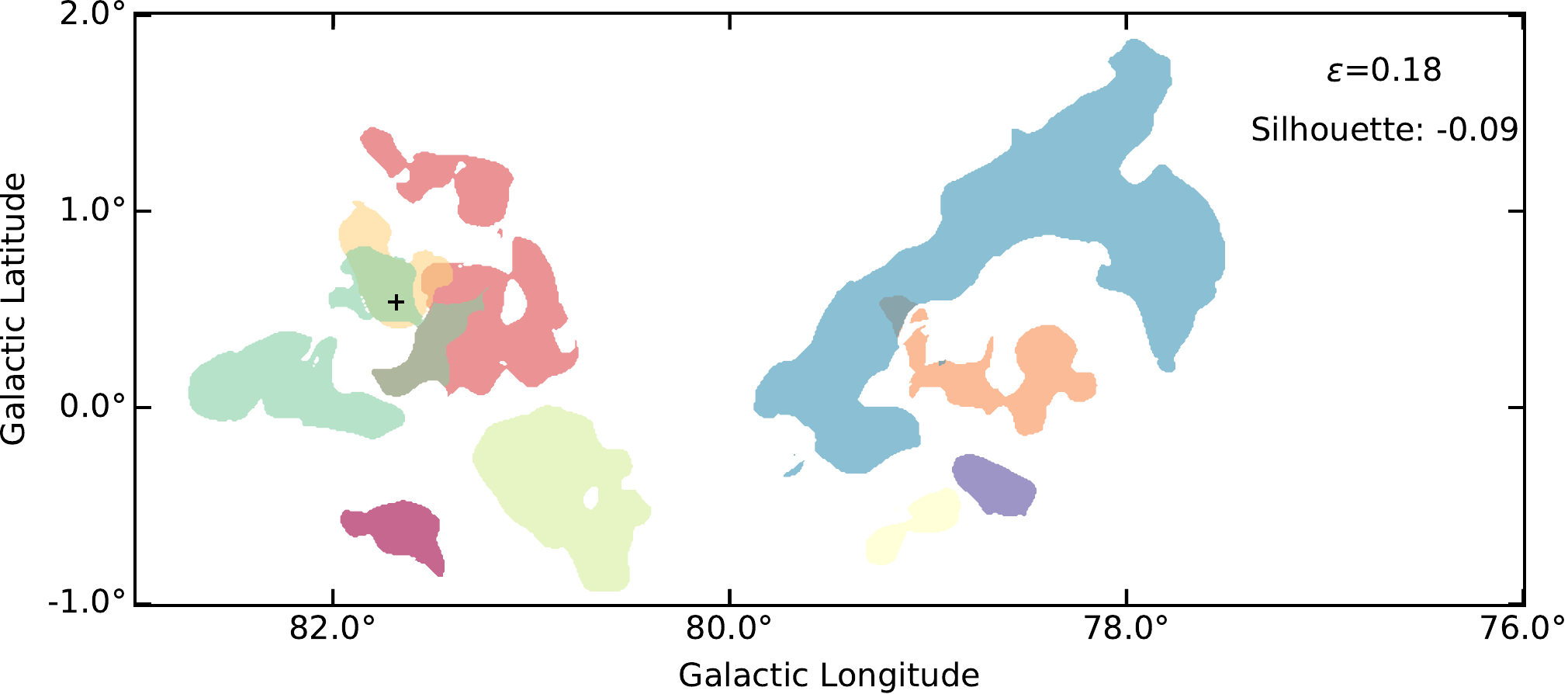}
\includegraphics[width = 0.49 \textwidth]{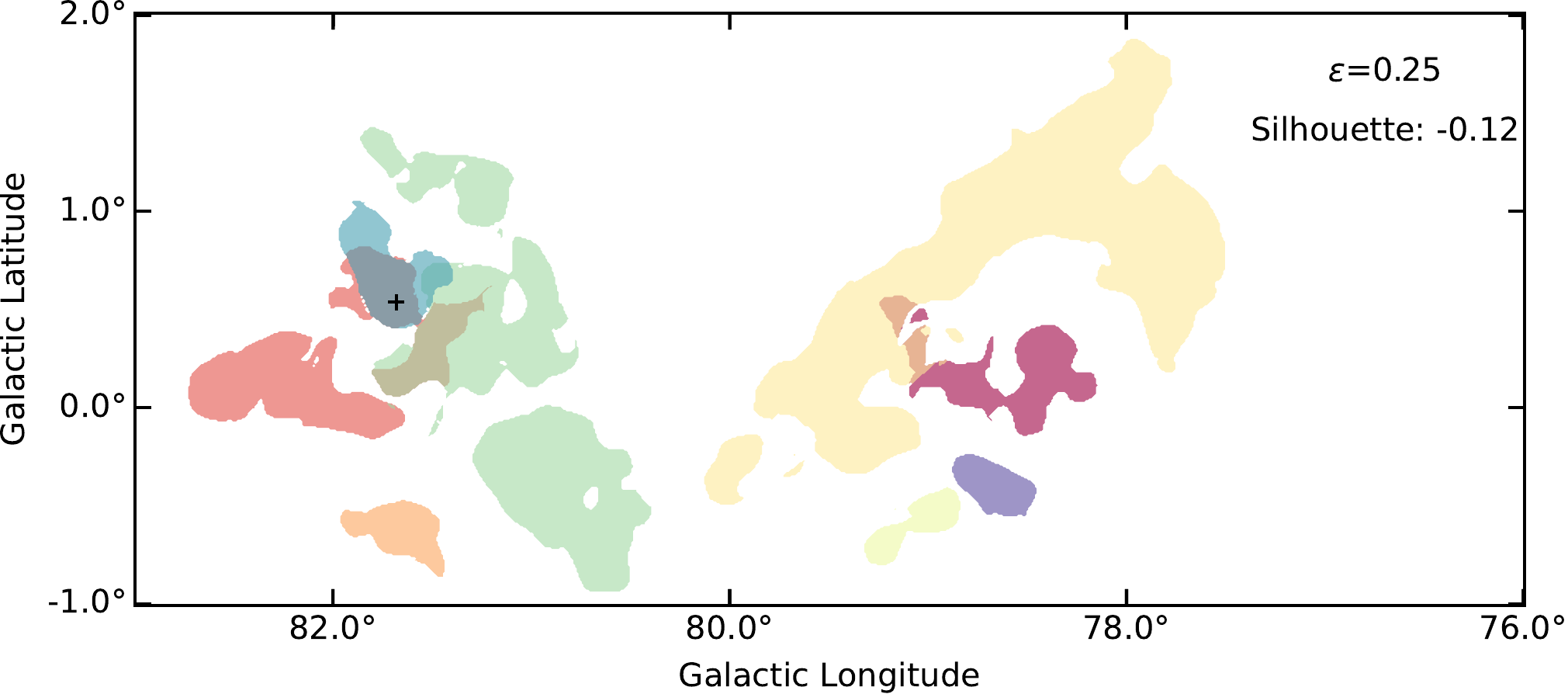}
\includegraphics[width = 0.49 \textwidth]{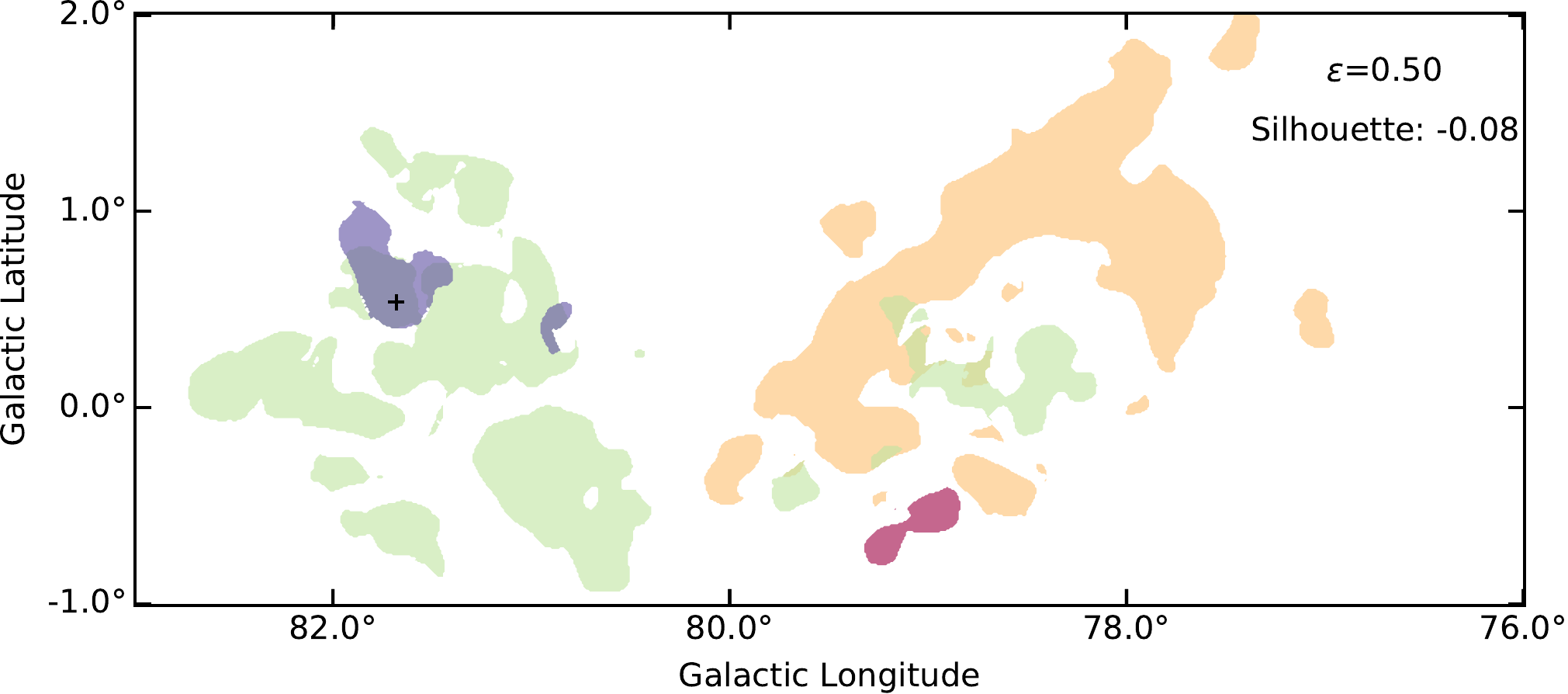}
\includegraphics[width = 0.49 \textwidth]{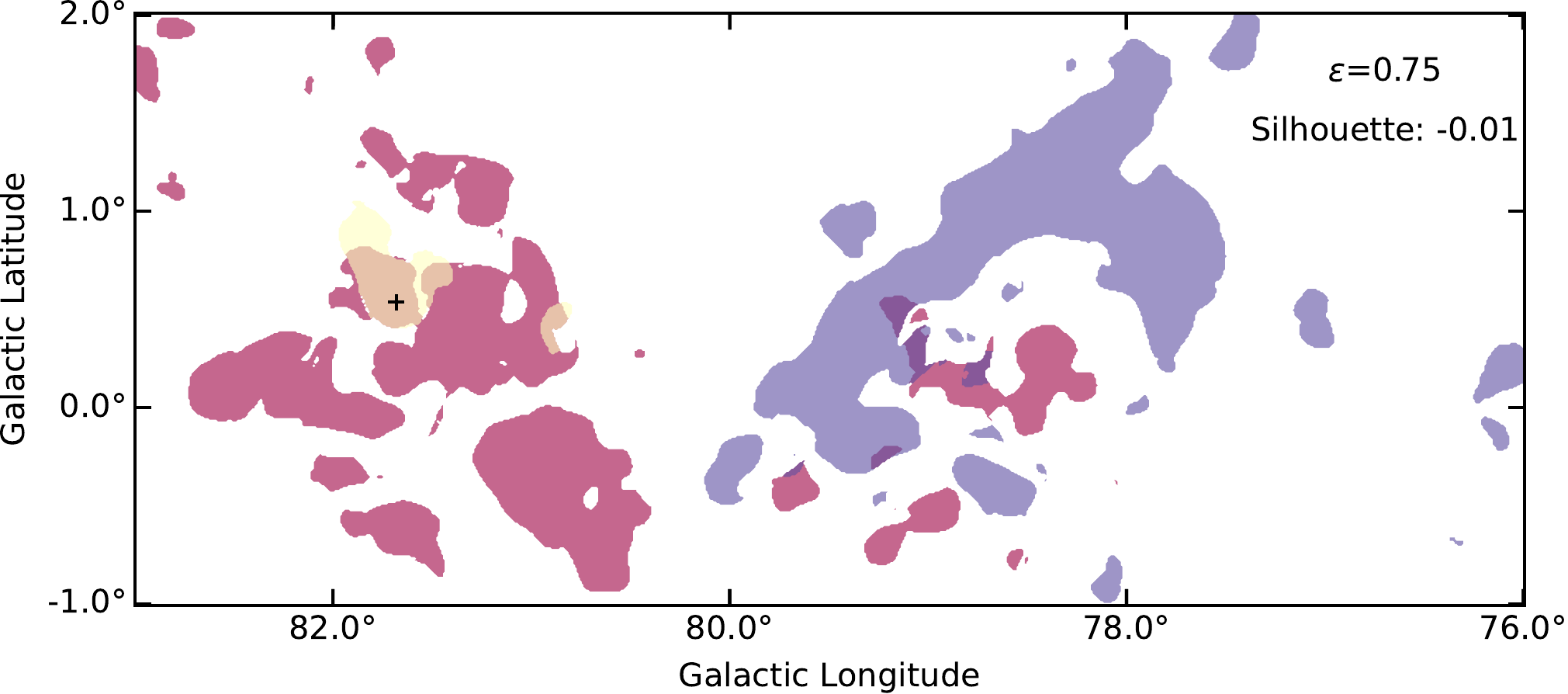}
\caption{{Cloud structures identified by the DBSCAN algorithm with different $\epsilon$ values. The $\epsilon$ value and silhouette score are indicated in the upper right corner of each panel. The different structures are marked with different colors.}\label{Fig:epsilon}}
\end{figure*}

Since the clustering results of the DBSCAN algorithm depend on $\epsilon$, we study the influences of the varied $\epsilon$ on the clustering results here. We run the algorithm with different values of $\epsilon$ to investigate the clustering results. As in Sect.~\ref{sec.decomp}, we discard the clustering structures with areas of less than three beams (10$\rlap{.}$\arcmin8). The results are presented in Fig.~\ref{Fig:epsilon}. Comparing the results, we find that the clustering method tends to result in more extended structure with higher $\epsilon$. Usually, higher silhouette scores indicate better results. We find that the silhouette scores converge to 0.37 for $\epsilon \gtrsim$1.25 when only one cloud structure is identified. Since the clustering results are mainly used to separate the line-of-sight velocity components to derive the local velocity gradients (see Sect.~\ref{dis:vg}), the clustering results of the high $\epsilon$ values are not suitable for our cases. Hence, we have to select the clustering results manually based on Fig.~\ref{Fig:epsilon}. A few large structures are missing in the clustering results with $\epsilon \leq$0.18, while many discrete and small structures with areas of less than three beams emerge for $\epsilon \geq$0.50. Furthermore, the line-of-sight overlapped velocity structure around $l=$81.5\degr, $b$=0.3\degr\,are well separated by $\epsilon=$0.25 but not by $\epsilon=$0.5.
In order to study the velocity fields of extended structures, we choose $\epsilon$=0.25 for our study. The local velocity gradients are exactly the same for the same pixels in the different clustering results except for boundary pixels which only make negligible difference in the velocity gradient maps. On the other side, the mainly affected results are the cloud areas, but the cloud areas are not used to reach any conclusion. Therefore, we conclude that our results with $\epsilon=$0.25 in this work should be valid. 

\section{The relationship between H$_{2}$CO (1$_{1,0}$--1$_{1,1}$) and $^{13}$CO (1--0)}\label{app.h2covsco}
\begin{figure}[!htbp]
\centering
\includegraphics[width = 0.45 \textwidth]{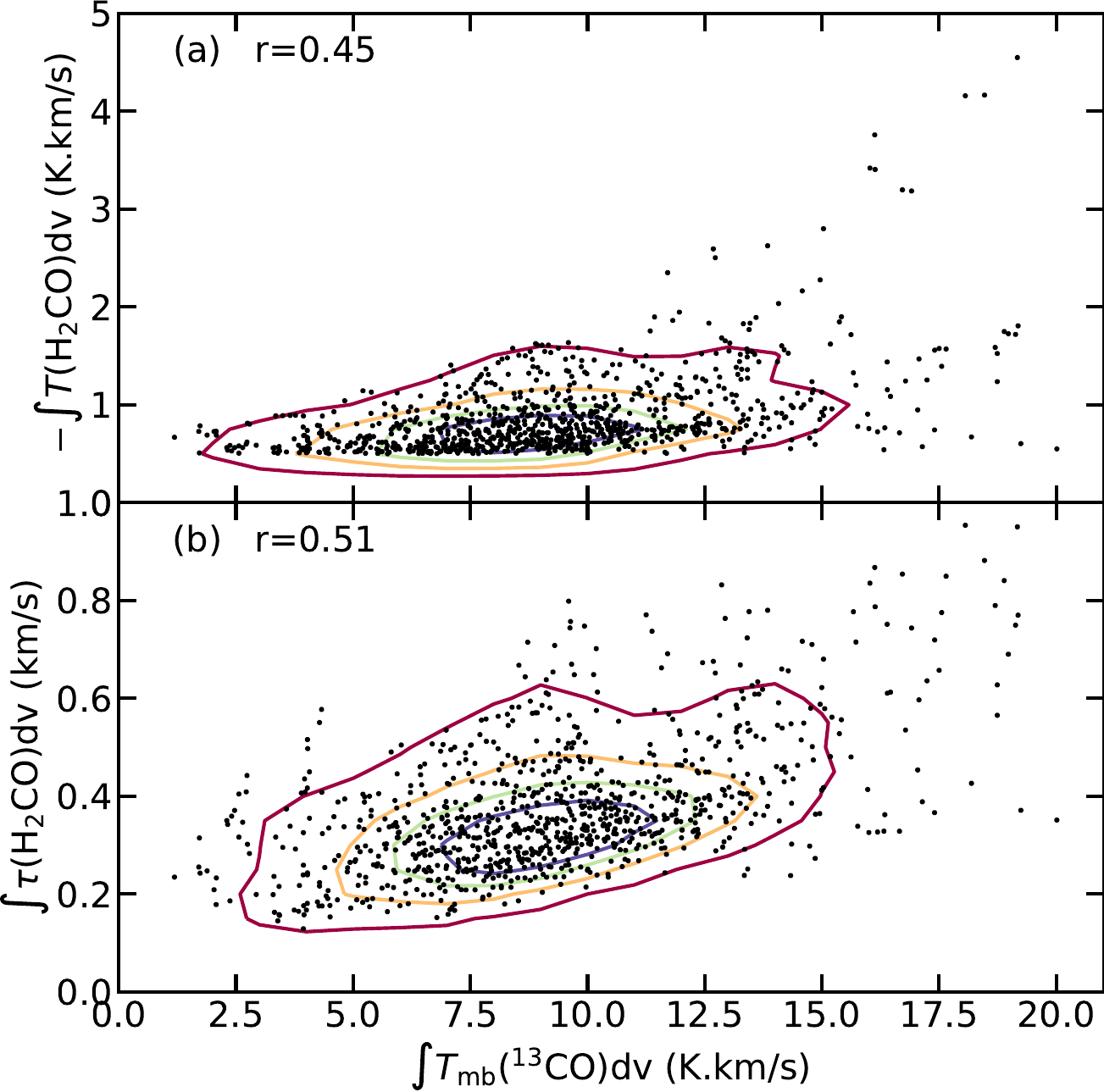}
\caption{{$^{13}$CO (1-0) integrated intensities as a function of the absolute values of the H$_{2}$CO integrated intensities (a) and the integrated optical depths of H$_{2}$CO (b).}\label{Fig:h2co-13co}}
\end{figure}

The relationship between H$_{2}$CO and CO isotopologues has been investigated by previous studies \citep[e.g.,][]{2013A&A...551A..28T,2014RAA....14..959T}. Our observations enable the comparison with a much larger sample size. Figure~\ref{Fig:h2co-13co} presents a comparison between the integrated properties of $^{13}$CO (1--0) and H$_{2}$CO (1$_{1,0}$--1$_{1,1}$). This comparison shows that the correlation coefficient (0.45) between $^{13}$CO (1--0) and H$_{2}$CO (1$_{1,0}$--1$_{1,1}$) integrated intensities is lower than the correlation coefficient (0.51) between $^{13}$CO (1--0) integrated intensities and the integrated optical depth of H$_{2}$CO (1$_{1,0}$--1$_{1,1}$). Furthermore, Figure~\ref{Fig:h2co-13co}a appears to show a higher degree of scattering than Figure~\ref{Fig:h2co-13co}b. Since the optical depths are derived by assuming that all radio continuum emission is located behind the molecular clouds, we speculate that most of the molecular gas should lie in front of the radio continuum emission. This is further supported by optical images that molecular clouds are seen as dark patches in Cygnus~X (see Fig.~1 in \citealt{2006A&A...458..855S} for instance). 

\section{Local velocity gradient maps}\label{app.vg}
\begin{figure*}[!htbp]
\centering
\includegraphics[width = 0.95 \textwidth]{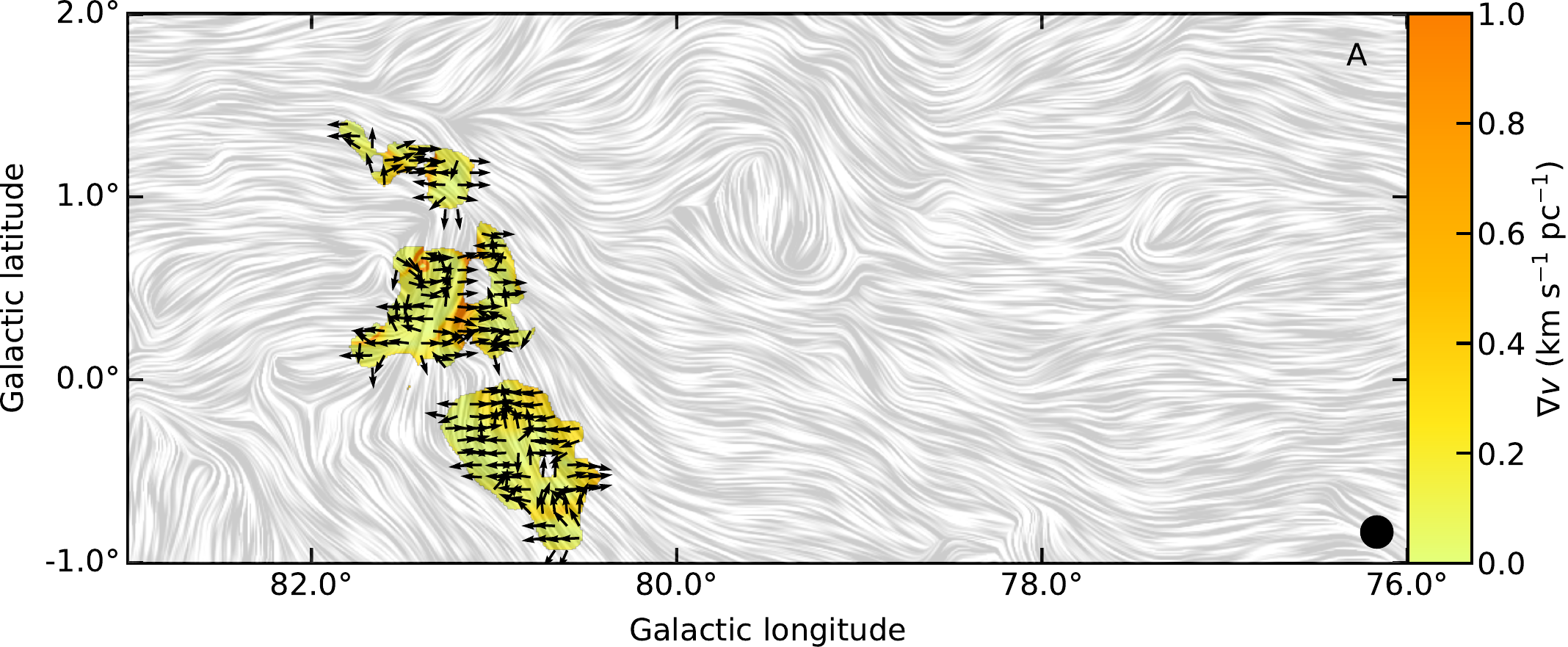}
\includegraphics[width = 0.95 \textwidth]{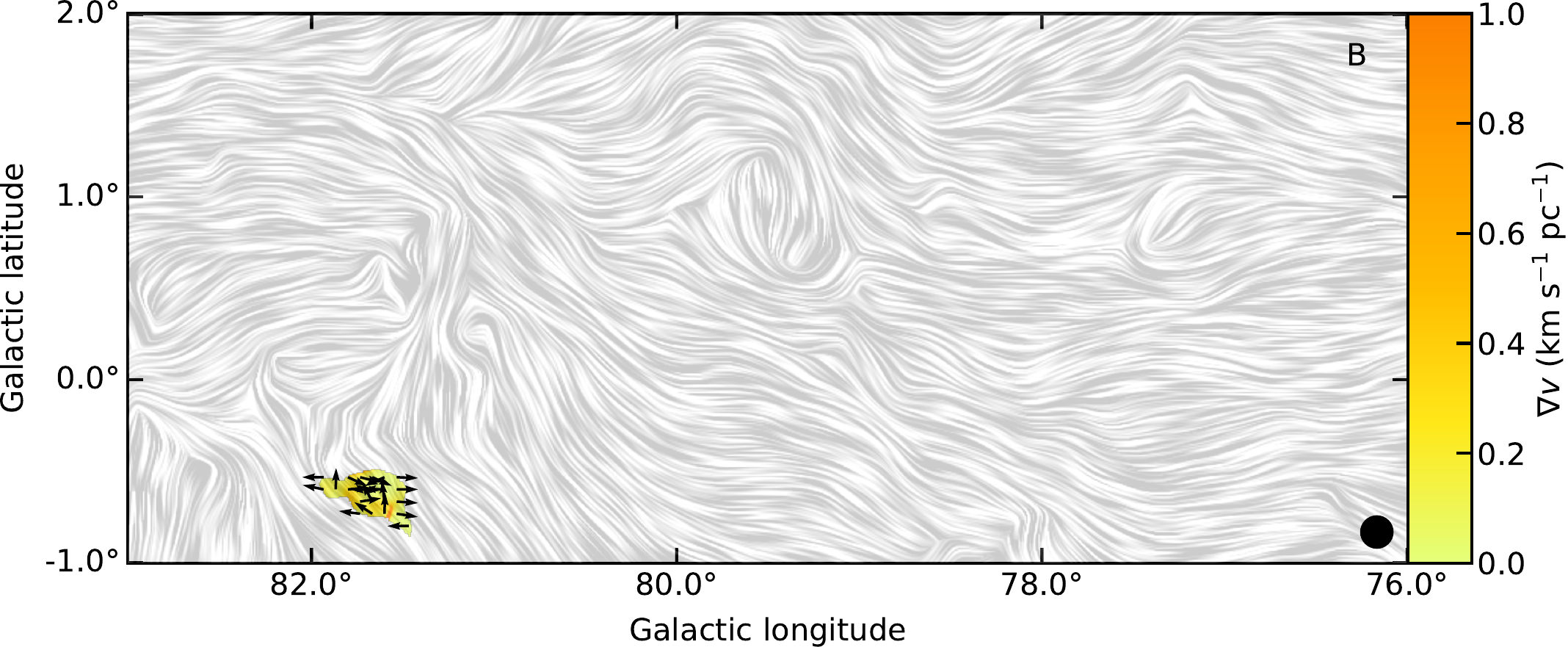}
\includegraphics[width = 0.95 \textwidth]{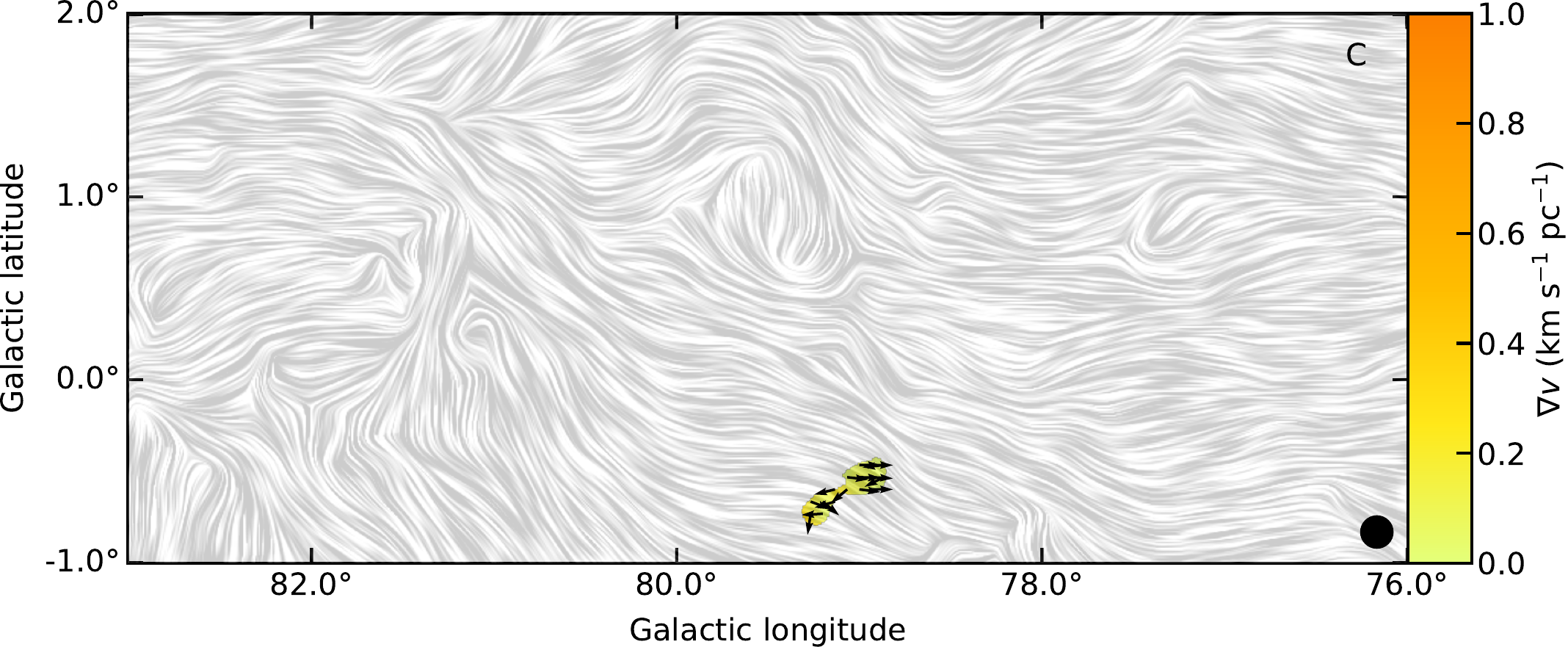}
\caption{{Same as Fig.~\ref{Fig:eff-vg} but for the other clouds.}\label{Fig:app-vg}}
\end{figure*}

The local velocity gradient map for cloud E has been presented in Sect.~\ref{sec.decomp}, and the local velocity gradient maps of the remaining seven clouds are shown in Fig.~\ref{Fig:app-vg}. 

\begin{figure*}[!htbp]
\centering
\includegraphics[width = 0.95 \textwidth]{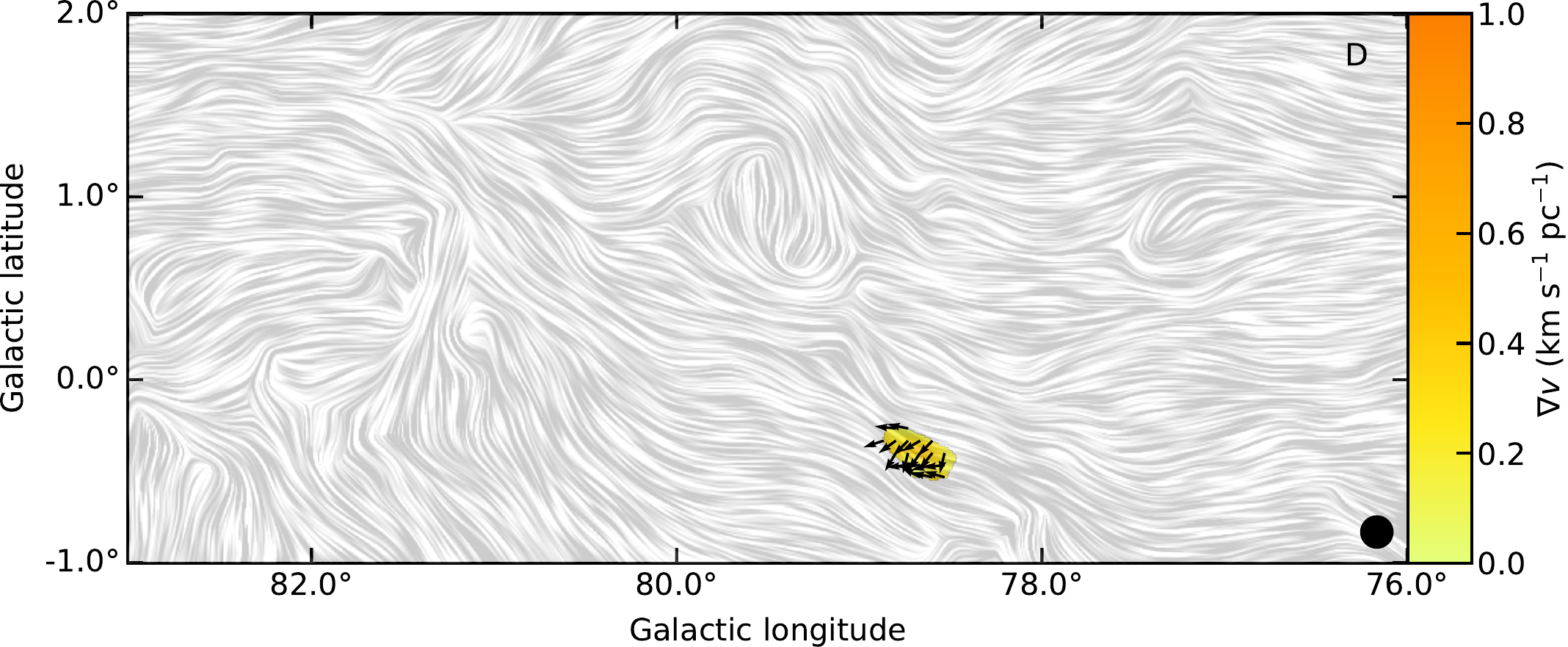}
\includegraphics[width = 0.95 \textwidth]{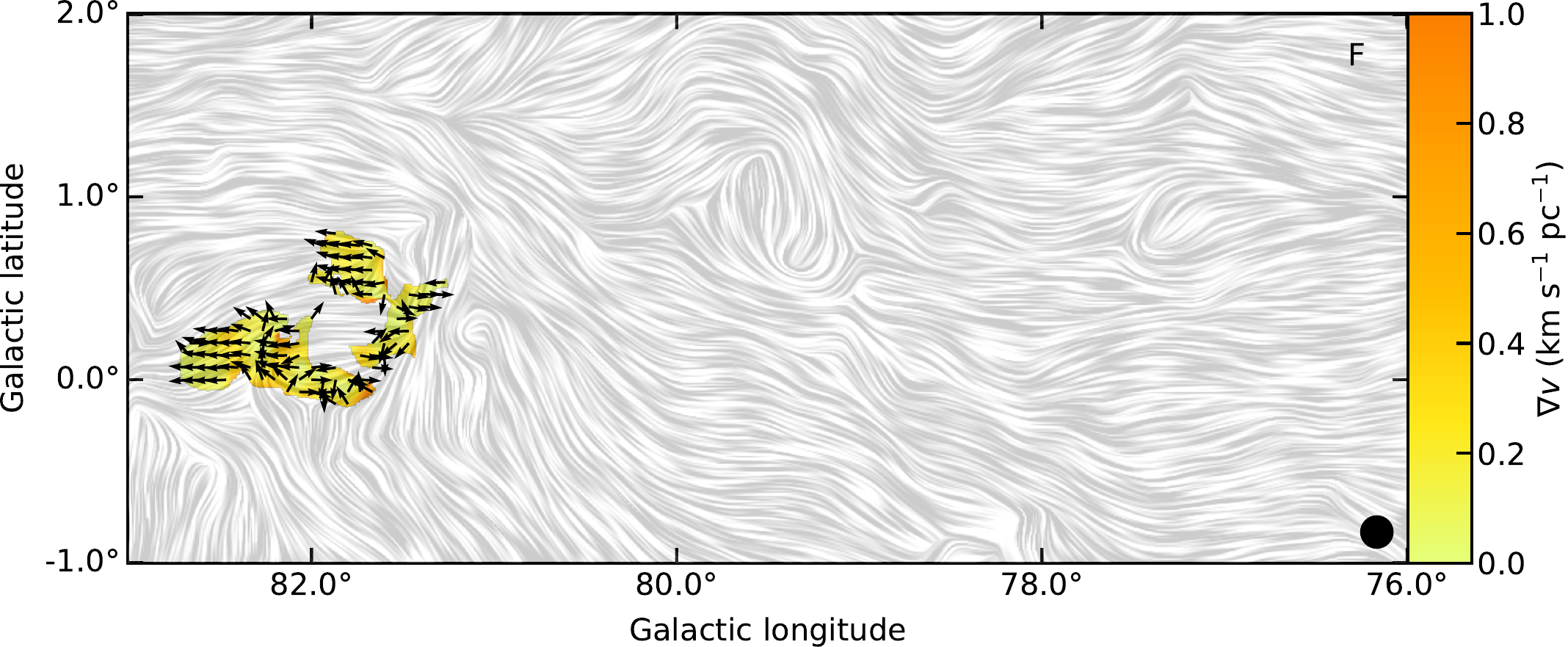}
\includegraphics[width = 0.95 \textwidth]{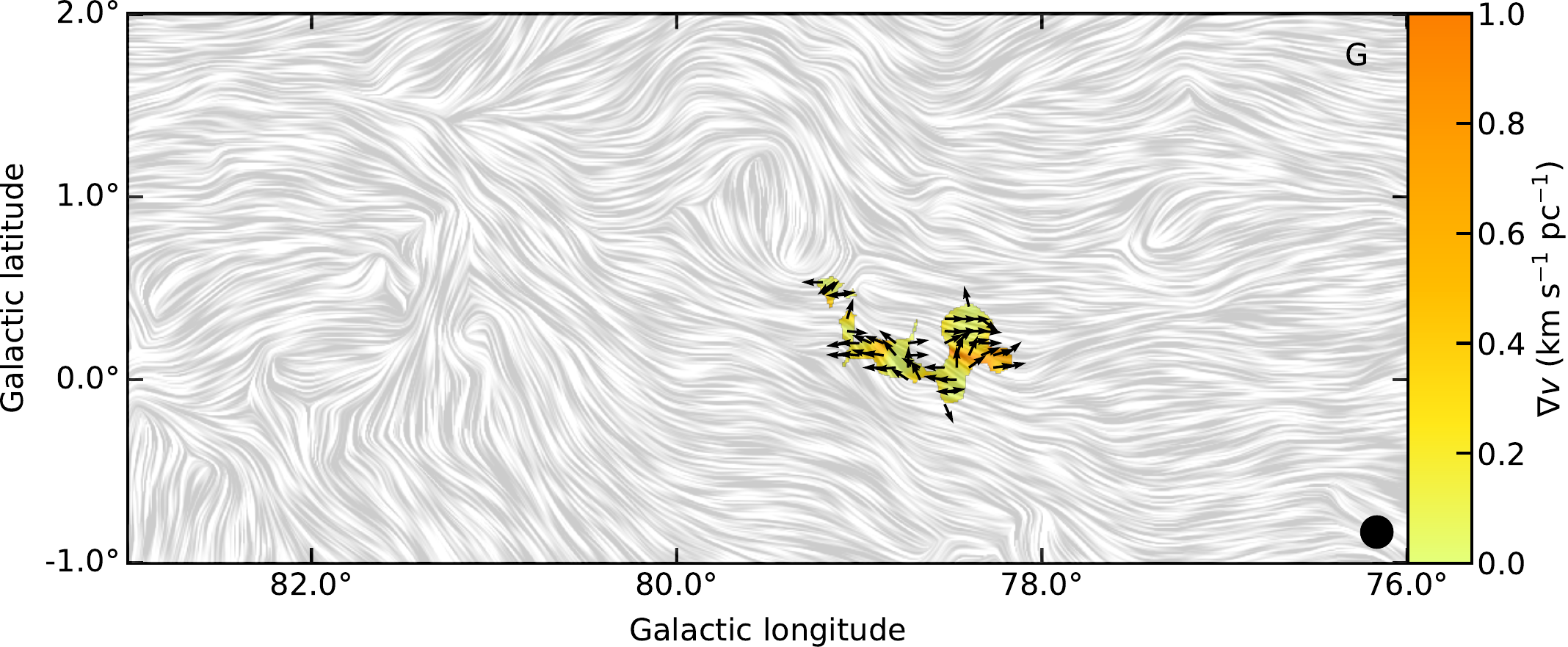}
\centerline{Fig.~\ref{Fig:app-vg}. --- Continued.}
\end{figure*}

\begin{figure*}[!htbp]
\centering
\includegraphics[width = 0.95 \textwidth]{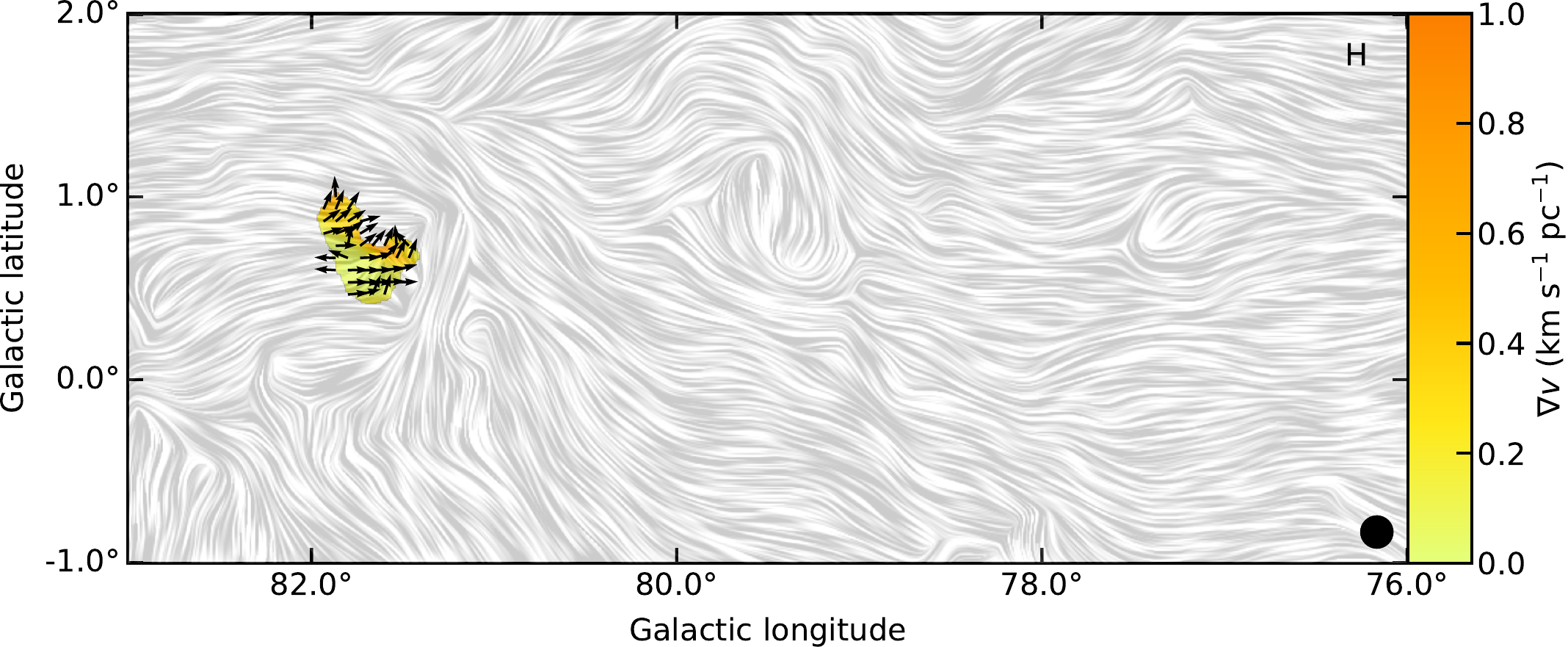}
\centerline{Fig.~\ref{Fig:app-vg}. --- Continued.}
\end{figure*}

\end{appendix}

\end{document}